\documentclass[fleqn,usenatbib]{mnras}

\usepackage{newtxtext,newtxmath}
\usepackage{orcidlink}

\usepackage[T1]{fontenc}

\DeclareRobustCommand{\VAN}[3]{#2}
\let\VANthebibliography\thebibliography
\def\thebibliography{\DeclareRobustCommand{\VAN}[3]{##3}\VANthebibliography}

\usepackage{graphicx}	
\usepackage{amsmath}	

\usepackage{hyperref}   
\usepackage{longtable}  

\usepackage{tcolorbox}  

\usepackage{xcolor, colortbl}

\newcommand{\cellr}{\rowcolor{lightgray}}  

\newcommand{\HI}{H\textsc{i}}  

\newcommand{\bea}{\begin{eqnarray}}
\newcommand{\eea}{\end{eqnarray}}

\usepackage{caption}      
\usepackage{subcaption}
\usepackage{multirow}

\title[RFI from RNSS in MeerKLASS L-band]{Radio Frequency Interference from Radio Navigation Satellite Systems: simulations and comparison to MeerKAT single-dish data}

\author[B. N. Engelbrecht et al.]{Brandon N. Engelbrecht,$^{1}$\thanks{E-mail: engelbrechtbn@gmail.com}
Mario G. Santos,$^{1,2}$
José Fonseca,$^{3,4,1}$
Yichao Li,$^{5,1}$
\newauthor Jingying Wang,$^{6,1}$
Melis O. Irfan,$^{7,1}$
Stuart E. Harper,$^{8}$
Keith Grainge,$^{8}$
Philip Bull,$^{8,1}$
\newauthor Isabella P. Carucci,$^{9,10}$
Steven Cunnington,$^{8}$
Alkistis Pourtsidou,$^{11,12,1}$
Marta Spinelli,$^{13,1}$
\newauthor Laura Wolz$^{8}$
\\
$^{1}$Department of Physics \& Astronomy, University of the Western Cape, Cape Town 7535, South Africa\\
$^{2}$South African Radio Astronomy Observatory (SARAO), 2 Fir Street, Cape Town, 7925, South Africa\\
$^3$Instituto de Astrof\'isica e Ci\^encias do Espa\c{c}o, Universidade do Porto CAUP, 4150-762 Porto, Portugal\\
$^4$Departamento de F\'isica e Astronomia, Faculdade de Ci\^{e}ncias, Universidade do Porto, Rua do Campo Alegre 687, PT4169-007 Porto, Portugal\\
$^{5}$Department of Physics, College of Sciences, Northeastern University, Wenhua Road, Shenyang, 11089, China\\
$^{6}$Shanghai Astronomical Observatory, Chinese Academy of Sciences, 80 Nandan Road, Shanghai, 200030, China\\
$^{7}$Institute of Astronomy, Madingley Road, Cambridge CB3 0HA, UK\\
$^{8}$Jodrell Bank Centre for Astrophysics, Department of Physics and Astronomy, The University of Manchester, Manchester M13 9PL, UK\\
$^{9}$INAF - Osservatorio Astronomico di Trieste, Via G.B. Tiepolo 11, 34131 Trieste, Italy \\
$^{10}$IFPU - Institute for Fundamental Physics of the Universe, Via Beirut 2, 34151 Trieste, Italy\\
$^{11}$Institute for Astronomy, The University of Edinburgh, Royal Observatory, Edinburgh EH9 3HJ, UK\\
$^{12}$Higgs Centre for Theoretical Physics, School of Physics and Astronomy, The University of Edinburgh, Edinburgh EH9 3FD, UK\\
$^{13}$Observatoire de la Côte d’Azur Laboratoire Lagrange, Bd de l’Observatoire, CS 34229, 06304 Nice cedex 4, France
}
\date{Accepted XXX. Received YYY; in original form ZZZ}

\pubyear{2023}

\begin{document}
\label{firstpage}
\pagerange{\pageref{firstpage}--\pageref{lastpage}}
\maketitle

\begin{abstract}

    Radio Frequency Interference (RFI) is emitted from various sources, terrestrial or orbital, and create a nuisance for ground-based 21cm experiments. In particular, single-dish 21cm intensity mapping experiments will be highly susceptible to contamination from these sources due to its wide primary beam and sensitivity. This work aims to simulate the contamination effects emitted from orbital sources in the Radio Navigational Satellite System within the 1100-1350 MHz frequency. This simulation can be split into two parts: (I) satellite positioning, emission power, and beam response on the telescope and (II) fitting of the satellite signal to data in order to improve the original model. We use previously observed single dish MeerKAT L-band data which needs to be specially calibrated to include data contaminated by satellite-based RFI. We find that due to non-linearity effects, it becomes non-trivial to fit the satellite power. However, when masking regions where this non-linearity is problematic,  we can recreate the satellite contamination with high accuracy around its peak frequencies. The simulation can predict satellite movements and signal for past and future observations, which can help in RFI avoidance and testing novel cleaning methods. The predicted signal from simulations sits below the noise in the target cosmology window for the L-band (970 - 1015 MHz) making it difficult to confirm any out-of-band emission from satellites. However, a power spectrum analysis shows that such signal can still contaminate the 21cm power spectrum at these frequencies. In our simulations, this contamination overwhelms the auto-power spectrum but still allows for a clean detection of the signal in cross-correlations with mild foreground cleaning. Whether such contamination does exist one will require further characterization of the satellite signals far away from their peak frequencies.

\end{abstract}




\section{Introduction}

    The integrated emission of the 21cm line of neutral hydrogen (\HI) can be used to map the Large Scale Structure (LSS) of the Universe.
    The method, known as intensity mapping (IM), allows us
    to scan large areas of the sky without the need to resolve individual galaxies \citep{Bharadwaj_2001, Chang_2008, Chang_2010, Masui_2013, Bull_2015}. Detecting this faint cosmological signal can be hindered by radio frequency interference (RFI) emitters which, 
    due to the technological evolution of communication devices \citep{Bentrum_2010}, have been increasingly contaminating the radio bands.
    These sources consist of terrestrial RFI, such as frequency modulation (FM) broadcasting towers, and orbital RFI, which emanates from the satellites \citep{Baan_2010}. 
    The main techniques applied to mitigate the eﬀects of RFI have been: avoidance, ﬁltering, and ﬂagging  \citep{Baan_2019}. 
    Radio telescopes are planned and
    operate within a designated radio quiet region, which aids in restricting contamination. However, as much as we locate radio facilities away from terrestrial RFI, 
    satellites in orbit around the Earth will always contaminate the signal and become
    one of its largest contaminants. This is of particular worry as satellites transmitting in the L-band frequency range between 1 and 2 GHz, contaminating regions of the spectrum important for {\HI} IM.
    The situation will only worsen due to the increasing number of satellites being launched annually \citep{itu_2021}. Real-time filtering schemes can prevent the leakage of strong RFI signals into the frequencies of interest before correlation \citep[e.g.]{ugmrt_buch_2023}. 
    The usual approach, however, is to flag (e.g. remove) post-correlation data contaminated by RFI \citep{Offringa_2012}.

As radio telescopes grow more sensitive and the sky becomes more contaminated with satellites, it becomes crucial to fully understand the radio signal emitted by them. Such a study serves three purposes: 1) to find parts of the sky and times when observations are less contaminated; 2) to understand how much the RFI contaminates the target frequencies for our science; and 3) to test if there are ways to further clean the signal from RFI on those target frequencies (away from the prominent peaks). There are several examples of follow-ups on satellite emissions and their impact on observations. The Green Bank Telescope (GBT)\footnote{\url{https://greenbankobservatory.org/rfi-scans-and-known-sources/}} has been characterizing the sources of RFI at different frequency bands. The Five Hundred Metre Telescope (FAST)\footnote{\url{https://fast.bao.ac.cn/}} has looked at estimating the power of various satellites by using an extra antenna to monitor nearby satellites \citep{yu_wang_2021}. The Parkes\footnote{\url{https://www.parkes.atnf.csiro.au/}} ultra-wide bandwidth receiver \citep{Hobbs_2020} contained navigational and telecommunication satellites within their observation. A detailed study of the MeerKAT RFI environment as a function of direction, frequency, time of day and baseline has also been done \footnote{\url{https://skaafrica.atlassian.net/wiki/spaces/ESDKB/pages/305332225/Radio+Frequency+Interference+RFI}}. 
    
Satellite emissions can be particularly adverse for single-dish 
HI IM surveys such as those planned for BINGO \citep{10.1093/mnras/stt1082} and FAST \citep{bigot_2016} as well as the MeerKAT \citep{Santos2017} and SKAO-MID \citep{Santos2015} arrays when used as a collection of single dishes. Although interferometers are also affected, the signal is further reduced due to time and frequency smearing (decorrelation) away from the delay centre. The more direct mitigation strategy is to consider frequencies far away from the allocated satellite transmission band.  
These allocations are set by the International Telecommunications Union (ITU), but 
there is still out-of-band transmission since the ITU only provides guidelines for the maximum out-of-band power that can leak out of these allocations \citep{itu_2004,itu_2015}. Given that the {\HI} 
 cosmological signal is less than a mK, one needs to understand how satellite RFI propagates outside the allocated bands, and gauge if their contamination is negligible.
 
With the cosmological surveys planned for MeerKAT and the Square Kilometre Array Observatory (SKAO), it is becoming increasingly important to study the impact of this emission at the SKAO site in South Africa. 
The situation should be better for surveys using MeerKAT's UHF band (580 -  1015 MHz) and SKAO-MID band 1 (350 - 1050 MHZ), but it will be more problematic for MeerKAT's L-band (900 - 1670 MHz) and SKAO-MID band 2 (950 - 1760 MHz). \cite{Harper2018} estimated the impact of global navigational satellite systems (GNSS) on a future SKAO-MID band 2 HI IM survey, using models for the total power and spectral structure of GNSS signals convolved with a model SKA beam. They conclude that for frequencies > 950 MHz, the emission from GNSS satellites will exceed the expected {\HI} signal for all angular scales. However, information is quite sparse on the technical specifications of satellites, namely
the signal's amplitude which introduces uncertainties in the modelling. Moreover, the impact on the final HI IM power spectrum depends on the scanning strategy used. Also, one could use cleaning methods to reduce the contamination level, or even subtract it as in \cite{finlay_2023}.

In this work, we will expand the analysis in \citet{Harper2018} (referred to as HD18 hereafter) using a complete catalog that incorporates radio navigational satellite systems (RNSS) extending beyond the focus of GNSS to include regional and augmented systems. We develop new simulations that are compared and fitted to 
calibrated MeerKAT data from \citet{Wang2020} ("W21" from now on). This data was used to make a recent detection of the cross-correlation power spectrum between {\HI} and galaxy surveys \citep{cunnington_2023b}. 
We focus on satellites with allocated emission bands below 2 GHZ, which should have the largest impact on the target cosmology window (970 - 1015 MHz). In total, we consider 73 satellites operating in MeerKAT's field-of-view during the scanning period of an observation commencing on 25-02-2019 02:40:11 SAST\footnote{South African Standard Time}. 
The new set of simulations can accurately estimate satellite contamination over a wide frequency range once properly calibrated with the data. In this work we will determine in which regimes we can calibrate and characterise the satellite signals. Additionally, we will use this simulation to understand the impact of satellites on the HI IM 3d power spectrum and test cleaning methods. These results will help in preparing observations with future surveys with MeerKAT and the SKAO and set the stage to understand the ultimate level of contamination on the science that will be extracted from such data.

The remainder of the paper is organized as follows: in \autoref{s:nvss}, we describe how we construct the navigational satellite signals and positions; in \autoref{s:oss}, we address the calibration of the observational data to include the previously flagged RFI region; in \autoref{s:results} we discuss the effectiveness of the model; in \autoref{s:application} we explore the applications of the model; and conclude in \autoref{s:con_fw}.

\section{Navigational Satellite Systems Simulation} \label{s:nvss}

 Artificial satellites address a host of objectives, ranging from communication, earth observation photography, navigational information, and so on. In this paper, we focus only on radio navigational satellite systems,
 including sub-categories such as global, regional, and augmented systems. The different systems within these sub-categories are usually referred to as constellations. These constellations cater to and have slight differences between civilian and military uses for a given country. 
 Currently, there are four global navigational satellite systems in orbit which provide positional information to users across the globe. These are the United States of America - NAVSTAR, referred to as \emph{GPS}, the European - Galileo system, which we will refer to as \emph{GAL}, the Russian - Global'naya Navigatsionnaya Sputnikovaya Sistema (GLONASS), which we will refer to as \emph{GLO}, and the Chinese Compass navigation system Beidou, which we will refer to as \emph{BEI}. 
 
On the other hand, Regional navigational systems operate across a specific country or fixed geography. Here we will consider the Indian Regional Navigation Satellite System NavIC, referred to as \emph{IRNSS} and the Japanese Quasi-Zenith Satellite System, referred to as \emph{QZS}. 
Another system incorporated into this work is the Satellite Based Augmentation System (SBAS).
This geostationary satellite system benefits the current GNSS by improving positioning accuracy. These constellations follow operational standards, such as allocated frequency bands, which follow the ITU recommendations \citep{itu_2004,itu_2015}. 

In \autoref{tab:sat_cat}, we list the navigational satellite systems considered in this work. This catalog contains information on the aforementioned constellations, including the relevant parameters required to model their spectral signal emission (see following sub-sections). We have used several sources in the literature to collect this information.
However some details are unavailable or quite uncertain, such as the satellite signal's transmitted power in some constellations. This has prompted us to allow some of these parameters to be free and later fitted to the data as explained in \autoref{ss:sat_power}.

In order to build a simulation of the observed satellite signal, the following information is needed:
\begin{itemize}
    \item The frequency range of the observation $\Delta\nu_j$ [MHz];
    \item The timestamps $t$ [sec] of the observation;
    \item Location of the telescope on the Earth [Lat \& Long.];
    \item Observational pointing per timestamp [RA \& Dec];
    \item an all-sky telescope beam model;
    \item positional information of the satellites
    \item the power emitted by the satellites as a function of frequency.
    \end{itemize}

The location information, telescope pointing, and beam model are ﬂexible and can be user-speciﬁc, allowing the simulation to be applied to various single-dish radio telescope experiments around the globe.
The following subsections explain in more detail the steps taken to simulate the observed satellite signal.

\subsection{Satellite position} \label{sec:sat_pos}

Each navigational constellation/system comprises several individual satellites. These individual satellites follow a predetermined course at a specific orbital distance. Although celestial mechanics could naively predict the position of the satellites, there are several effects that make this difficult to calculate over long time scales and up to date information is required. 
The positional vector information for each satellite
is contained within a two-line element (TLE) dataset on the CelesTrak\footnote{\url{www.celestrak.com}} database. The data is saved as snapshots for the various satellite constellations. This catalogue is updated every 24 hours and does not store previous TLE information. 
Historical TLE information from months to years back can be accessed via the Wayback Machine\footnote{\url{https://archive.org/web/}}. The Wayback Machine stores snapshot from various days but not concurrently. However, when comparing historical and present snapshots, we have found that a TLE can reliably describe a satellite position within 3 months. Therefore,
when selecting historical snapshots, we simply use the one closest to the 
observation dates.

Once the necessary TLE information has been acquired, we apply the python package \texttt{Skyfield}\footnote{\url{https://rhodesmill.org/skyfield/}} \citep{skyfield} to load the TLE vectors and retrieve the orbital positions of the individual satellite, such as latitude \& longitude; right ascension (RA) \& declination (DEC) or azimuth \& altitude (elevation) for a reference position on Earth. In our case, we took the position to be the central location of the MeerKAT array: Longitude: 21$^\circ$ 26$^{'}$ 38$^{''}$ E and Latitude: 30$^{\circ}$ 42$^{'}$ 47.41$^{''}$ S. Given that the diameter of the MeerKAT array is much smaller than the orbital height of the navigational satellites, each antenna will observe the satellites at approximately the same angular distance to the beam centre.


The final ingredient is the angular distance, $\theta$, between the satellite and the pointing direction of the telescope, which can be calculated through \citep{Duffett-Smith_Zwart_2017}:
\begin{equation}
    cos(\theta) = \rm \sin(al_p)\sin(al_s)+ \cos(al_p)\cos(al_s)\cos(az_p - az_s)\\
    \label{eq:angular_sep}
\end{equation}
where $\textnormal{az}$/$\textnormal{al}$ is the azimuth/altitude and the subscripts $\textnormal{p}$/$\textnormal{s}$ represent the telescope pointing and satellite position respectively. The telescope pointing can be provided from a scanning strategy or taken directly from the observation data files. The satellite position and telescope pointing are calculated for every time stamp in the scan (2-second resolution in this work).

\begin{figure}
                \centering
                \includegraphics[width=0.5\textwidth]{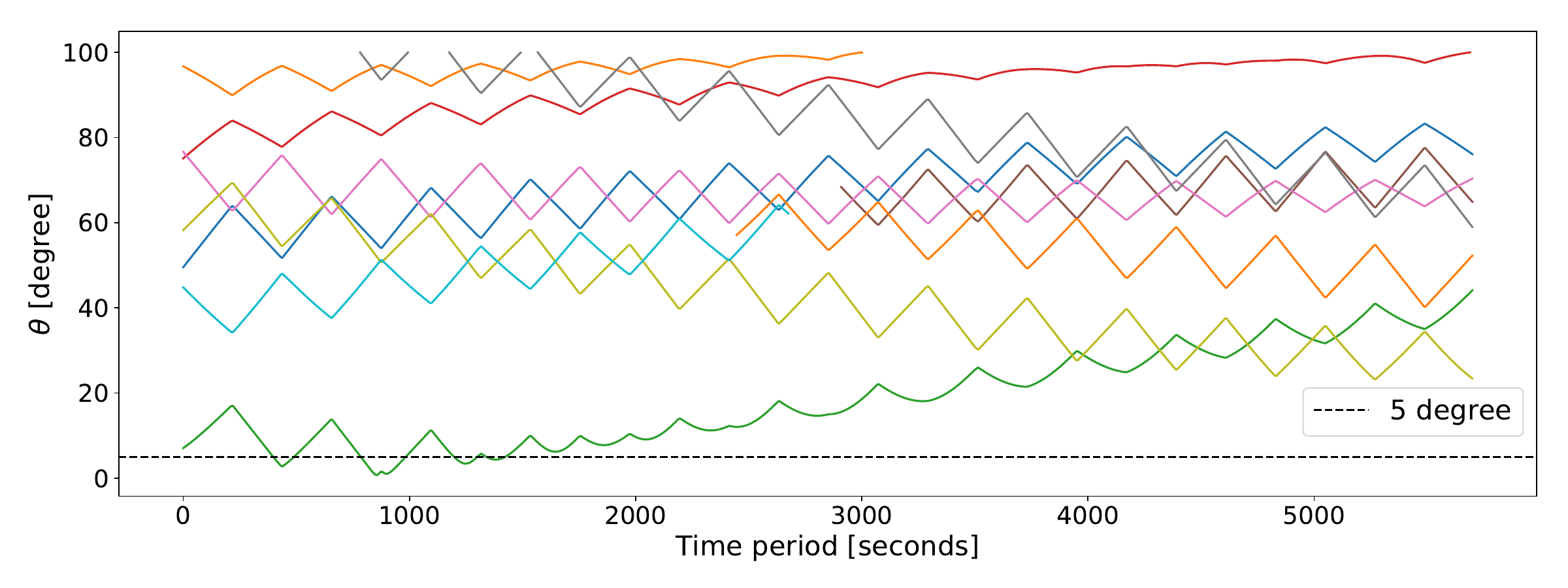}
                \caption{The angular distance to the telescope pointing direction (\autoref{eq:angular_sep}) for various satellites in the GALILEO constellation (solid lines represent). The black dashed line represents an angular distance of $5^{\circ}$.
                }
                \label{fig:ang_sep_gal}
            \end{figure}

 In \autoref{fig:ang_sep_gal}, we show the angular distances of various satellites within the GAL constellation for one of the observational scans taken from actual MeerKAT data. The seesaw structure is related to the fast scanning strategy adopted for this observation. Individual satellites can move below the telescope horizon and are then ignored. The appearance or disappearance of satellite information within the observation window illustrates this. We see that a single satellite from the GAL constellation passes directly through the pointing centre of the telescope.


 \begin{figure}
     \centering
     \includegraphics[width=0.5\textwidth]{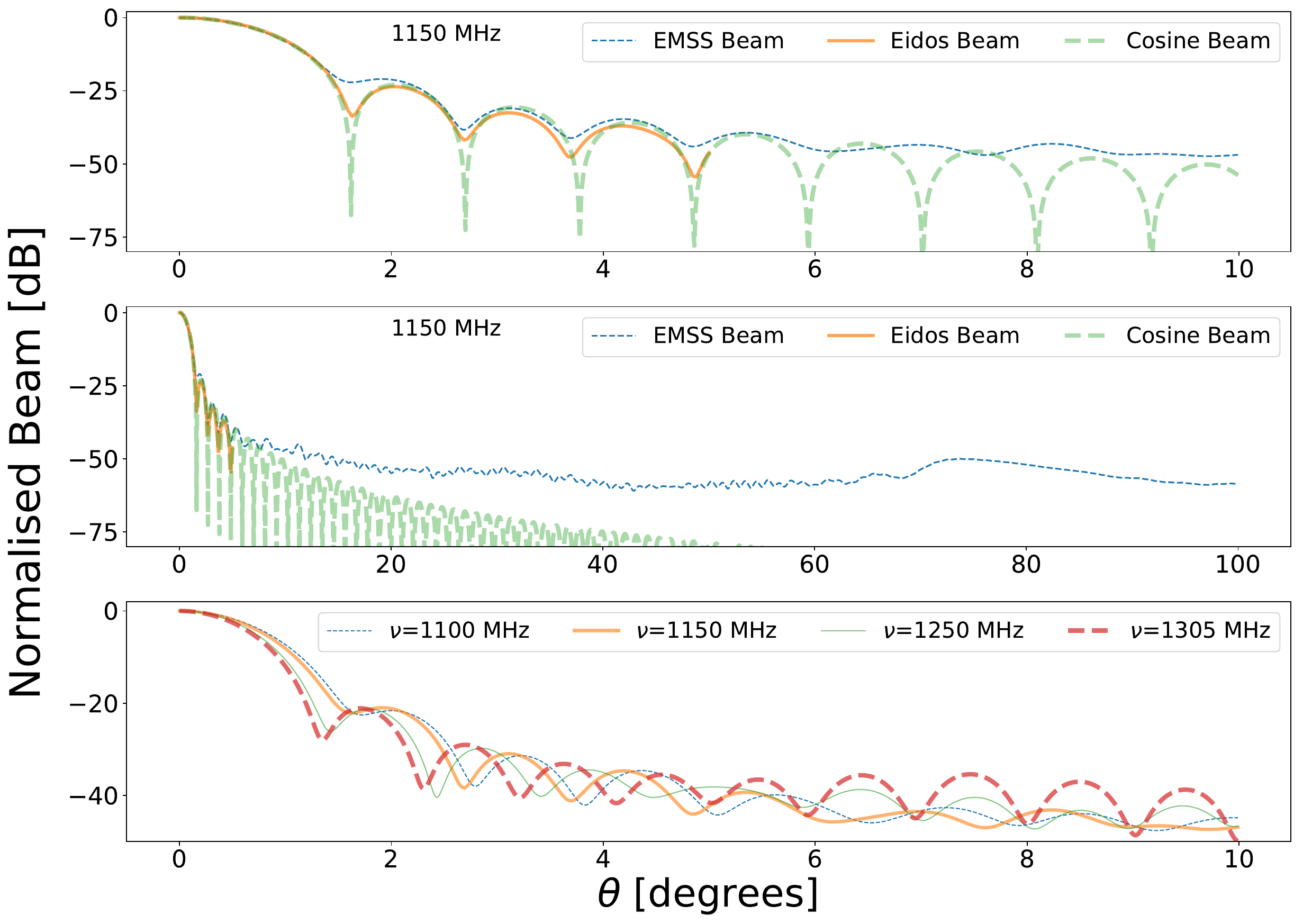}
    \caption{The different beam models available in the simulation. In the $1^{\textnormal{st}}\ \&\ 2^{\textnormal{nd}}$ panels we show the beam models: EMSS (blue thin dashed), Eidos (orange solid) and Cosine (green thick dashed) at 1150 MHz, with the 
     $2{\textnormal{nd}}$ panel up to
    higher angular separation ($\theta$). The $3^{\textnormal{rd}}$ panel shows the EMSS beam at different frequencies: 1100 MHz (blue thin dashed), 1150 MHz (orange thick solid), 1250 MHz (thin green solid), and 1305 MHz (red thick dashed).}
    \label{fig:beam_model_diff}
\end{figure}

\subsection{Beam response}
To effectively simulate the observed signal of the satellites present in the data, we require an accurate description of the MeerKAT beam response.    
An accurate beam model will inform us not only of the main lobe of the telescope but, more importantly, of the side lobe structure as well as its frequency dependence. In \cite{asad_2021}, the MeerKAT beam model was measured and fitted to a radius of $5^{\circ}$. However, satellite emission beyond this radius can still affect our observations \citep{Harper2018}. Therefore we decided to use a MeerKAT L-band beam model based on detailed electromagnetic simulations provided by the company EMSS Antennas\footnote{\url{https://www.emssantennas.com/}}, which offers the response of the MeerKAT antenna in the horizontal HH and vertical VV polarizations. These simulations cover a frequency range of 900-1670MHz and an angular range of $0^{\circ}-100^{\circ}$. Above that, we set the beam response to zero. 

Note that we only consider the spherically averaged beam, so there is only dependence on the angular distance from the beam centre (the polar angle, "$\theta$"). In reality, beams are asymmetrical. The MeerKAT beam was measured to deviate from this symmetry at less than $1\%$ within the main lobe, although the deviation can become much larger at the second sidelobe and above \citep[see][]{asad_2021, 2023AJ....165...78D}.
The provided EM simulations have an angular resolution of $0.1^{\circ}$ and a frequency resolution of 2 MHz, so interpolation was done both in angle and frequency.
The code also provides the holographic beam model from \cite{asad_2021}, which we refer to as "Eidos", and the analytical "Cosine beam" which is a reasonable fit to the MeerKAT beam \citep{2020ApJ...888...61M}. The top and middle panel of \autoref{fig:beam_model_diff} shows a comparison between these models. We see that above $10^{\circ}$ the Cosine model starts deviating from the simulation. 

In this paper, we only consider total intensity measurements. Although satellite emission can be polarised, modelling that is difficult and so we combine both polarisations both in the simulation and in the observed data (usually called horizontal, HH, and vertical VV, polarisations). The beam model is also normalised to be one at its maximum.
The bottom panel of \autoref{fig:beam_model_diff} highlights the EMSS beam model as a function of the angular separation for different frequencies.

\begin{figure*}
            \centering
            \includegraphics[width=0.9\textwidth]{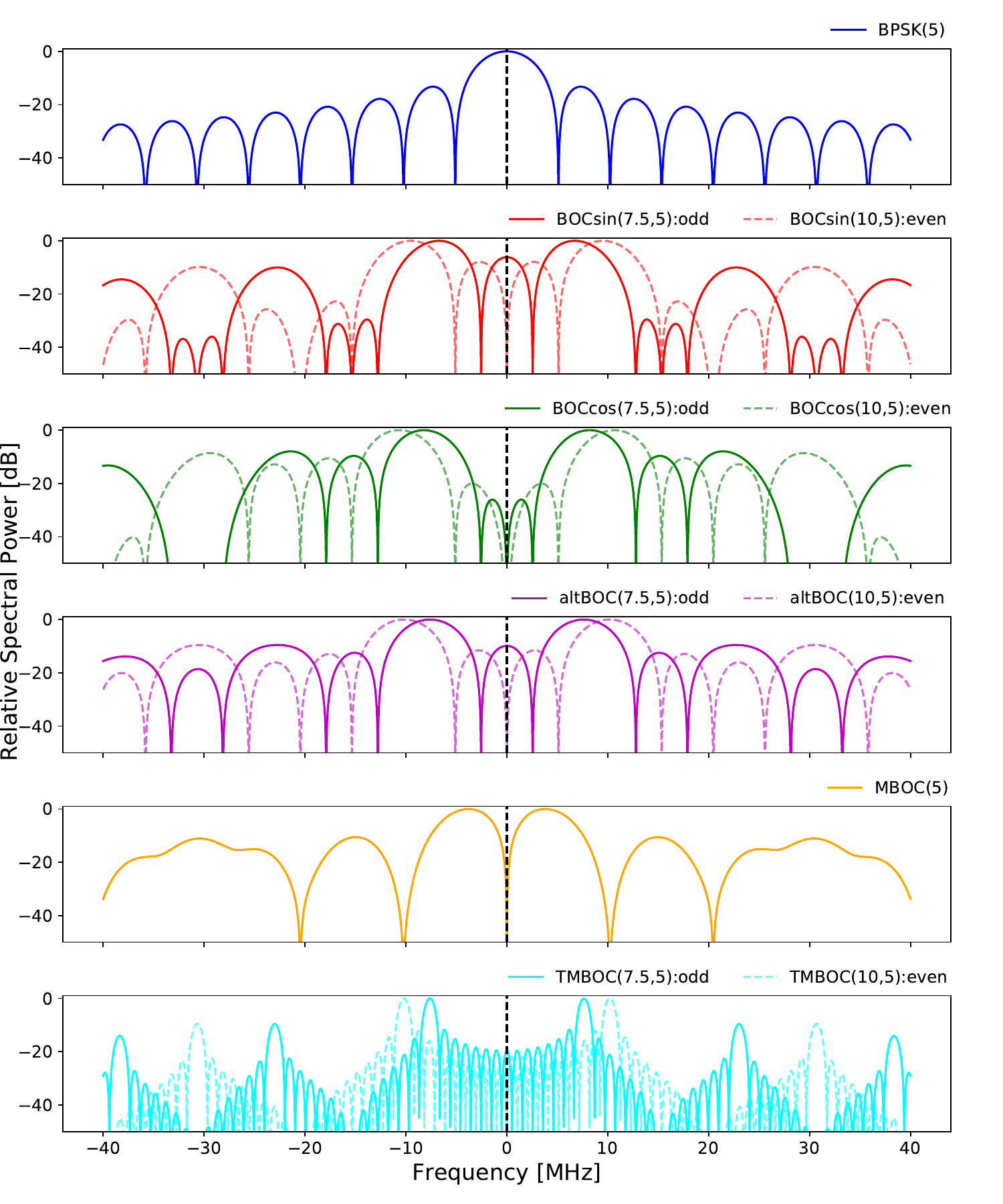}
            \caption{Normalised Power Spectrum Density of transmission patterns. 
            Solid lines represent odd values of $n$ and dashed lines even ones. 
            Starting from the top panel : BPSK; sine BOC; cosine BOC; altBOC; MBOC and TMBOC. {In \autoref{tab:sat_cat}, we highlight the various modulation types employed by the GNSS.}}
            \label{fig:psd}
        \end{figure*}

    \subsection{Power-Spectrum Density} \label{ss:psd}
        The power spectrum density (PSD) is the spectral response of 
        the transmitted signal by the satellite. The signal along frequency can be described as a combination of sine and cosine waves, which we refer to as the modulation of the signal. 
        The simplest of these modulations is a constant pulse known as binary phase shift keying (BPSK), which, when Fourier transformed, is described by:   
        \begin{equation}
            P_{\textnormal{BPSK}_{(n_c)}}(\nu) = \dfrac{\textnormal{sinc} \big( \pi (\nu - \nu_c) /[n_c f_0] \big)}{\sqrt{n_c f_0}},
            \label{eq:bpsk}
        \end{equation}
        where $f_0$=1.023 MHz is the reference frequency, $n_c$ is the chip rate, which refers to the rate at which a phase change occurs, and $\nu_c$ is the central frequency of the signal.
       
        The signal's PSD is then given by the auto-correlation \citep{g._montenbruck_2017_modulation}:
        \begin{equation}
            \textnormal{S}_{\textnormal{xx}}(\nu) = |P_{\textnormal{xx}}(\nu)|^2\,,
        \end{equation}
        where $xx$ is the modulation type {and $\textnormal{S}_{\textnormal{xx}}(\nu)$ is normalised to integrate to 1 over frequency}.
        In \autoref{fig:psd} we show other, more complex 
        signal modulation types 
        used in the simulation (see appendix~\ref{app:psd}).
        Currently, our model can simulate binary-phase shift keying (BPSK); binary offset carrier (BOC); alternative BOC (altBOC), multiplexed BOC (MBOC) and time-MBOC (TMBOC). 
        We list all the modulations used in this paper and the variables needed for them in table~\ref{tab:sat_cat}.
        These signals will be present at all frequencies, although the out-of-band emission will be smaller. It is possible that there is extra dampening of the signal away from the main peak but such information is not available and so we assume these modulations are a good representation of the signal at all frequencies. 
        Estimating if the level of out-of-band emission is correct is one of the goals of this paper and will be addressed in a later section.

    \subsection{Signal temperature} \label{ss:sat_power}
    
        Each satellite can emit a combination of different signals with specific modulations as described in the previous section. For each of these signals, $i$, the observed brightness temperature from one satellite is calculated as:
        \begin{equation}
            \textnormal{T}_{\textnormal{sat, i}}(t, \nu) = \textnormal{B}(\theta, \nu)\  \textnormal{S}_{xx_i}(\nu) \dfrac{\textnormal{P}_i}{{r}^2} \dfrac{c^2}{4 \pi \nu^2 k_b}\,,
            \label{eq:sat_tod}
        \end{equation}
        where $\nu$ is the observed frequency, $k_b$ is the Boltzmann constant and $c$ the speed of light in the vacuum. The beam response is given by $B(\theta, \nu)$, where again, $\theta(t)$ represents the angular separation between the satellite position and the pointing centre of the telescope and is a function of time.
        The radial distance between satellites and the telescope on Earth $r(t)$ is also a function of time and is calculated with the \emph{Skyfield} package \citep{skyfield}. Because the channel width for MeerKAT is very small, we neglect variations in $\textnormal{S}_{xx}$ and just take the value at the central frequency of the channel, $\nu$.
        
        P$_i$ represents the total emitting power for the signal $i$, which is defined using the transmitter antenna gain $\textnormal{G}_{\textnormal{t}_i}$ (dBi) and the transmitted power $\textnormal{P}_{\textnormal{t}_i}$ (dBW):
        \begin{equation}
            \textnormal{P}_i = \dfrac{\textnormal{G}_{\textnormal{t}_i} \textnormal{P}_{\textnormal{t}_i}}{4\pi}\,.
            \label{eq:sat_power}
        \end{equation}
        The spectral energy distribution, SED (units of power per frequency) for each signal from a satellite is given by the product of P$_i$ and $S_{xx_i}$.
        In some instances, the signal power values are not known to the public, and therefore, we selected a value of  $\textnormal{P}_{\textnormal{t}}=$10 dBW and $\textnormal{G}_{\textnormal{t}} = 10$dBi where required. 
         
         Note that we neglect the satellite beam in the modeling, e.g. we assume the satellite is always pointing towards the telescope. This beam is supposed to be quite wide and so any directional effect should be negligible, specially for satellites not far away from the telescope zenith. The situation will be more uncertain for satellites far away from the telescope position as they will be pointing down, not towards the telescope. 
         
        The values used for each signal can be found in \autoref{tab:sat_cat}. The first column corresponds to the index $i$ used to identify the signal. Not all signals in the table are used in our simulation because their peak emission is quite far from our target frequency band and their contribution should be negligible compared to the others. 
        
        For simplicity, we will assume that every satellite of a given constellation emits all the signals listed for that constellation (highlighted in grey in \autoref{tab:sat_cat}). This might not be necessarily true, as we know that of the listed signals some belong to specific generations of a constellation. As an example younger satellites could have additional signals on board compared to their predecessors. However, such information is again not publicly available in most circumstances. So, in the simulation, we sum over all satellite signals for a given constellation and constellations as described in the beginning of \autoref{s:results}.
        The terms $\textnormal{B}(\theta, \nu)$ and $\textnormal{r}$ are then calculated for each satellite individually. Note again that if the satellite is below the horizon at the telescope site or is more than $100^{\circ}$ away from the pointing centre we remove it from the calculation.

    \begin{figure*}
        \centering
        \includegraphics[width=0.9\textwidth]{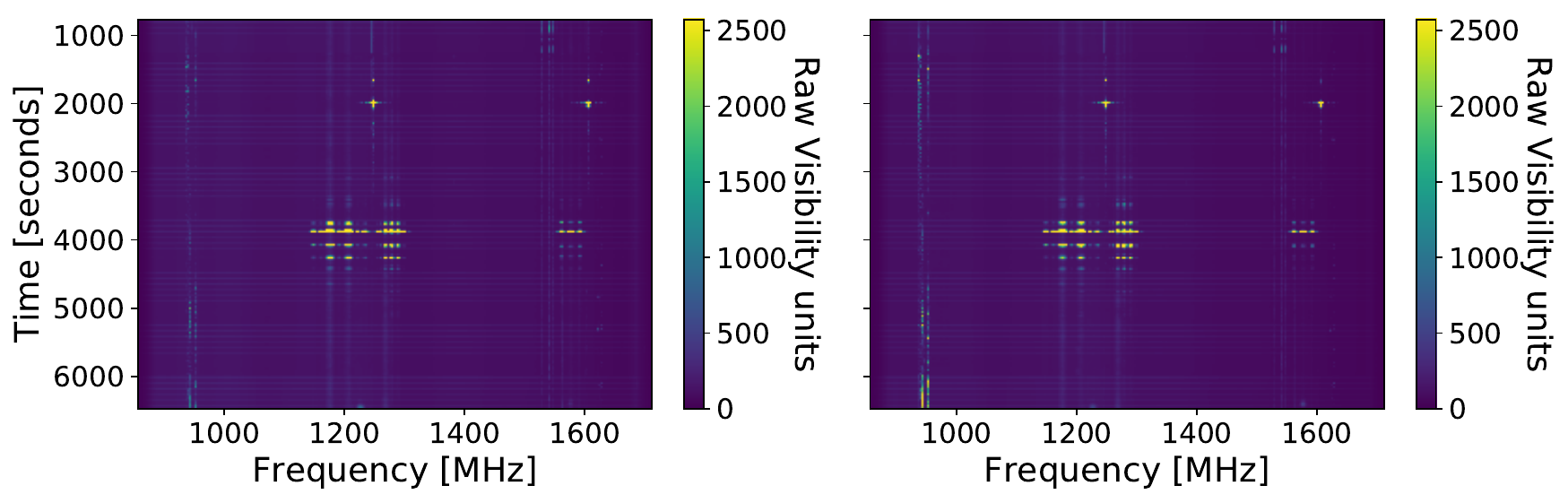}
        \caption{Raw visibility maps of antenna m000 for HH (right) and VV (left) polarizations for the observation scanning period as a function of frequency and time. The bright spots correspond to saturation in the receiver due to strong RFI, such as when satellites pass close to the pointing (see \autoref{fig:ang_sep_gal}).
        }
        \label{fig:raw_vis_map}
    \end{figure*}

\section{Observed Satellite Signal} \label{s:oss}
    To build and test our GNSS satellite simulation, we used the MeerKAT observational data 
    described in W21. This data is a single-dish HI IM pilot survey that mapped a region of around 200 $\deg^2$ centred at ra $\sim 11$h and dec $\sim 0^{\circ}$ over the L-band, which ranges between $\approx$980-1700 MHz over a span of 4096 channels and a frequency resolution of 0.2 MHz (see \autoref{fig:raw_vis_map}). The observations are split into blocks of about 1.5h, each covering the same area with all dishes pointing in the same direction. The data is taken in both linear polarisations, horizontal (HH) and vertical (VV).
    A standard single-dish flux/bandpass calibrator is observed before and after each block, and noise diodes are fired every 20 seconds in order to calibrate gain fluctuations in time. The observation blocks used in this analysis are listed in \autoref{tab:observational_blocks}. In the next sections, we show the results for block 1551055211, dish m000 only, unless stated otherwise.

    \begin{table}
        \centering
        \begin{tabular}{c|c|c}
             Observation ID & Antenna no. & Date/Time (UCT)  \\
             \hline
             \hline
             \textbf{1551055211} & \textbf{m000} & \textbf{2019-02-25 00:40:11} \\
             1553966342 & m000 & 2019-03-30 17:19:02 \\
             1554156377 & m000 & 2019-04-01 22:06:17 \\
             1556138397 & m000 & 2019-04-24 20:39:57 \\
             1562857793 & m004 & 2019-07-11 15:09:53 \\

        \end{tabular}
        \caption{The 5 observational blocks used in the analysis. The block ID, antenna number and date and time of the observation are listed. The results, unless stated otherwise, are shown for the highlighted observation.}
        \label{tab:observational_blocks}
    \end{table}

\begin{figure}
        \centering
        \includegraphics[width=\columnwidth]{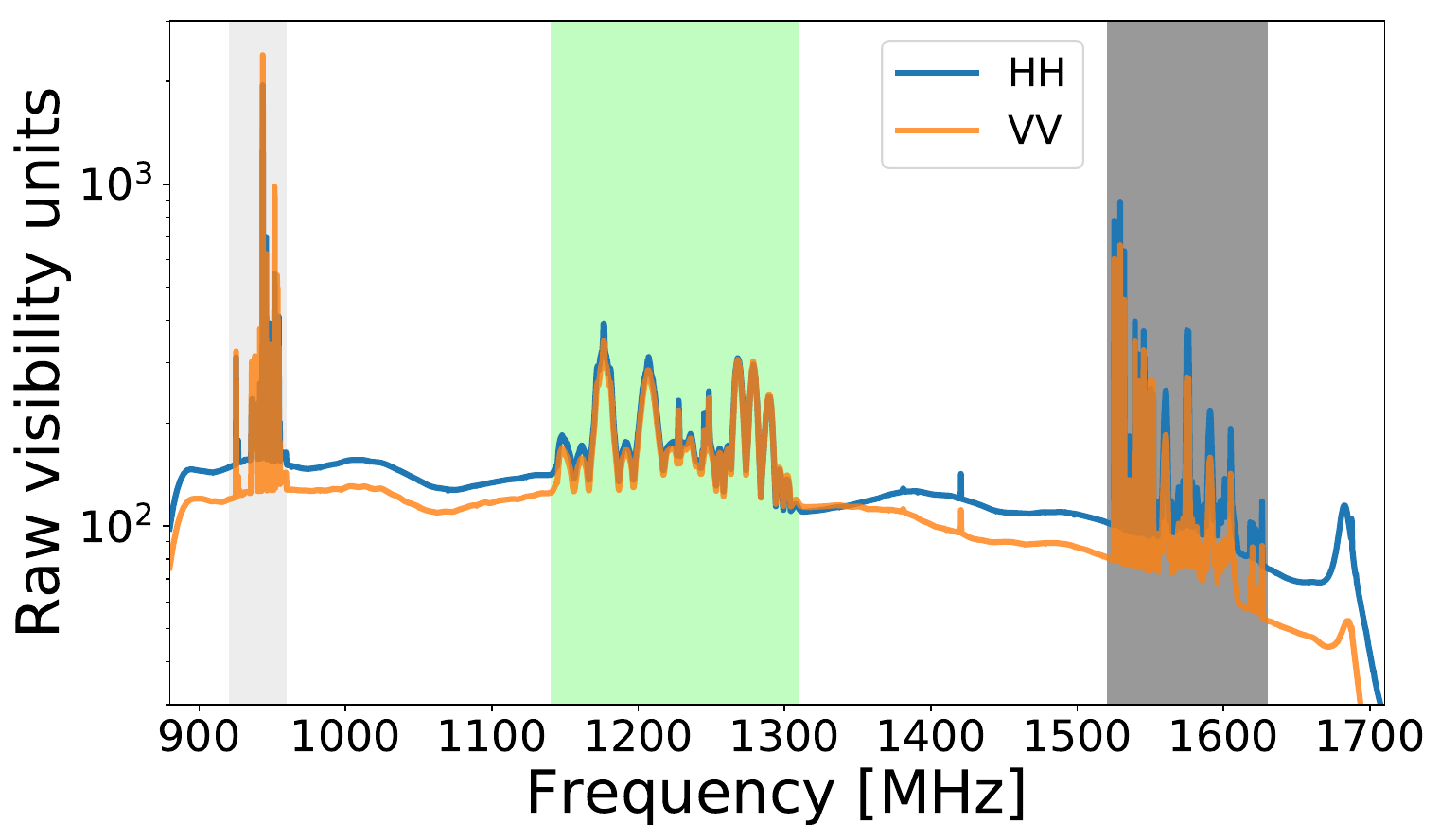}
        \caption{The time-average of \autoref{fig:raw_vis_map}, with the HH (blue curve) and VV (orange curve) polarizations shown for the m000 antenna. The light-grey (920-960 MHz) shaded region represents the low-end RFI, which stems from {Global System for Mobiles (GSM)} and aeroplane transponders \citep{2020_isaac}. The green-shaded region (1140-1310 MHz) represents our frequency range of interest dominated by the navigational satellite RFI. The dark grey (1520-1630 MHz) shaded region relates to the higher frequency RFI in the L-band,  which contains signals from a few navigational services but is dominated by telecommunication services.} 
        \label{fig:raw_vis_1d_map}
    \end{figure}

    In \autoref{fig:raw_vis_1d_map}, we show different frequency regions contaminated by RFI for one of the blocks. The light-grey shaded region (920-960 MHz) corresponds to the Global System for Mobiles (GSM), aeroplane transponders, etc. \citep{2020_isaac}. The dark grey shaded region at higher frequencies (1520-1630 MHz) corresponds to telecommunication services and a few navigational satellites. The region shaded in green corresponds to our frequencies of interest (1140-1310 MHz) which navigational satellites' RFI dominates. As a reference, our target cosmology window used to make a cross-correlation detection of the power spectrum with the same data \citep{cunnington_2023b} is at 973-1015 MHz.
    As mentioned before, we ignore the navigational satellites present in the high L-band frequency range since their contribution should be subdominant at 1140-1310 MHz compared to the satellites emitting there (and even less in the target cosmology band).

    \subsection{Calibration of the RFI-contaminated region} \label{ss:re_cal}

\begin{figure}
        \centering
        \includegraphics[width=\columnwidth]{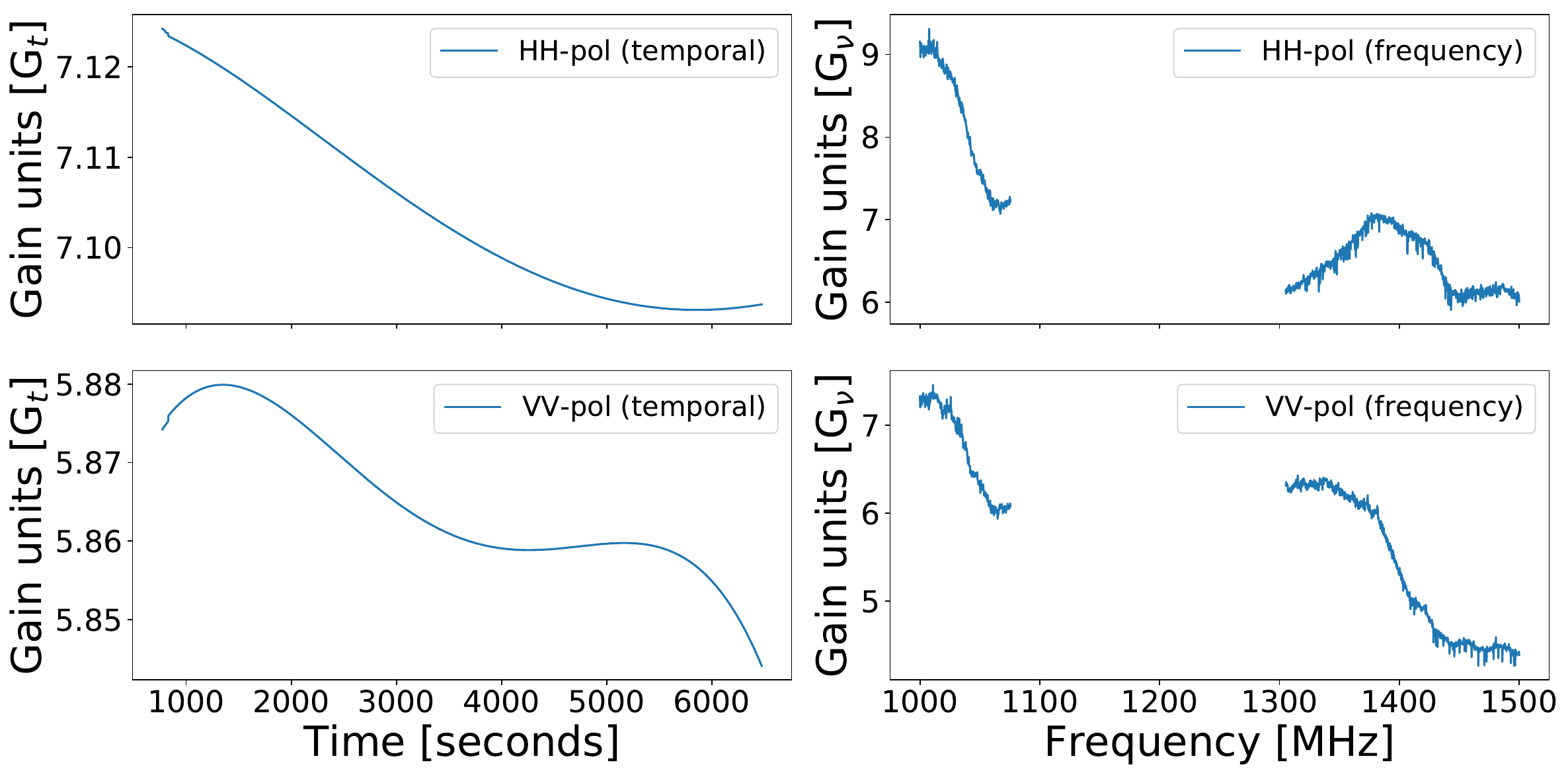}
        \caption{Functional form of the gain as a function of time (left panels) and frequency (right panels) for m000 antenna \citep{Wang2020}. The top panels are for HH polarization and the bottom ones for VV.
        RFI flagging results in the missing information in frequency.}
        \label{fig:gain_map_t_v}
    \end{figure}

    In principle, one would like to use the gain solutions derived in W21. From that, we could calibrate the TOD (time-ordered data) for each block, dish and polarisation using:
    \begin{equation}
            \textnormal{T}_{\textnormal{cal}}(t, \nu) = \textnormal{T}_{\textnormal{raw}}(t, \nu)/  g(t, \nu),
            \label{eq:t_cal}
        \end{equation}
         where T$_{\textnormal{cal}}$ is the calibrated temperature in Kelvin, T$_{\textnormal{raw}}(t, \nu)$ the raw visibility in correlator units, and $g(t, \nu)$ is the gain.
    Unfortunately, the calibration pipeline used in W21 immediately flagged the 1100-1350 MHz region for all times due to the strong RFI. 
     This poses a problem for us since we need the data to be calibrated in that frequency interval in order to compare to our simulations.
    In \autoref{fig:gain_map_t_v}, we show the derived gain from W21 for one block and one dish and the HH and VV polarisations (top and bottom, respectively). Left panels show the gain as a function of time, averaged over frequency and right panels the time average as a function of frequency (only non flagged times and frequencies are included in these averages).

     Using a point source to calibrate the flagged region is quite challenging due to the very strong RFI signal. Noise diodes could be better, but their frequency dependence is not very well known from the lab and again, calibrating the noise diodes beforehand with a point source would have the same contamination issues in the target RFI contaminated region. Instead, we took a different approach. We first assume that the gain can be broken into the temporal ($g_t$) and frequency ($g_{\nu}$) components, e.g. $g(t, \nu) \approx g_t(t)g_{\nu}(\nu)$. This basically means that the bandpass shape is assumed constant throughout the observation, with only an overall amplitude variation in time. Such assumption is well supported by the analysis in W21, with frequency-dependent variations measured to be less than 1\%. 
     The gain can then be written as
    \begin{equation}
                \tilde g(t, \nu) = \frac{\langle g(t,\nu)\rangle_{\nu} \langle g(t,\nu)\rangle_t}{\langle g(t,\nu) \rangle_{t,\nu} },
                \label{eq:reconstruct_gain_map}
    \end{equation}
    where $\tilde g$ is used to represent the new derived gain.
     The averages, $\langle\, \rangle$, are over the non flagged times and frequencies. This immediately gives us the time variation of the gain as in \autoref{fig:gain_map_t_v}. 
    
    Obtaining $g_{\nu}$ is more complicated due to the missing information. A simple solution would be to interpolate the frequency curve in \autoref{fig:gain_map_t_v}, but the gap is too wide. To obtain more points for the interpolation, we assumed that the frequency dependence of the gain follows the frequency dependence of the raw visibilities without the satellites. This implies that the input temperature should be flat across the frequency gap, except for the satellites. In reality, the ground pickup, receiver temperature and galactic synchrotron have a slope. However, this effect will be small in the interval considered and can be corrected later when matching to the gain solution outside the RFI band.
   Below we describe in detail the steps we followed to complete the missing frequency information in the gain maps:

        \begin{itemize}
                \item We start from the raw visibility of both HH \& VV polarizations before RFI flagging. We then select the minimum visibility values across time for each frequency channel, which we take as an estimate
                of the floor level of the channel when satellites are not present. We do this in
                the frequency range spanning 1000-1500 MHz, which is wider than the RFI-contaminated band, but it allows us to obtain the tail-end information for the frequency area we wish to calibrate; 
                

            \item The next step is to remove all the peaks in the data caused by the satellite contamination,
            to get a smooth curve of the floor level. We start by doing a spline fit to this minimum raw visibility which provides a smooth curve in frequency. 
            We then take the difference between the spline and minimum raw visibility curves. From the residual, a sigma clip\footnote{\url{https://docs.astropy.org/en/stable/api/astropy.stats.sigma_clip.html}} is then applied to remove the peaks that stem from the satellite emission. 
            We then interactively applied this procedure to the clipped curve until no more clipping was needed (after five interactions).
            We then obtained a smooth overlay spanning the affected frequency channels as seen in \autoref{fig:min_raw_visibility_smooth_4};
    
            \begin{figure}
                \centering
                \includegraphics[width=0.48\textwidth]{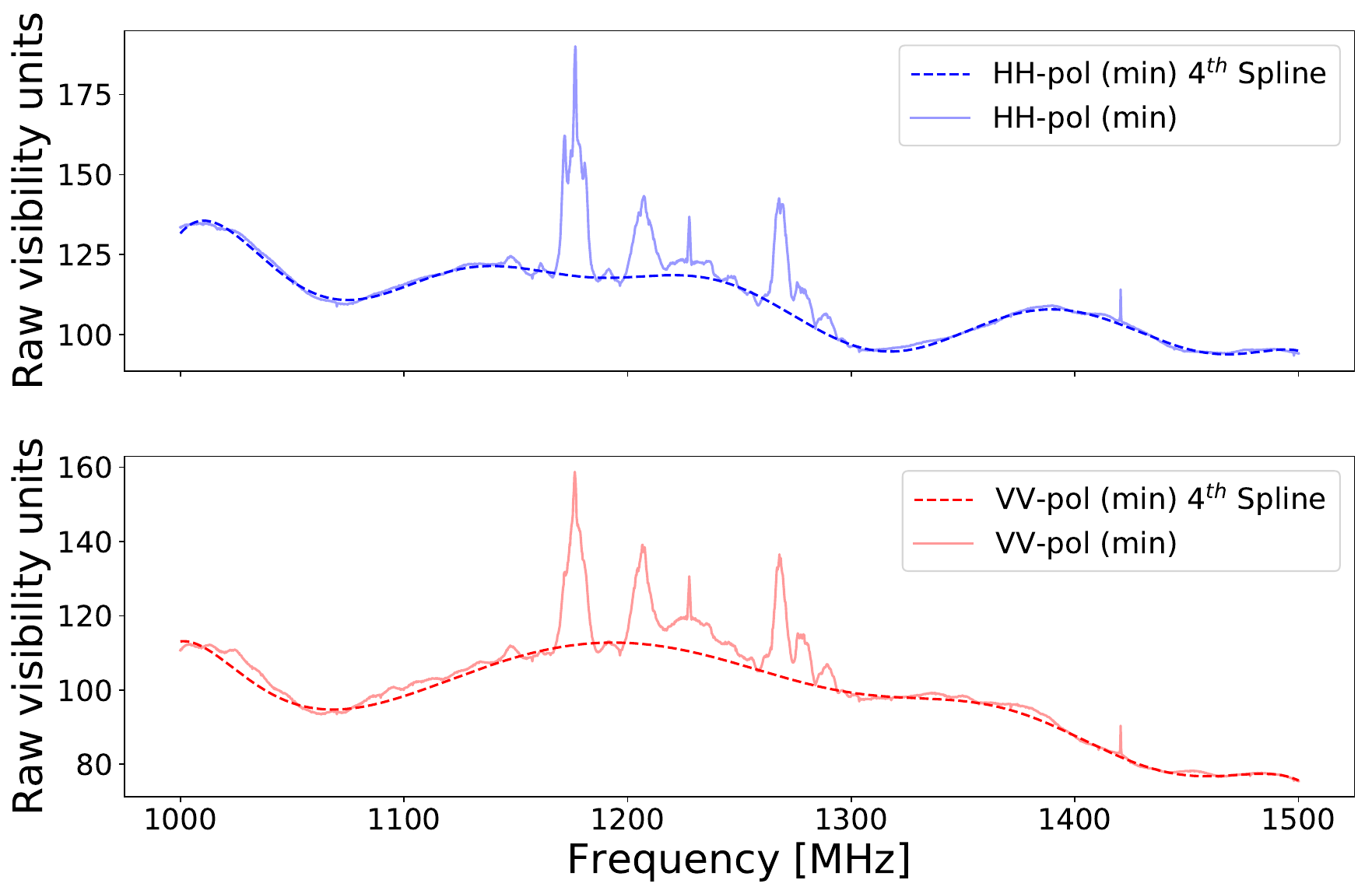}
                \caption{Estimate of the base raw visibility without satellites after the $4^{\textnormal{th}}$ iterative clipping (dashed line) overlaid on antenna m000 minimum raw visibility (solid line) for the HH (top) and VV (bottom) polarizations.}
                \label{fig:min_raw_visibility_smooth_4}
            \end{figure}
    
          \item The resulting smooth curve represents the underlying base's structure of the raw visibility and we assume that the gain will be proportional to this.
          We then normalize the smooth curve by matching it to the known gain from W21 outside the satellite's frequency region. We did this between 1138MHz and 1307MHz for the HH polarization and between 1100MHz and 1314MHz for the VV polarization. Finally, we create a smooth function (spline) that becomes the final frequency component of the gain map, as seen in \autoref{fig:min_raw_visibility_concatenate}.
            
            \begin{figure}
                \centering
                \includegraphics[width=0.5\textwidth]{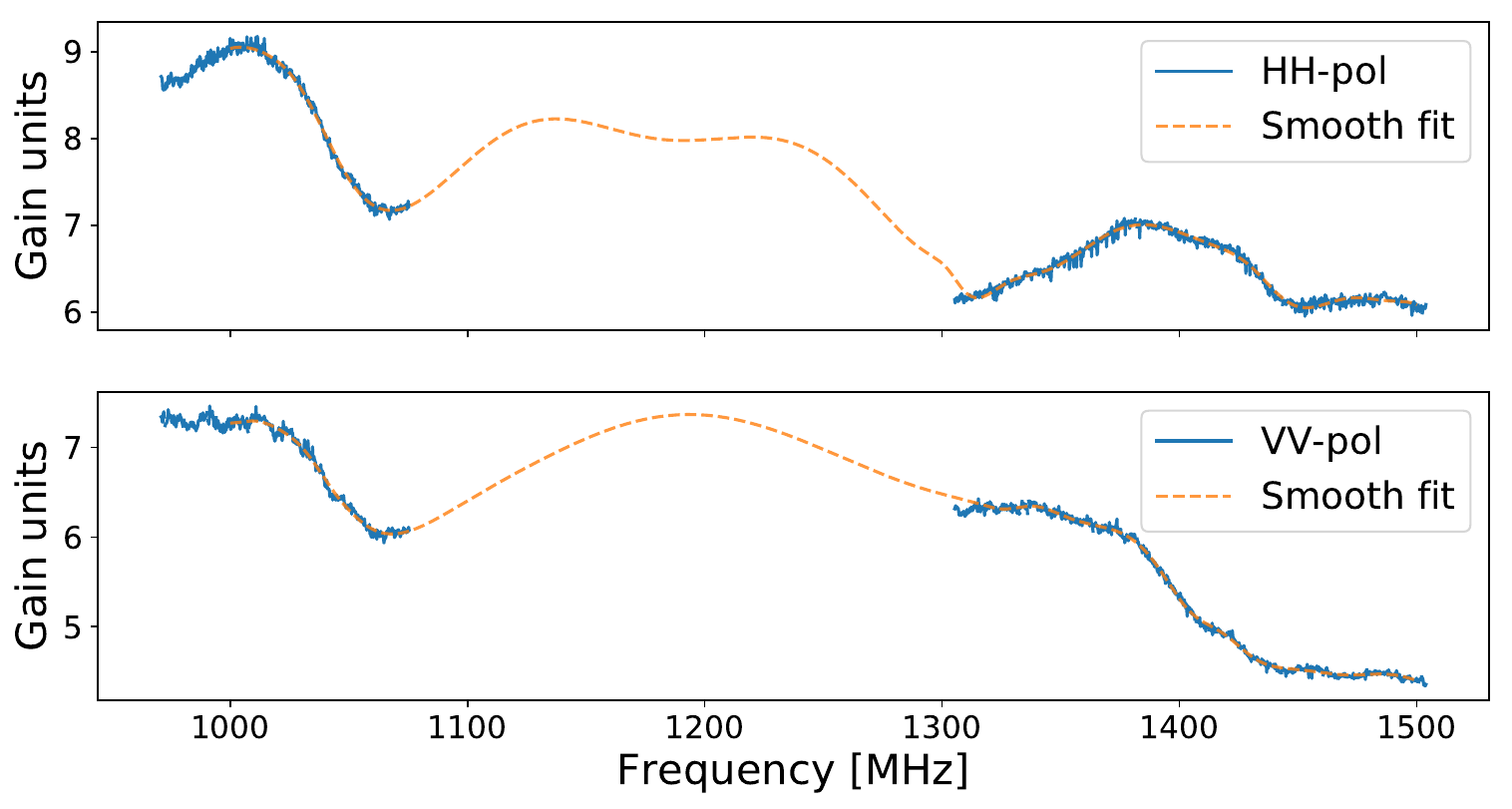}
                \caption{The bandpass of the gain for the m000 antenna in the HH (top panel) and VV (bottom panel) polarization. The blue curve represents the available gain information from W21; the orange dashed curve represents the best estimate of the gain across the frequency region of interest.}
                \label{fig:min_raw_visibility_concatenate}
            \end{figure}

    \end{itemize}

    The final temperature map from the data will be the average of the calibrated temperatures, $\textnormal{T}_{\textnormal{cal}}(t, \nu)$, from each polarisation.

    \subsection{Background model} \label{ss:bg_model}

    In order to compare the satellite simulation to the data, we also need to factor in the other temperature contributions, which we call the background model. This background is much smaller than the satellite emission and reasonably flat in frequency, making up the baseline level we see in \autoref{fig:raw_vis_1d_map}.
     We follow W21 to model this background temperature ($\textnormal{T}_{\textnormal{BG}}$) as
    \begin{equation}
            \textnormal{T}_{\textnormal{BG}}(t, \nu) = \textnormal{T}_{\textnormal{rec}}(t, \nu) + \textnormal{T}_{\textnormal{el}}(t, \nu) + \textnormal{T}_{\textnormal{gal}}(t, \nu) + \textnormal{T}_{\textnormal{CMB}}\,,
            \label{eq:bg_model}
        \end{equation}
    where T$_{\textnormal{rec}}$ is the receiver temperature, T$_{\textnormal{el}}$ depends on the elevation and includes contributions from the ground spill and atmosphere, T$_{\textnormal{gal}}$ is the synchrotron emission from the Galaxy and T$_{\textnormal{CMB}}$ is Cosmic Microwave Background (CMB) temperature. 
    
\begin{figure}
            \centering
            \includegraphics[width=0.5\textwidth]{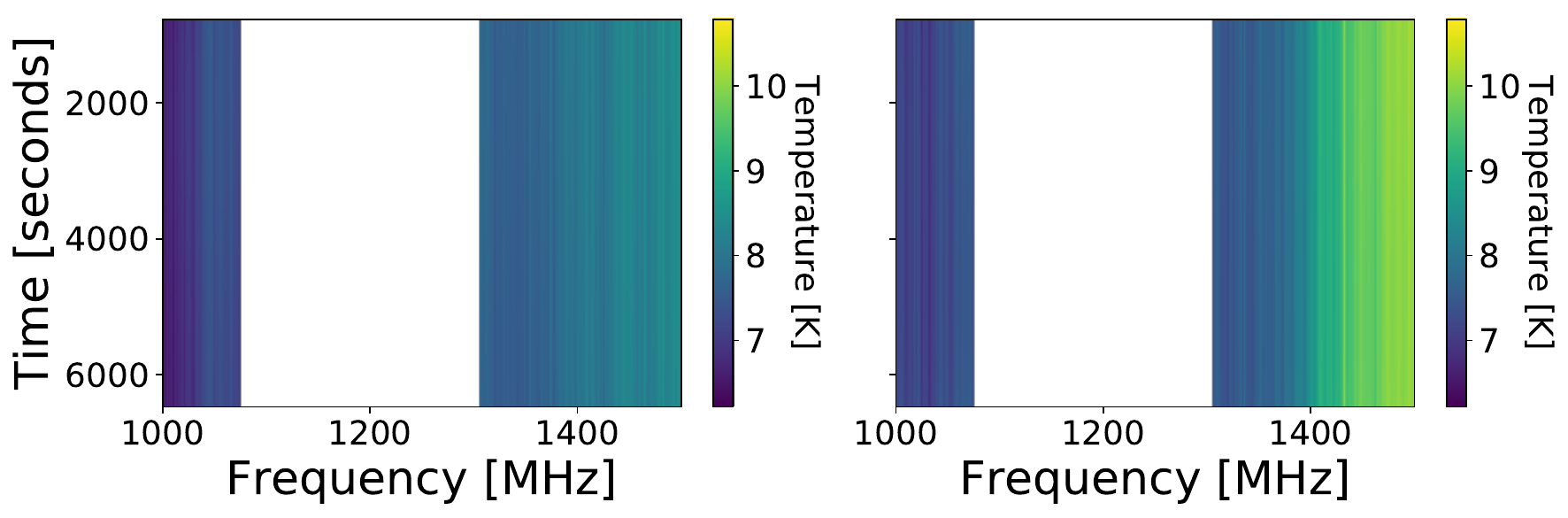}
            \caption{Receiver temperature for the m000 antenna in HH (left panel) and VV (right panel) polarization. RFI cleaning results in missing channel frequency information from 1075 to 1305 MHz}
            \label{fig:t_rec}
        \end{figure}

    T$_{\textnormal{CMB}}$ is constant, and since the observations were done at constant elevation, $\textnormal{T}_{\textnormal{el}}$ should be constant in time. The same is expected from $\textnormal{T}_{\textnormal{rec}}$ since the instrument is quite stable over the course of the 1.5h observation. $\textnormal{T}_{\textnormal{gal}}$ will change as the telescope pointing moves through the sky but these fluctuations will be at $\sim 0.1$ K level. So we could in principle just remove most of this background contribution by subtracting a constant in time. We would then have to do the same to the satellite simulation for a proper comparison. This would mean we wouldn't be able to fit for the satellites DC level which can be useful for other situations. So in this paper, we didn't remove any time offset and instead included $\textnormal{T}_{\textnormal{BG}}$ in the simulation. For this, we use the same models as in W21. 
    However, since the receiver temperature was calculated as part of the calibration process in W21, the frequencies corresponding to satellite emission are also flagged (\autoref{fig:t_rec}). 

\begin{figure}
            \centering
            \includegraphics[width=0.5\textwidth]{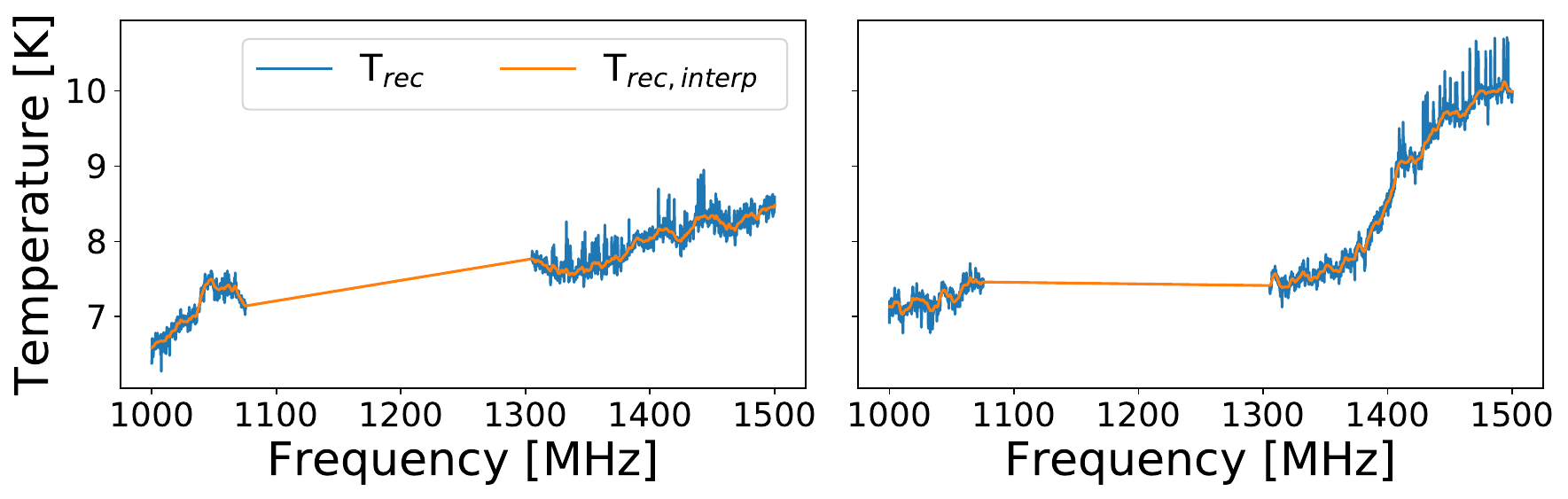}
            \caption{A time-averaged representation of the receiver temperature along frequency for the m000 antenna with HH (left) and VV (right) polarization. The blue curve is the receiver temperature from the model, whilst the orange dash curve is the RBF interpolation overlaid. In the lower and higher frequencies, the RBF follows the data and makes the best attempt to connect the high and low-frequency values.}
            \label{fig:t_rec_1d_interp}
        \end{figure}

\begin{figure}
            \centering
            \includegraphics[width=0.5\textwidth]{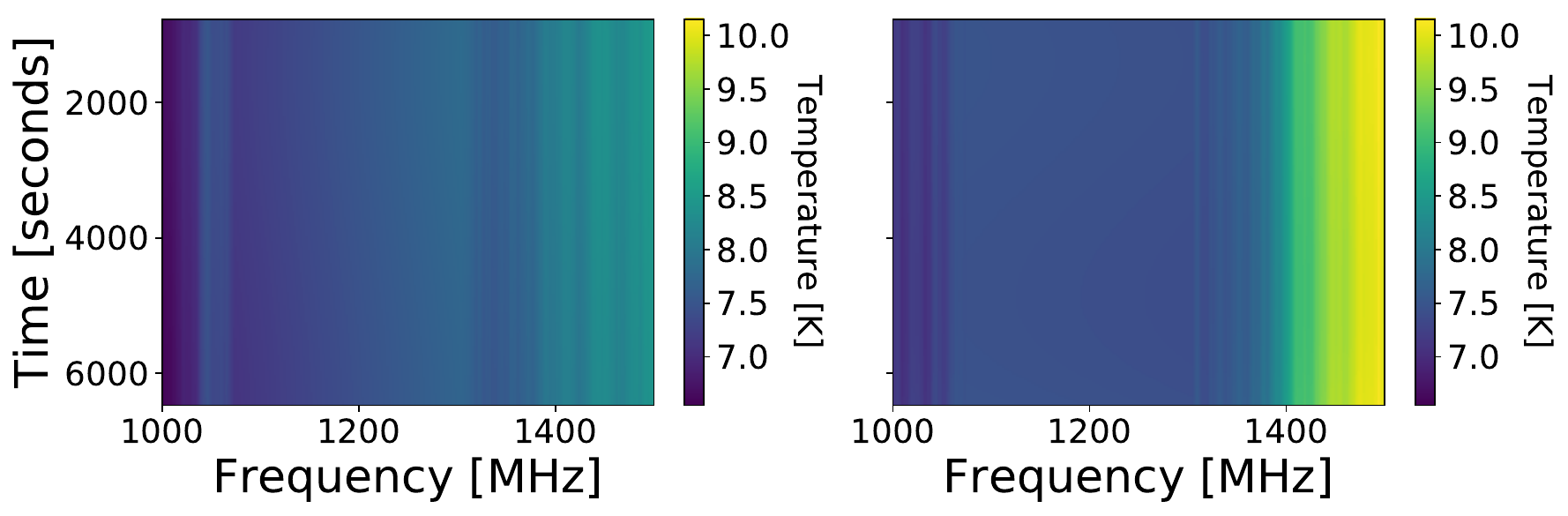}
            \caption{Interpolated receiver temperature for the m000 antenna in the HH (left panel) and VV (right panel) polarization.}
            \label{fig:t_rec_interp}
        \end{figure}

    To restore the masked information, we will follow the same approach as the one used to estimate the gain in the RFI-contaminated band. We assume that the receiver temperature is a separable function in time and frequency, which is then proportional to the frequency and time average of $\textnormal{T}_{\textnormal{rec}}(t, \nu)$, respectively. These averages do not use any of the flagged data, and this is done for each block, dish and polarization. Again, the function in time is already continuous. 
        For the frequency dependence, we interpolate the known data points (solid blue in \autoref{fig:t_rec_1d_interp}) using a radial basis function (RBF\footnote{\url{https://docs.scipy.org/doc/scipy/reference/generated/scipy.interpolate.Rbf.html}}). This results in a curve (orange line) covering the whole frequency range (\autoref{fig:t_rec_1d_interp}). In this instance, we do not have any data points in the RFI-flagged region, which makes the interpolation more uncertain. However, looking at the fluctuations in \autoref{fig:t_rec_1d_interp}, one should expect any deviations to be less than 0.5 K. Since this is an additive quantity to the far stronger satellite signal, such errors should have little impact in the analysis.  We plot the final receiver temperature used in \autoref{fig:t_rec_interp}.

\begin{figure*}
                 \centering
                 \begin{subfigure}[b]{0.78\textwidth}
                     \centering
                     \includegraphics[width=\textwidth]{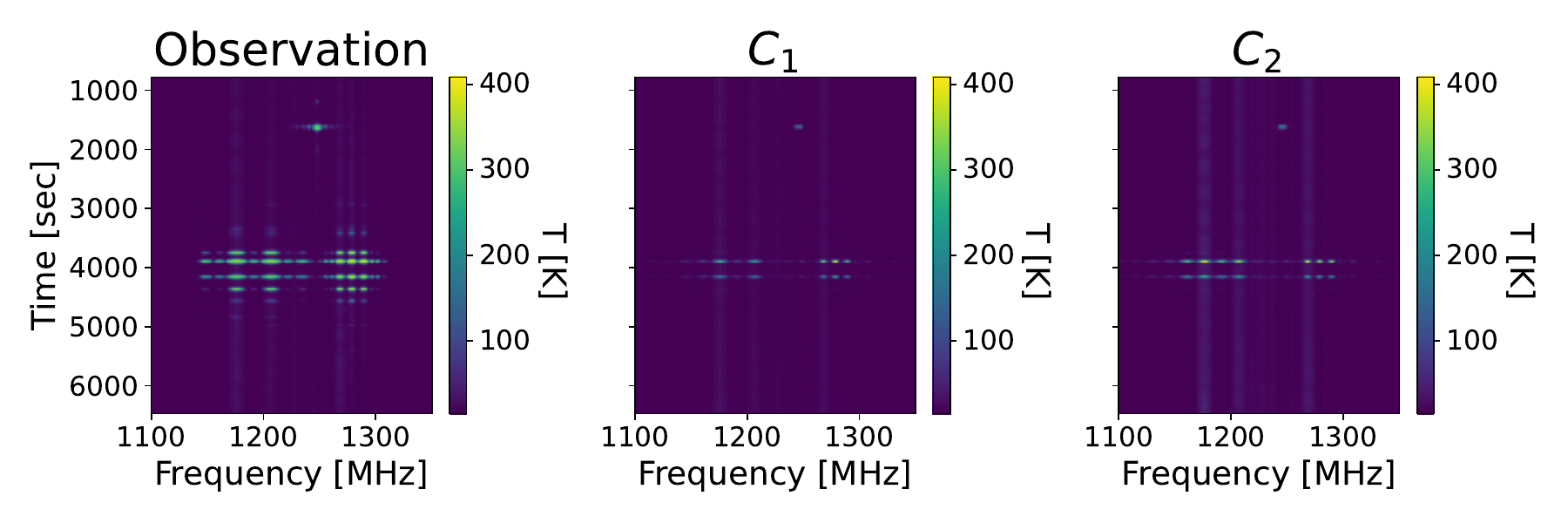}
                     \label{fig:no_mask_waterfall}
                 \end{subfigure}
                 \hfill
                 \begin{subfigure}[b]{0.78\textwidth}
                     \centering
                     \includegraphics[width=\textwidth]{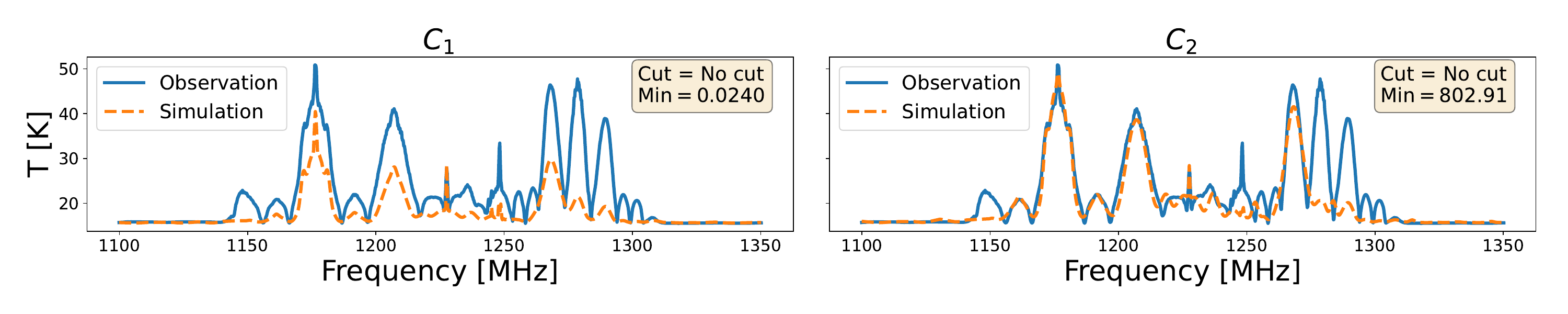}
                     \label{fig:no_mask_1d}
                 \end{subfigure}
             \caption{Best-fit comparison for the \emph{no-mask scenario}. 
             The $1^{\textnormal{st}}$ row shows the waterfall images for the observation (left) and the simulation:  case 1 (centre) and case 2 (right). The temperature colour bar was set at the maximum and minimum values corresponding to the observation. The $2^{\textnormal{nd}}$ row shows the time-averaged frequency plane of the waterfall plots, with the left panel being the $C_1$ case and the right being the $C_2$ case. The solid blue curve represents the observation, and the dashed orange curve represents the simulation's best fit for the two cases.}
             \label{fig:no_mask_case}
            \end{figure*}

\section{Fitting the satellite simulation to observations} \label{s:results}
    
    In this section, we aim to determine how well the simulation performs in recreating the navigational satellite signals we see in the observational data of the MeerKAT telescope. We compare the simulated satellite signal T$_{\textnormal{sat}}(t, \nu)$ to the calibrated observations from each block. In this regard, we have to combine the background model of \autoref{ss:bg_model} to that of T$_{\textnormal{sat}}(t, \nu)$. This allows us to obtain the same temperature base component between the observation and simulation. Our results, which are centred around 1100-1350 MHz, extend toward the tail of the spectrum, where the background model dominates.

    The signal modelling presents a few uncertainties, namely, in the overall amplitude of the temperature of a signal from a constellation. The emitting power and gain of the satellite are not constant with time, and we expect it to be reduced as the satellite ages. Moreover, in some cases, the values tabulated in the literature appear as upper limits or are not available. Uncertainties in the telescope primary beam or calibration errors can also affect the results.
    Because of these issues, we introduce a set of $\alpha_i$ parameters associated with the overall amplitudes of each signal within a given constellation:
    \begin{equation}
    \textnormal{T}_{\textnormal{S,i}} (t,\nu)=  \alpha_{\textnormal{sat, i}}\ \textnormal{T}_{\textnormal{sat, i}}(t, \nu) \, ,
    \end{equation}
    where $\textnormal{T}_{\textnormal{S,i}}$ is the simulated temperature of the \emph{i-th} signal and $\textnormal{T}_{\textnormal{sat, i}}$ is given by \autoref{eq:sat_tod}. Each signal in  \autoref{tab:sat_cat} has an associated $\alpha_{\textnormal{sat, i}}$.
    This amplitude is assumed constant in frequency and time, although we do fit it to different time ranges in the sections below. 
    Note that if there is a good match between the model and the observations, these $\alpha$s should equal one. The number of parameters introduced depends on the number of relevant signals that fall within our frequency range of interest and which constellations are visible to the telescope. The total simulated signal is then:
    \begin{equation}        
    \textnormal{T}_{\textnormal{S}}(t,\nu) = \sum_{\rm constellation}\sum_{i j} \alpha_{\textnormal{sat, i}} \textnormal{T}^j_{\textnormal{sat, i}}(t, \nu) + T_{\textnormal{BG}}(t,\nu)\, ,
    \end{equation}
    where we sum over all constellations, $i$ runs over the signals present in each constellation and $j$ runs over the satellites in that constellation.
    

    \subsection{Fitting the $\alpha_i$ amplitudes}\label{ss:fit_amplitudes}


        In order to fit for the $\alpha_i$, we minimize a cost function (CF) which we define as
        \begin{equation}
            {\rm CF} = \frac{1}{N}\sum\dfrac{(\textnormal{T}_\textnormal{O} - \textnormal{T}_\textnormal{S})^2}{\sigma_\textnormal{D}^2},
        \end{equation}
        where $\textnormal{T}_\textnormal{O}$ is the adjusted calibrated observational data and $\textnormal{T}_\textnormal{S}$ is the simulated temperature response. $N$ is the total number of points used in the fit and is used just for normalisation when we compare cases with different number of data points. The sum is over all times and frequencies of the time-ordered data.        
     $\sigma_{\textnormal{D}}^2$ can be seen as a weight for each data point, and we consider two distinct cases:
        \[\sigma_{\textnormal{D}}= \begin{cases} 
              \textnormal{T}_{\textnormal{O}}\ [\textnormal{case 1 }(C_1)] \\
              \textnormal{or}\\
              1\ \ \ \ [\textnormal{case 2 }(C_2)]
           \end{cases}
        \]
        Case 1, denoted as $C_1$ below, uses the temperature itself, which is proportional to the expected thermal noise as in the radiometer equation. Essentially, this would correspond to the standard $\chi^2$. This can also be seen as minimizing the percentage difference between the model and the data. The problem with this cost function is that the uncertainties in the model should dominate the expected thermal noise, since at 2 seconds and 0.2 MHz resolution, the thermal noise gives an error of about 0.1\% for each data point. This case basically down-weights the data in the region where the satellite's signal is the brightest so that the fitting will be dominated by the (many) small temperature values in the region where the background model dominates, making it harder to fit the rather important bright satellite peaks.
        For this reason, we also consider case 2 (denoted as $C_2$ below), where all the data points are equally weighted, i.e., $\sigma_{\textnormal{D}}=1$. 

        The minimisation of the cost function is accomplished with the \emph{scipy.optimize.minimize} package\footnote{\url{https://docs.scipy.org/doc/scipy/reference/generated/scipy.optimize.minimize.html}} using the \textit{Powell} method. The initial conditions for the $\alpha$ parameters are set to zero for the first iteration; a lower bound of zero and an unconstrained upper bound are placed on the parameters as boundary conditions. A tolerance threshold (tol) of 1e-6 is applied to the minimisation as a convergence limit. If a signal is never used in a given observation (because the corresponding satellites are too far away from the pointing direction) then the $\alpha$ will stay at zero.

         
        
        The results of the fitting to block 1551055211 are shown in \autoref{fig:no_mask_case}. We see that although some of the main features are fitted, the amplitude of the simulation signal is smaller and misses some of the peaks. This is better seen in the bottom panel, where we take a time average of the waterfall plot. This is true for both cost functions, although case 2 performs better. The solid blue represents the observational data, while the dashed orange line represents the simulation best-fit results for the $C_1$ (left) and $C_2$ (right) cases, with a minimum value of the cost function of 0.0240 and 802.91 respectively.
             \begin{figure}
                 \centering
                 \begin{subfigure}[b]{0.48\textwidth}
                     \centering
                     \includegraphics[width=\textwidth]{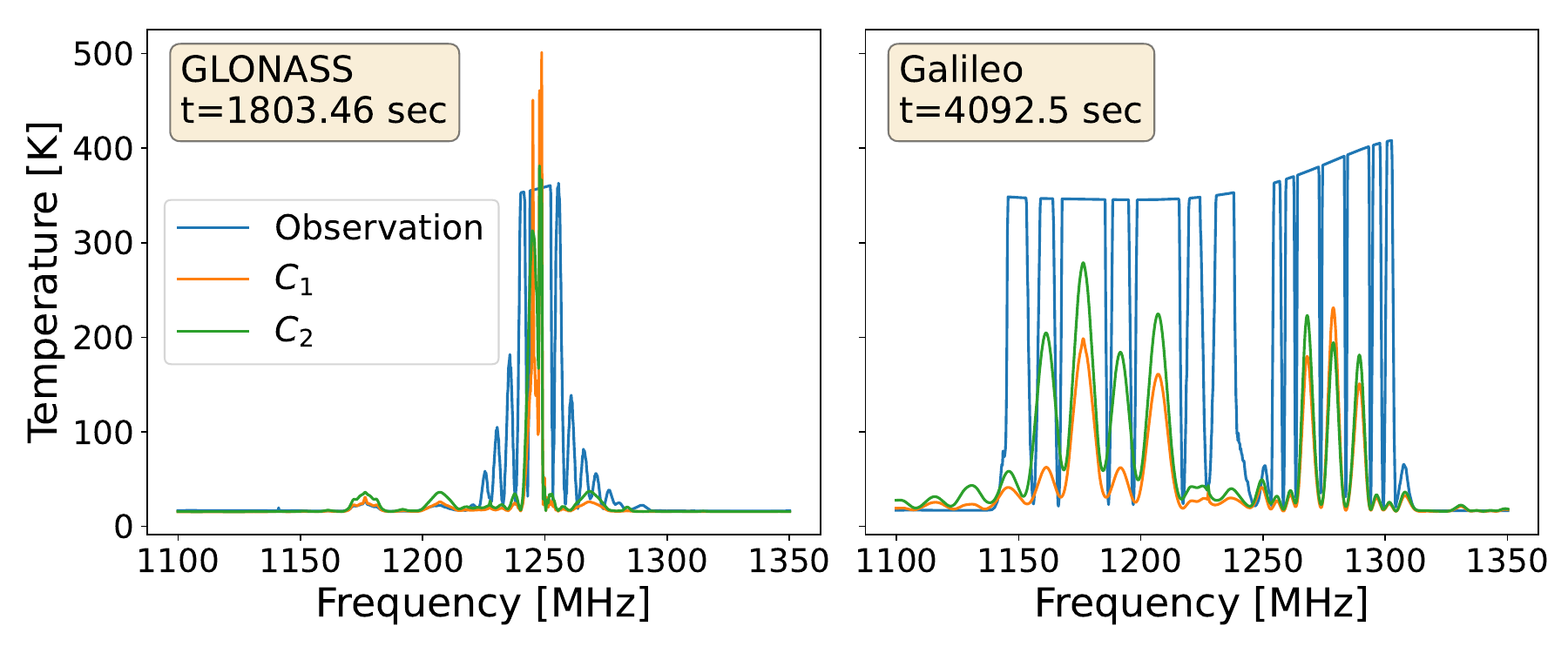}
                 \end{subfigure}
                 \hfill
                 \begin{subfigure}[b]{0.48\textwidth}
                     \centering
                     \includegraphics[width=\textwidth]{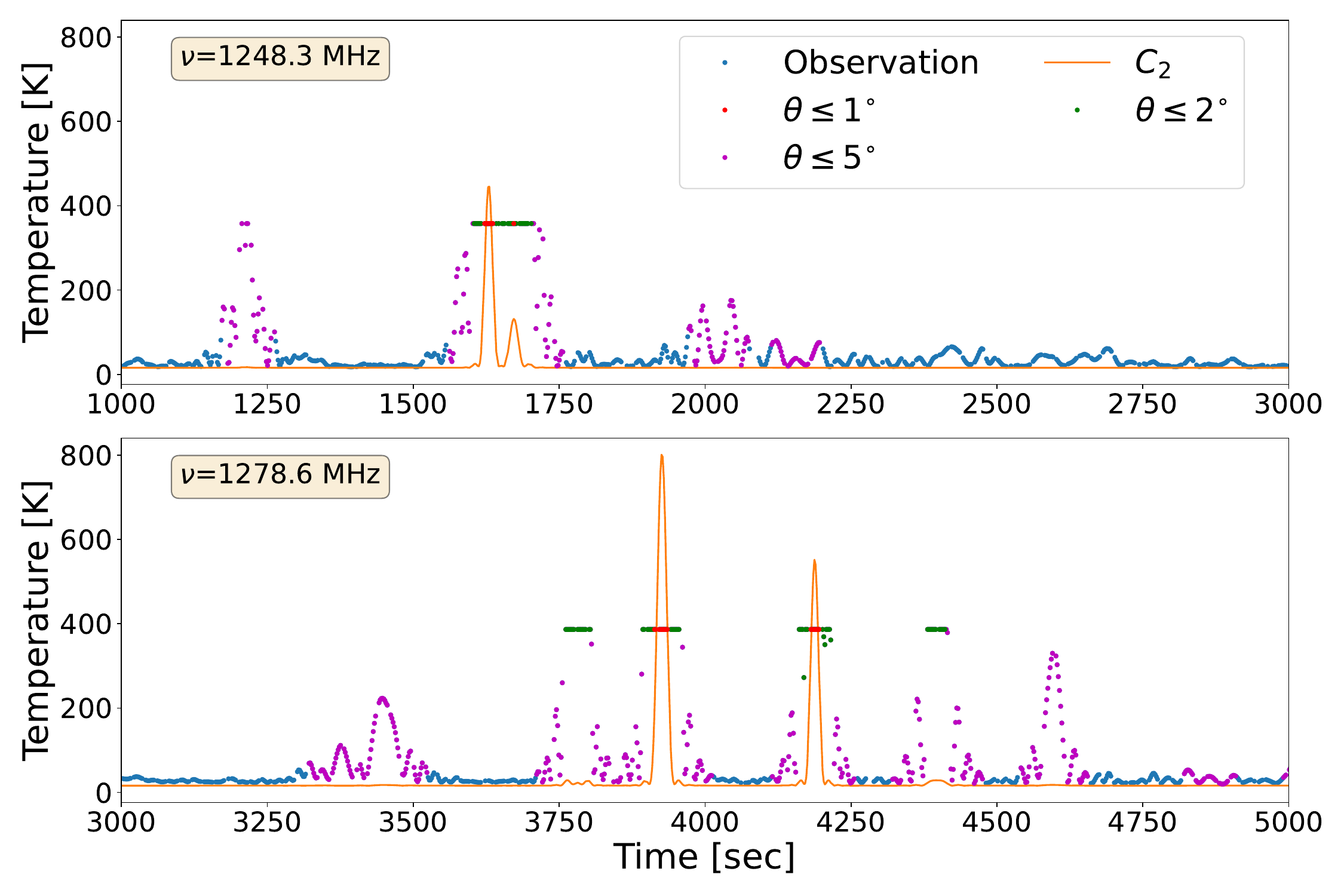}
                 \end{subfigure}
                 \hfill
                 \begin{subfigure}[b]{0.48\textwidth}
                     \centering
                     \includegraphics[width=\textwidth]{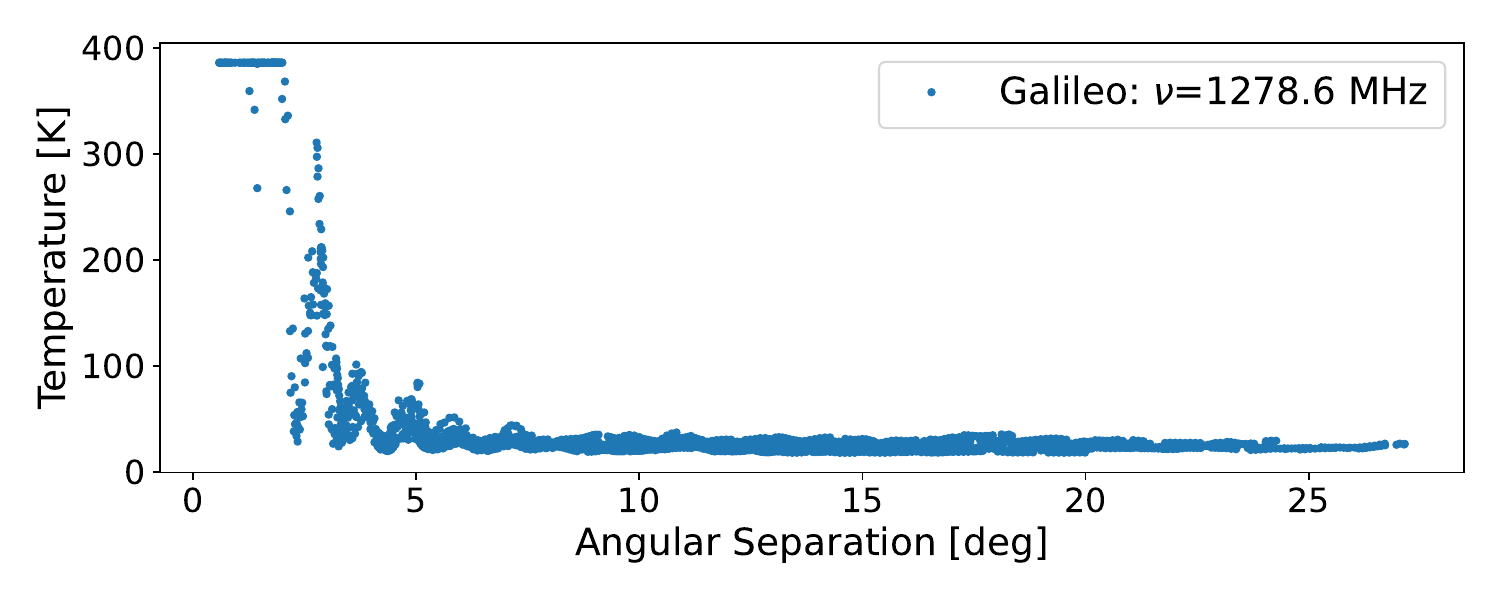}
                 \end{subfigure}
                \caption{A snapshot of the data along different axes: frequency ($1^{\textnormal{st}}$ panel), time ($2^{\textnormal{nd}}$ \& $3^{\textnormal{rd}}$panels) 
                and along the angular separation of a Galileo satellite: GSAT0209-(PRN-E09) ($4^{\textnormal{th}}$ panel). In the $1^{\textnormal{st}}$ panel, the data (blue), $C_1$ case (orange), and $C_2$ case (green) at two different timestamps correspond to incursions from the GLONASS (left panel) and Galileo (right panel) constellations. The snapshot is when $\theta \leq 1^{\circ}$ for both constellations; it also reveals the peak frequency of GLONASS at $\sim$1248 MHz. However, in the case of the Galileo constellation, the peak is hidden due to the saturation of the receiver. In the $2^{\textnormal{nd}}$ \& $3^{\textnormal{rd}}$panels, based on the information in the $1^{\textnormal{st}}$ panel, we select the peak frequencies from  \autoref{tab:sat_cat}. The $2^{\textnormal{nd}}$ panel is at 1248.3 MHz (GLONASS), and the $3^{\textnormal{rd}}$ panel at 1278.6 MHz (Galileo). The observational data are shown with blue dots, the $C_2$ case is shown in orange, and the limits of incursion by the satellites are shown for the $1^{\circ}, 2^{\circ}, 5^{\circ}$ in red, green and magenta respectively. In the $4^{\textnormal{th}}$ panel, we show the observed temperature versus the angular separation values for the Galileo satellites at the frequency selected around which the satellite operates.}
                \label{fig:no_mask_case_2}
            \end{figure}
            The reason for this is that the observation saturates when the satellites get too close to the telescope, which makes our model break down. 
            In this block, two satellites passed through the telescope's pointing. The first belongs to the GLONASS constellation and breached the $\theta \leq 1^{\circ}$ between 1622.39-1672.37 seconds. While the second satellite belongs to the Galileo constellation (see  \autoref{fig:ang_sep_gal}), and was within the $\theta \leq 1^{\circ}$ between 3913.426-4195.30 seconds. This saturation is not clear in the bottom panel of \autoref{fig:no_mask_case} because we averaged over time and only a small percentage of the time is saturated. This is better seen in \autoref{fig:no_mask_case_2}. The top panel shows the signal across frequency for the snapshot at the time when the signal is stronger (satellites close to the pointing). Note that the saturation/cut-off level is not constant across frequency. The two middle panels show the total signal along time for the peak frequencies belonging to the GLONASS constellation (left panel) at 1248.3 MHz, and the Galileo constellation (right panel) at 1278.75 MHz. Although other signals (and satellites) may contribute to these frequencies, these are the dominant ones. The fluctuations in time should depend on the angular distance to the pointing direction and follow the primary beam shape. The solid orange line represents the simulation best-fit results for the $C_2$ case only while the dots represent the observational data, with different colors corresponding to different angular distances. Below $2^{\circ}$ the telescope system completely saturates at these frequencies.
            
            The bottom panel of \autoref{fig:no_mask_case_2} shows the calibrated temperature from the observed block at the frequency of 1278.6 MHz against the expected movement of the Galileo satellite GSAT02029-(PRN-0E09) \footnote{\url{https://www.euspa.europa.eu/european-space/galileo/programme}}. The position is calculated exactly for the observation date and timestamps and the angular distance is derived with respect to the dish pointing centre, taking into account the movement of the dishes as well. Although other signals will "contaminate" this frequency, we can clearly see the shape of the MeerKAT primary beam as expected. Besides contamination from other satellites (at other angular distances) there are other reasons why the comparison to the primary beam isn't perfect, in particular the saturation of the signal and small errors in the satellite position. These errors have a greater impact as the satellite approaches the dish pointing. Finally we note that the saturation not only imposes a cutoff on the observed signal but also makes the gain non-linear. This non-linearity can affect most frequencies when the signal becomes strong, even outside the peak frequency, and can show up even before saturation. 
            However our model was built assuming the telescope response was linear even in the presence of navigation satellite RFI. As we can see, this is not always true and will affect our fit to the data. Thus, this poses the question of addressing the saturation issue by masking pixels/time stamps where the telescope response becomes non-linear. I.e., can we improve the model that recreates 
            the satellite's emissions? We will explore these possibilities in the following section.

\subsection{Testing limits of validity} \label{ss:masking_thres} 

From now on we will only consider the $C_2$ case to test the limits of
validity of our approach. In all masking scenarios we have compared $C_1$
with $C_2$ and concluded that $C_2$ gives better overall fits.

        \subsubsection{Angular Masking}

\begin{figure}
                \centering
                \begin{subfigure}[b]{0.48\textwidth}
                     \centering
                    \includegraphics[width=\textwidth]{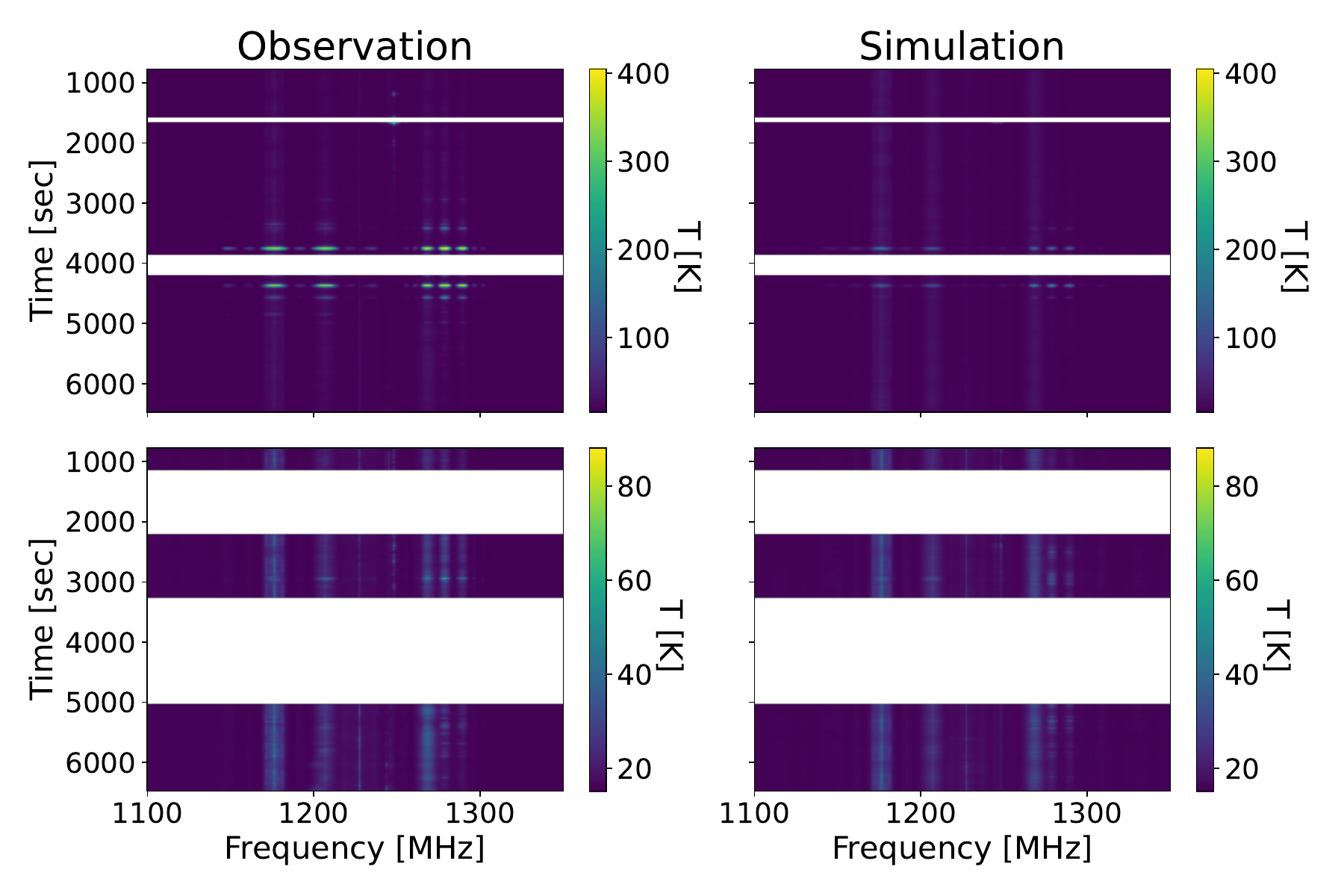}
                    \label{fig:chi_fitting_degree_waterfall}
                \end{subfigure}
                \hfill
                \begin{subfigure}[b]{0.48\textwidth}
                    \centering
                    \includegraphics[width=\textwidth]{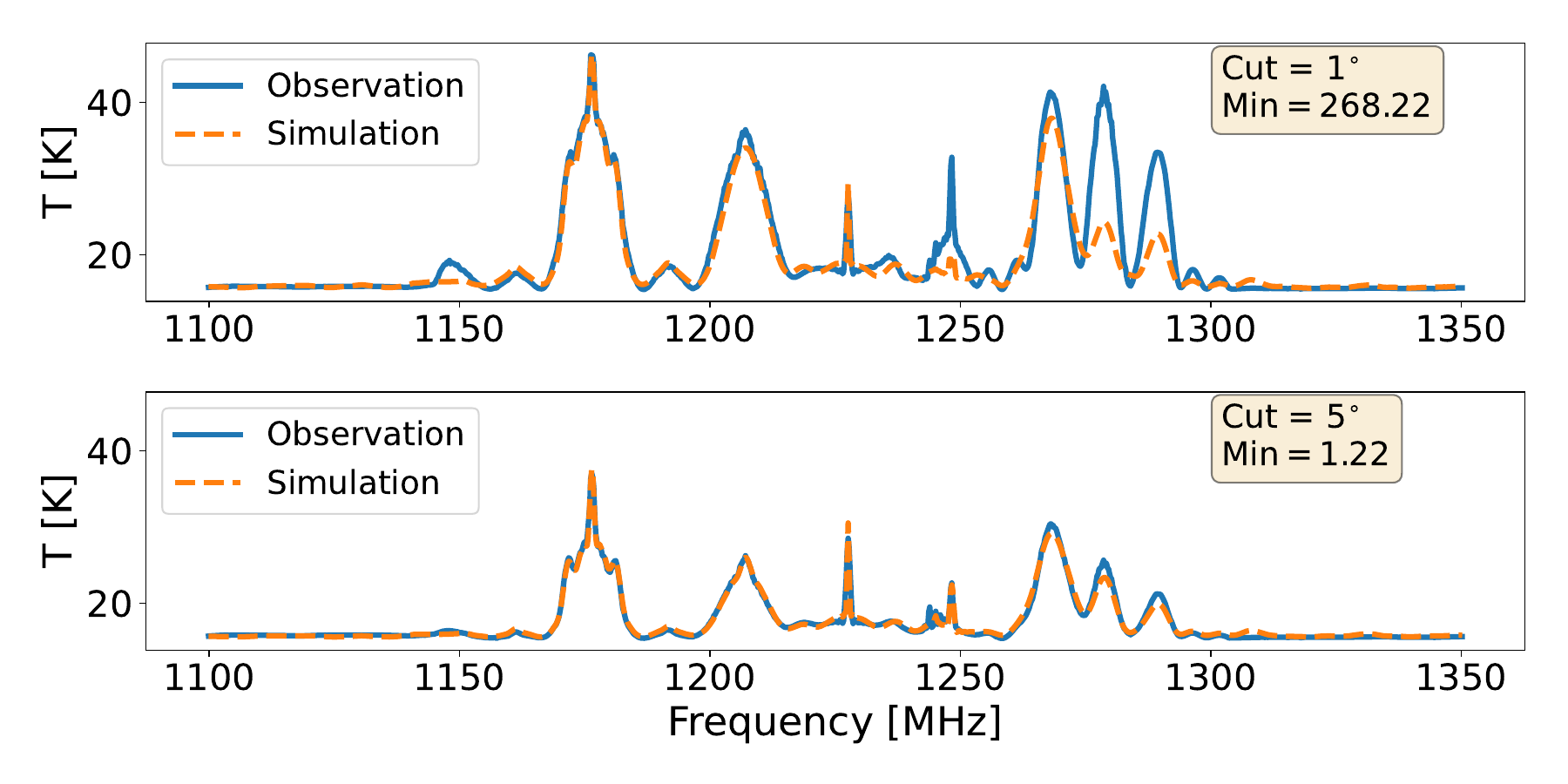}
                    \label{fig:chi_fitting_degree}
                    \end{subfigure}
                \caption{The best-fit comparison for the \emph{angular masking scenario}. The waterfall images ($1^{\textnormal{st}}$ \& $2^{\textnormal{nd}}$ row) for the various angular are displayed: the observation (left) and the simulation (right). The $1^{\textnormal{st}}$ row shows the $1^{\circ}$, and the $2^{\textnormal{nd}}$ row shows the $5^{\circ}$ angular masking cut, respectively. In the $3^{\textnormal{rd}}$ \& $4^{\textnormal{th}}$ row are the time averaged results as a function of frequency for the two angular cuts. The observation is in solid blue curve and the simulation in dashed orange.
                }
                \label{fig:angular_mask_case}           
            \end{figure}

            In order to address the saturation/non-linearity of the data which affects the fitting, we start by trying to remove any points that correspond to satellites close to the observation direction. Using a simulation with the same observation specifications, we calculate the positions of all the satellites as in \autoref{fig:ang_sep_gal} and remove timestamps corresponding to an angular distance below a certain threshold. This is done on both the observation data and simulations. We used an angular cut of $1^{\circ}$ and $5^{\circ}$. The results can be seen in \autoref{fig:angular_mask_case}. 
            A satellite with $\theta \leq1^{\circ}$ would be found inside the main lobe of the MeerKAT beam and saturate the signal, while $\theta \leq5^{\circ}$ should also remove the contribution of satellites that can cause non-linearity effects on the receivers without necessarily saturating the signal. We can see that indeed the fitting performs much better for the cut with $5^{\circ}$. On the other hand, the $5^{\circ}$ cut means that any satellite with a weaker signal will probably be below the error and won't be fitted for. Moreover, this cut is quite aggressive, removing up to 48\% of the data while the $1^{\circ}$ will remove about 6\%. Note that these implications for the fit refer to the main peaks in the data. The effect at lower frequencies where we do our cosmological analysis is very hard to see and doesn't impact the fit.

        \subsubsection{Thermal Masking}

\begin{figure}
                \centering
                \begin{subfigure}[b]{0.48\textwidth}
                     \centering
                    \includegraphics[width=\textwidth]{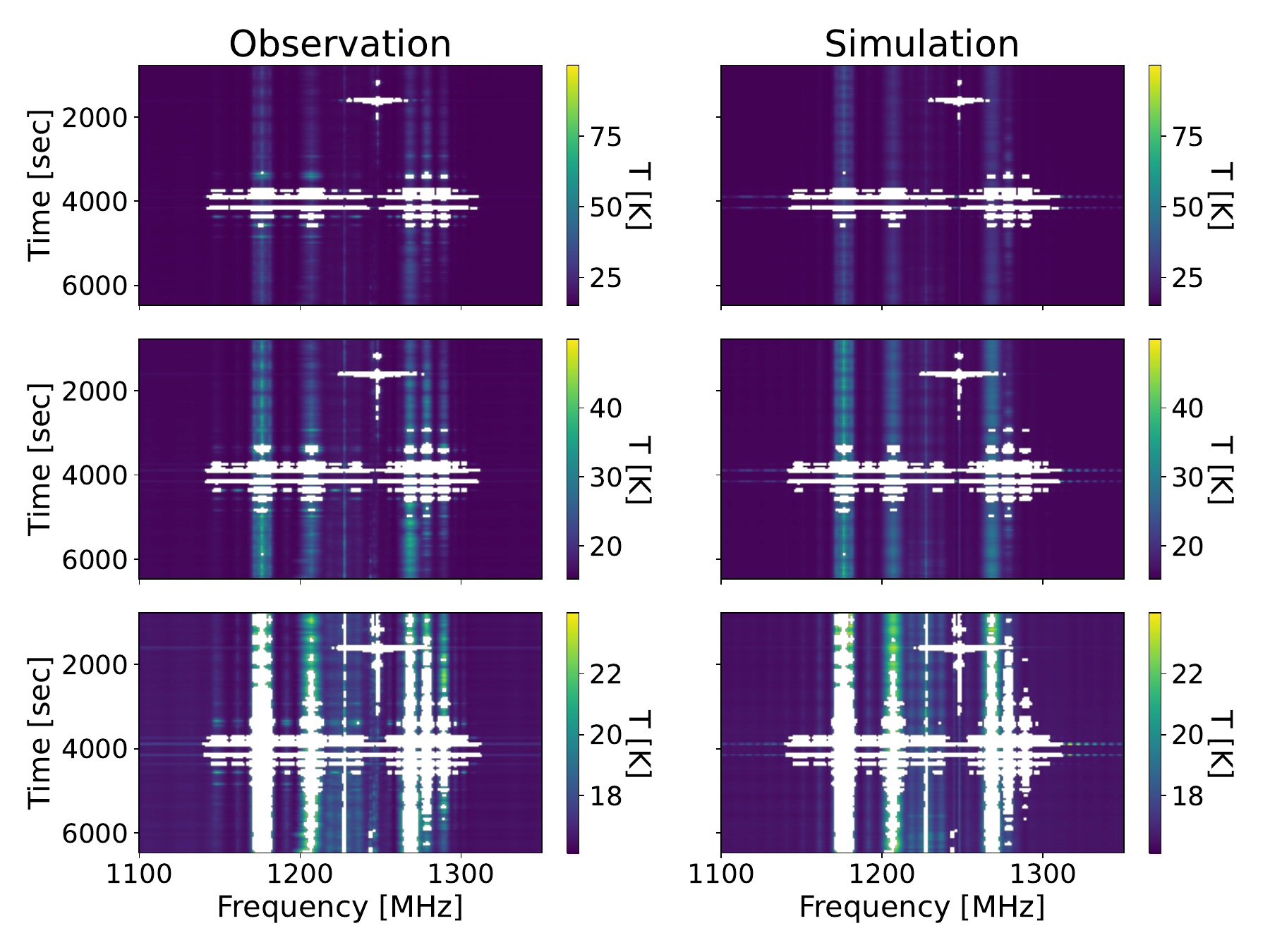}
                    \label{fig:chi_fitting_temporal_waterfal}
                \end{subfigure}
                \hfill
                \begin{subfigure}[b]{0.48\textwidth}
                    \centering
                    \includegraphics[width=\textwidth]{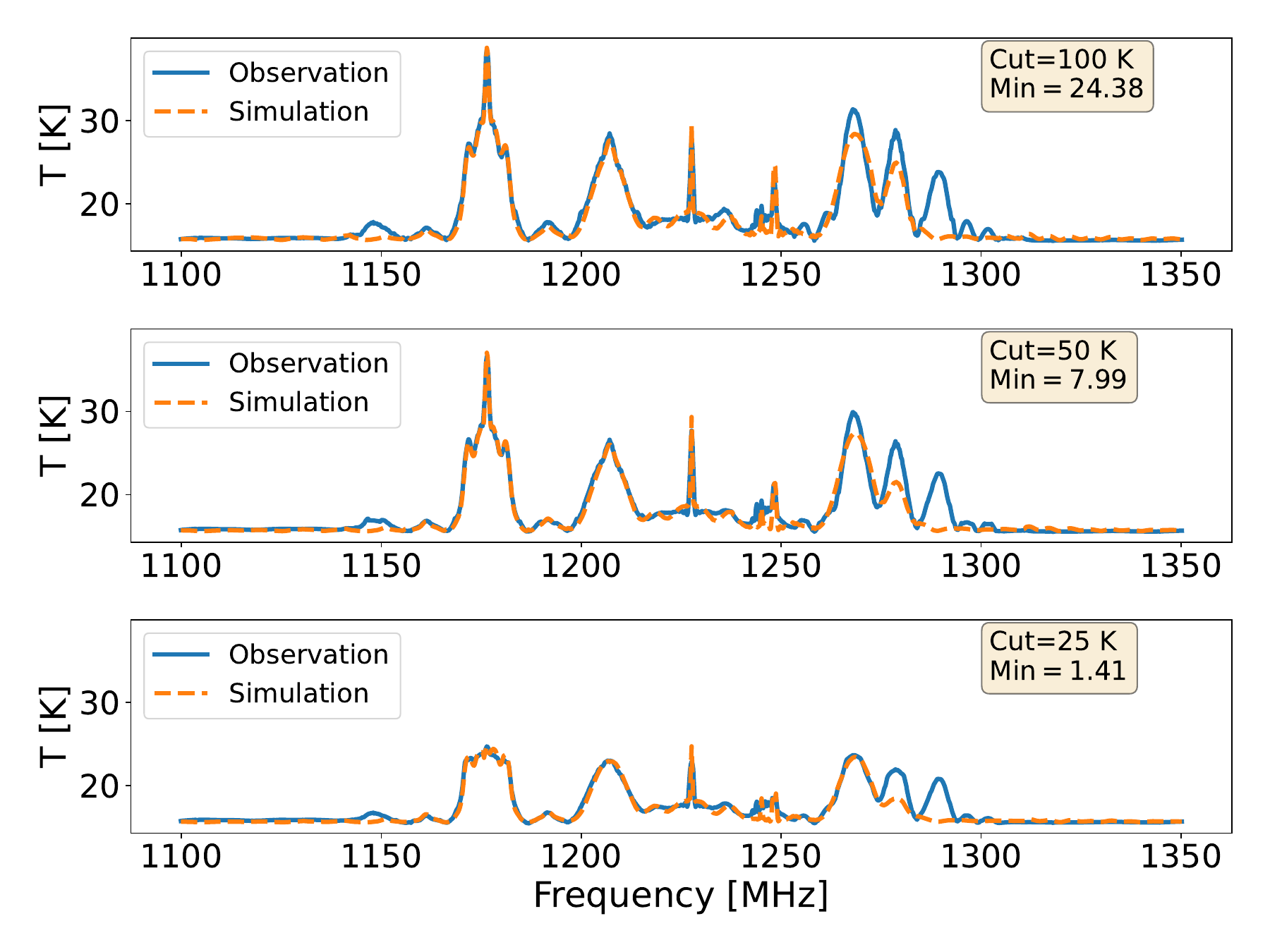}
                    \label{fig:chi_fitting_thermal}
                    \end{subfigure}
                \caption{The best-fit comparison for the \emph{thermal masking scenario}. The waterfall images ($1^{\textnormal{st}}, 2^{\textnormal{nd}}\ \&\ 3^{\textnormal{rd}}$ rows) with the observation (left), and the simulation (right). The time-averaged results below in the $4^{\textnormal{th}}, 5^{\textnormal{th}}\ \&\ 6^{\textnormal{th}}$ rows, with the observation (solid blue curve) and simulation (dashed orange curve). The various thermal thresholds given are: 100 K ($1^{\textnormal{st}}$ row), 50 K ($2^{\textnormal{nd}}$ row), and 25 K ($3^{\textnormal{rd}}$ row).}
                \label{fig:chi_thermal_fit}           
            \end{figure}

            This approach tries to address the issue of saturation directly, by making cuts when the temperature goes above a certain threshold. This is applied to the observation data both across time and frequency and the mask is then transferred to the simulation.  The threshold values for the thermal cut are 25 K, 50 K, and 100 K, corresponding to 19\%, 2\% and 1\% of data loss respectively. The results are shown in \autoref{fig:chi_thermal_fit}. The fitting improves as more of the saturated pixels are removed. However, this is still not as good as the $5^{\circ}$ scenario, even for the 25K cut, as can be seen by the frequency plots or the minimum values of the cost function. This is more obvious for the emission above 1250 MHz, which is a contribution from the Galileo E6 and Beidou B1-2 \& B3 signals. Also, on the lower end of the frequency scale, around 1150 MHz, the simulation does not fit for the smaller peaks, in particular the second side lobe emanating from the central peak at $\sim$1175MHz. 
            It is also interesting to note that the thermal cut of 25K limits the peak contributions from satellite signals, removing more signal there than the $5^{\circ}$ cut.

        \subsubsection{Full Thermal Masking}
        
\begin{figure}
                \centering
                \begin{subfigure}[b]{0.48\textwidth}
                     \centering
                    \includegraphics[width=\textwidth]{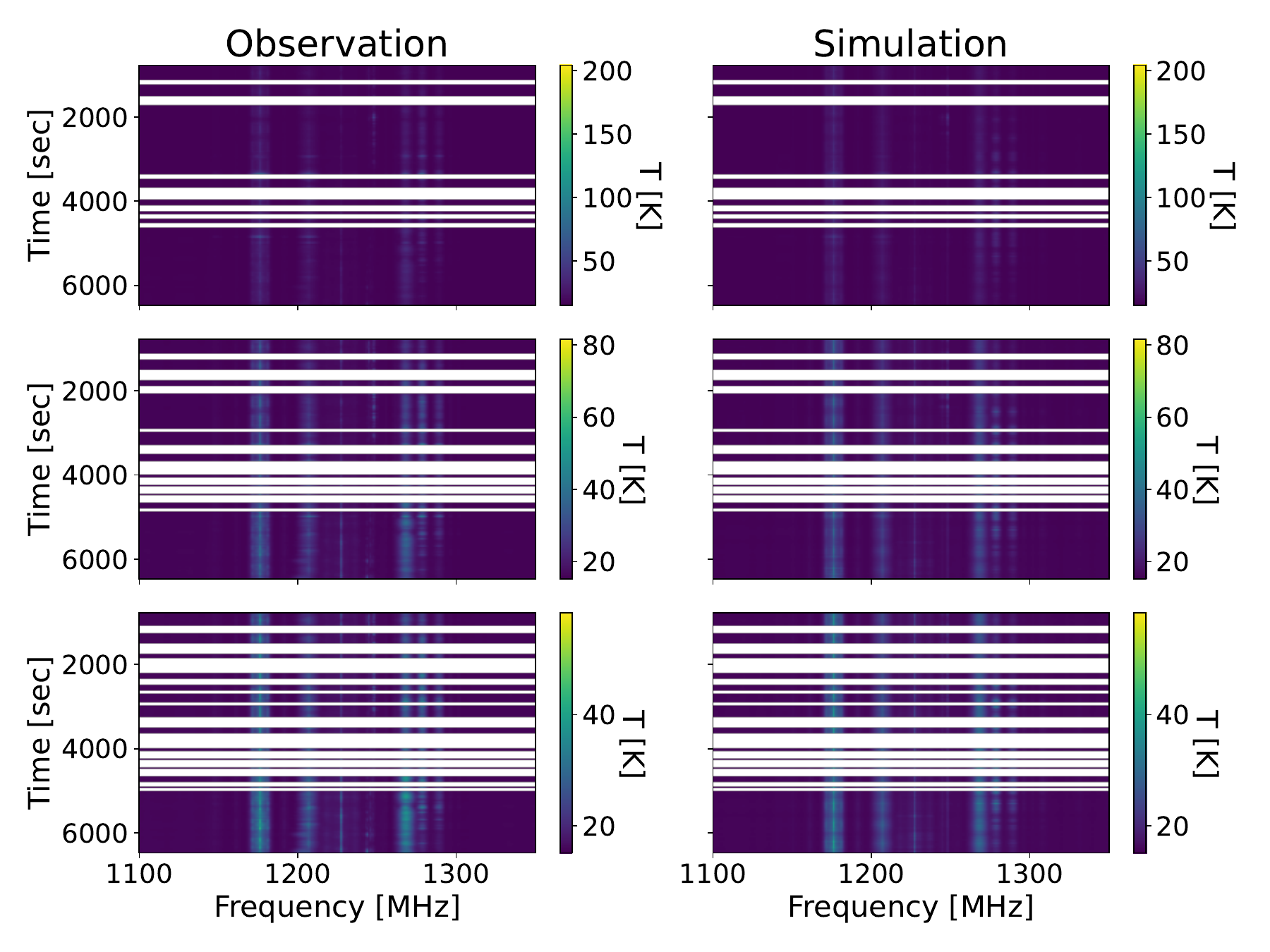}
                    \label{fig:chi_fitting_pixtime_waterfall}
                \end{subfigure}
                \hfill
                \begin{subfigure}[b]{0.48\textwidth}
                    \centering
                    \includegraphics[width=\textwidth]{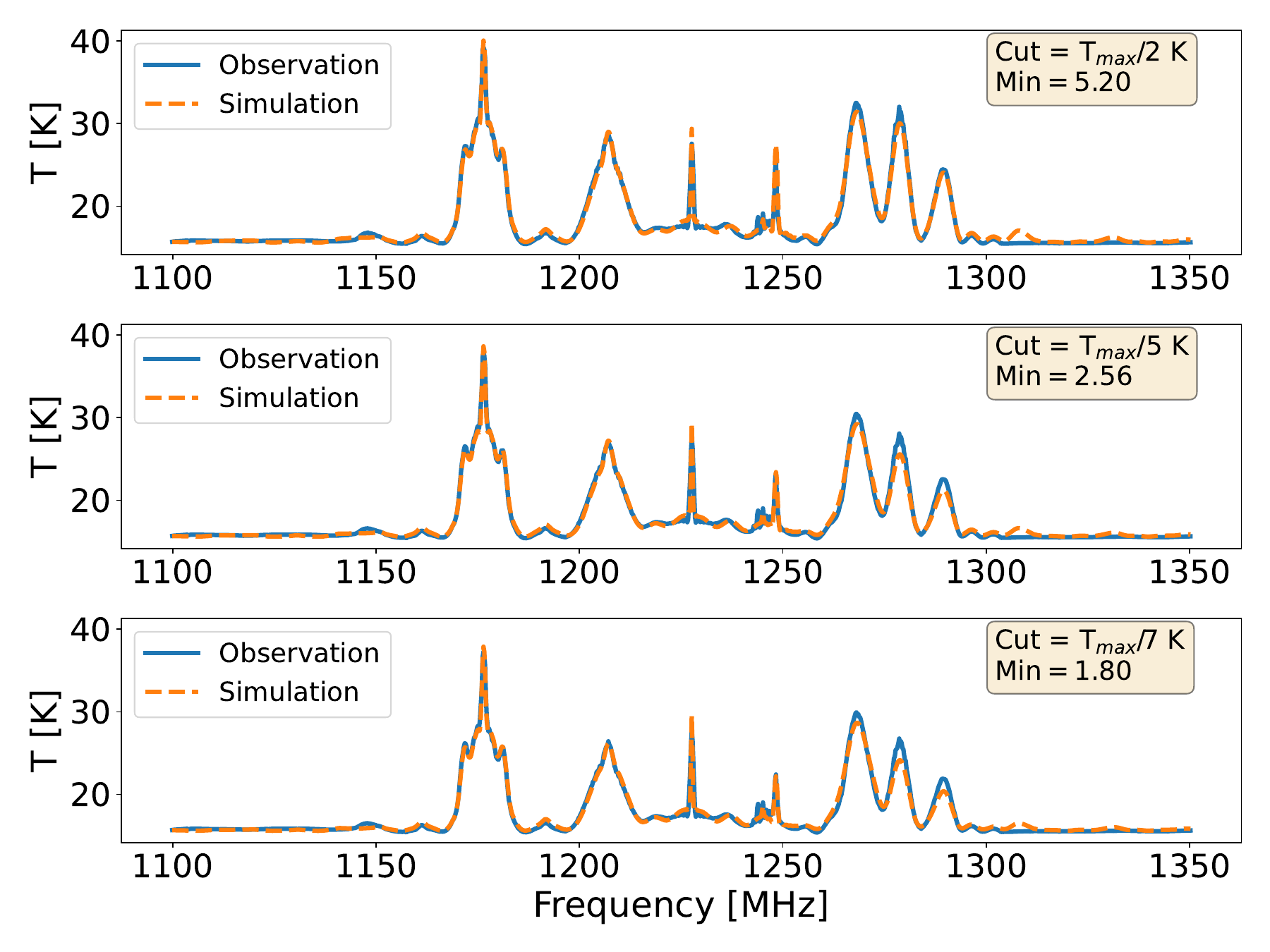}
                    \label{fig:fitting_pixtime}
                    \end{subfigure}
                \caption{The best-fit comparison for the \emph{full thermal masking scenario}. The waterfall images ($1^{\textnormal{st}}, 2^{\textnormal{nd}}\ \&\ 3^{\textnormal{rd}}$ rows) with the observational (left), and the simulation (right). The time-averaged results below in the $4^{\textnormal{th}}, 5^{\textnormal{th}}\ \&\ 6^{\textnormal{th}}$ rows, with the observation (solid blue curve) and simulation (dashed orange curve). The various pixel thermal limit are set at: 50.00\%  ($1^{\textnormal{st}}\ \&\ 4^{\textnormal{th}}$ row), 20.00\% ($2^{\textnormal{nd}}\ \&\ 5^{\textnormal{th}}$ row), and 14.29\% ($3^{\textnormal{rd}}\ \&\ 6^{\textnormal{th}}$ row) of the maximum temperature.}
                \label{fig:chi_fitting_pixtime}           
            \end{figure}

            In the previous scenario, only the "pixels" in time and frequency that exceeded a given threshold were flagged. We now consider a case where all the points across frequency are flagged for a given time as long as one of the values exceeds that cut-off. The reasoning is that once the receiver system saturates, the gain will become non-linear, affecting all frequencies, not just those where the signal hits the limit. Since the saturation limit seems to change between blocks, we decided to use a percentage of the maximum pixel value from each block instead of the threshold.
            We applied masking thresholds of : $T_{\rm max}/2$, $T_{\rm max}/5$, and $T_{\rm max}/7$ and show the results in \autoref{fig:chi_fitting_pixtime}. For this block (ending 211) and antenna m000, the maximum temperature $T_{\rm max}$ is $\approx 400$ K, which correlates to temperature thresholds of $\approx 200$ K, $\approx 80$ K and $\approx 60$ K respectively.
            As seen from the time average plots in the bottom panels, the fits are quite good, even for the higher-level thresholds. 
            In particular, they are clearly better than the equivalent temperature cut in the previous scenario. This shows that there is indeed unmodeled contamination away from the peak frequencies, probably due to non-linearities, which is removed by flagging the full-time dump. At the same time, the level of data loss is smaller than, say, the $5^{\circ}$, with 7\%, 14\% and 20\% for each threshold.


\autoref{tab:chi2_best_fit} summarises the results from the different fitting approaches. There is a clear trend of improvement as we remove more of the strong peaks, which should remove the saturated data points in the system. Other things could contribute to this improvement, such as reducing the out-of-band signal which is hard to model. In the next sections, we discuss possible issues in our model that could affect this fitting besides the expected saturation/non-linearity.
        \begin{table}
            \centering
            \begin{tabular}{|c|c|c|}
                \hline
                Masking Type & $C_2$ & \% flagged data \\
                \hline
                \hline
                \multicolumn{3}{|c|}{\textbf{Angular}} \\
                \hline
                $1^{\circ}$ & 261.24 & 5.4 \\ 
                $5^{\circ}$ & 1.21 & 47.55 \\
                \hline
                \multicolumn{3}{|c|}{\textbf{Thermal}} \\
                \hline
                100 K & 24.47 & 1.39 \\
                50 K & 7.99 & 2.25 \\
                25 K & 1.41 & 9.23 \\
                \hline
                \multicolumn{3}{|c|}{\textbf{Full Thermal}} \\
                \hline
                T$_{\rm max}/2$ & 5.20 & 7.49 \\
                T$_{\rm max}/5$ & 2.56 & 14.25 \\
                T$_{\rm max}/7$ & 1.80 & 20.1 \\
                \hline
            \end{tabular}
            \caption{The minimum cost function ($C_2$) values for various masking scenarios and the \% of data that is flagged per scenario.            
            }
            \label{tab:chi2_best_fit}
        \end{table}

        \subsubsection{Time variation of the $\alpha$ parameters}

    The $\alpha$ parameters are used to renormalise the satellite total power amplitude. In that sense, they should have little to no time evolution since the satellite signal should be stable in time. Three situations can change this. First, uncertainties in the telescope's primary beam, such as asymmetries, could be absorbed into the $\alpha$s, which can then change depending on the position of the satellite with respect to the telescope pointing. Second, as the observed signal is always a mix of different satellites, it won't be a surprise if different values of the same $\alpha$s provide a better fit depending on what satellites appear in the field of view. But such variations should be small when one single signal dominates the data. A third reason for time evolution (and difficulties in the fitting) is that individual satellites from a given constellation might not be emitting the exact same signals as we assume. The $\alpha$ amplitudes will try to correct for this and might change as these satellites from the same constellation go in and out of the field of view.

\begin{figure}
                \centering
                \begin{subfigure}[b]{0.48\textwidth}
                     \centering
                    \includegraphics[width=\textwidth]{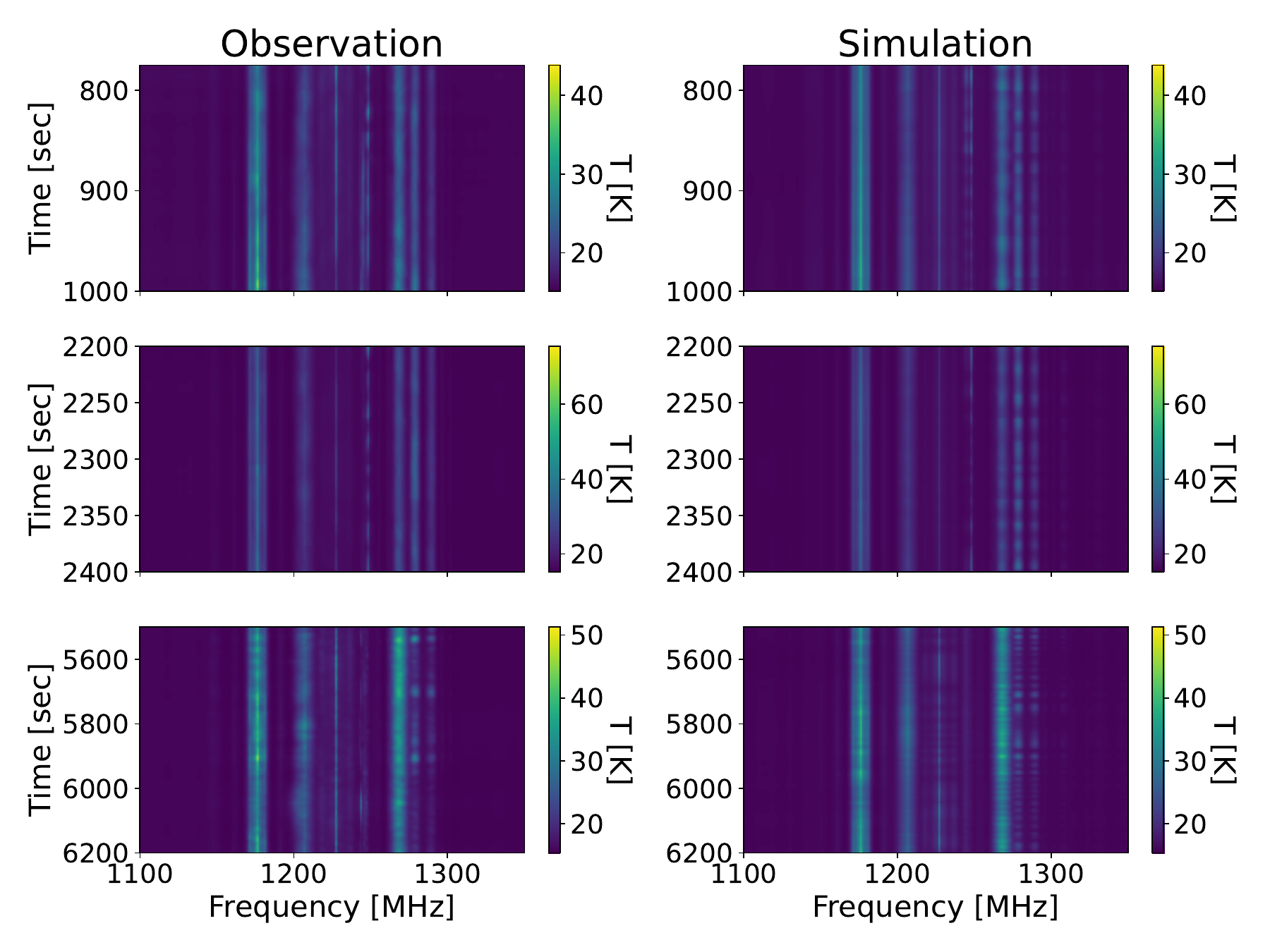}
                    \label{fig:chi_fitting_temporal_waterfall}
                \end{subfigure}
                \hfill
                \begin{subfigure}[b]{0.48\textwidth}
                    \centering
                    \includegraphics[width=\textwidth]{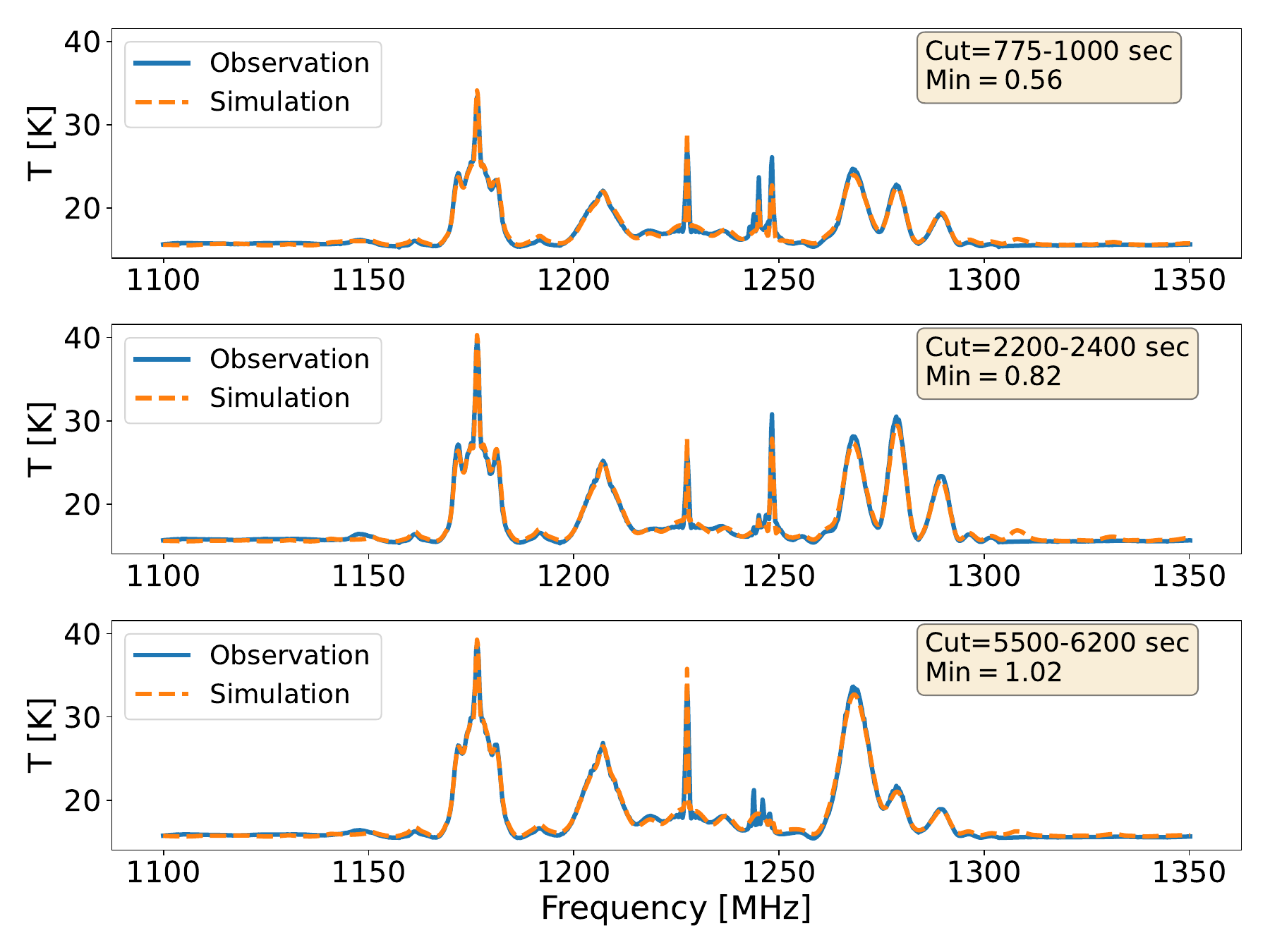}
                    \label{fig:fitting_temporal}
                    \end{subfigure}
                \caption{The best-fit comparison for different time windows. The waterfall images ($1^{\textnormal{st}}, 2^{\textnormal{nd}}\ \&\ 3^{\textnormal{rd}}$ rows) with the observation (left), and the $C_2$ case (right). The time-averaged results below in the $4^{\textnormal{th}}, 5^{\textnormal{th}}\ \&\ 6^{\textnormal{th}}$ rows, with the observation (solid blue curve) and simulation (dashed orange curve). The various temporal periods are set at 775-1000 sec ($1^{\textnormal{st}}\ \&\ 4^{\textnormal{th}}$ row), 2200-2400 sec ($2^{\textnormal{nd}}\ \&\ 5^{\textnormal{th}}$ row), and 5500-6200 sec ($3^{\textnormal{rd}}\ \&\ 6^{\textnormal{th}}$ row).}
                \label{fig:chi_fitting_temporal}           
            \end{figure}

\begin{figure}
                \centering
                \includegraphics[width=0.45\textwidth]{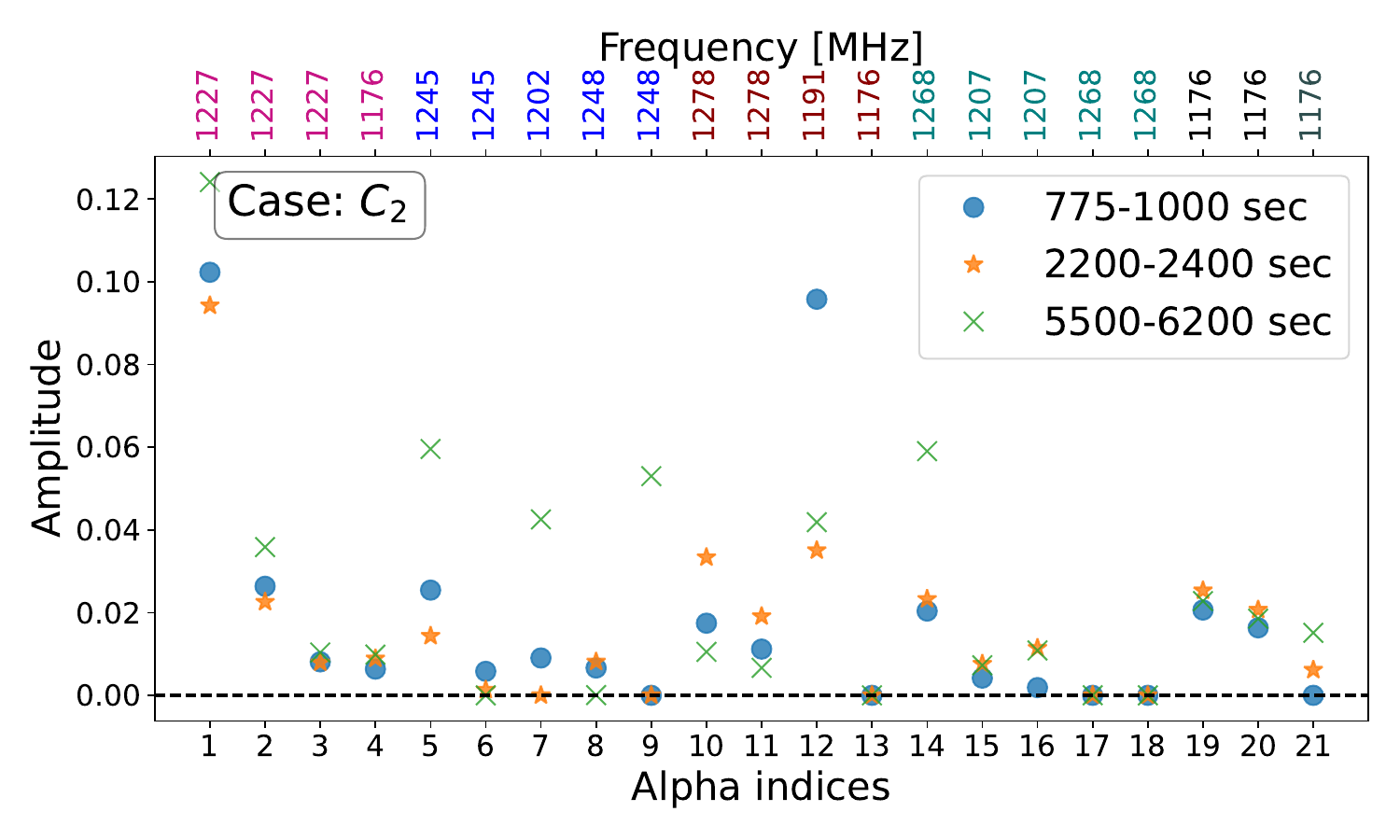}
                \caption{The $\alpha$ amplitudes fitted to each temporal window (\autoref{fig:chi_fitting_temporal}). The different periods are 775-1000 (blue $\bullet$), 2200-2400 (orange $\star$) and 5500-6200 (green \textbf{x}). The central frequency of each $\alpha$ signal is shown on the top y-axis. For the constellations: GPS-violet ($\alpha_1 - \alpha_4$), Glonass-blue ($\alpha_5 - \alpha_9$), Galileo-dark red ($\alpha_{10} - \alpha_{13}$), Beidou-teal ($\alpha_{14} - \alpha_{18}$), IRNSS-black ($\alpha_{19} - \alpha_{20}$) and SBAS-dark grey ($\alpha_{21}$).}
                \label{fig:chi_fitting_temporal_alpha}
            \end{figure}

    In \autoref{fig:chi_fitting_temporal}, we show the observation and fitting for three different time windows, corresponding to 775-1000 seconds, 2200-2400 seconds and 5500-6200 seconds. These time periods were picked so that no satellites enter the $5^{\circ}$ zone to avoid saturation/non-linearities. \autoref{fig:chi_fitting_temporal_alpha} shows the fitted $\alpha$ parameters. We can see a non-negligible variation in some of the $\alpha$s. In particular, $i=5, 7, 9, 12, 14$, which is probably due to satellites from Galileo and GLONASS entering the field of view. We note again that since the starting values for the $\alpha$s before the fit are set to zero, an $\alpha$ moving from zero just means that the satellites with the corresponding signal have moved into the field of view. However, once an $\alpha$ becomes non-zero, it should keep that value, if not for the abovementioned reasons.
    
\begin{figure}
            \centering
            \begin{subfigure}[b]{0.48\textwidth}
                 \centering
                \includegraphics[width=\textwidth]{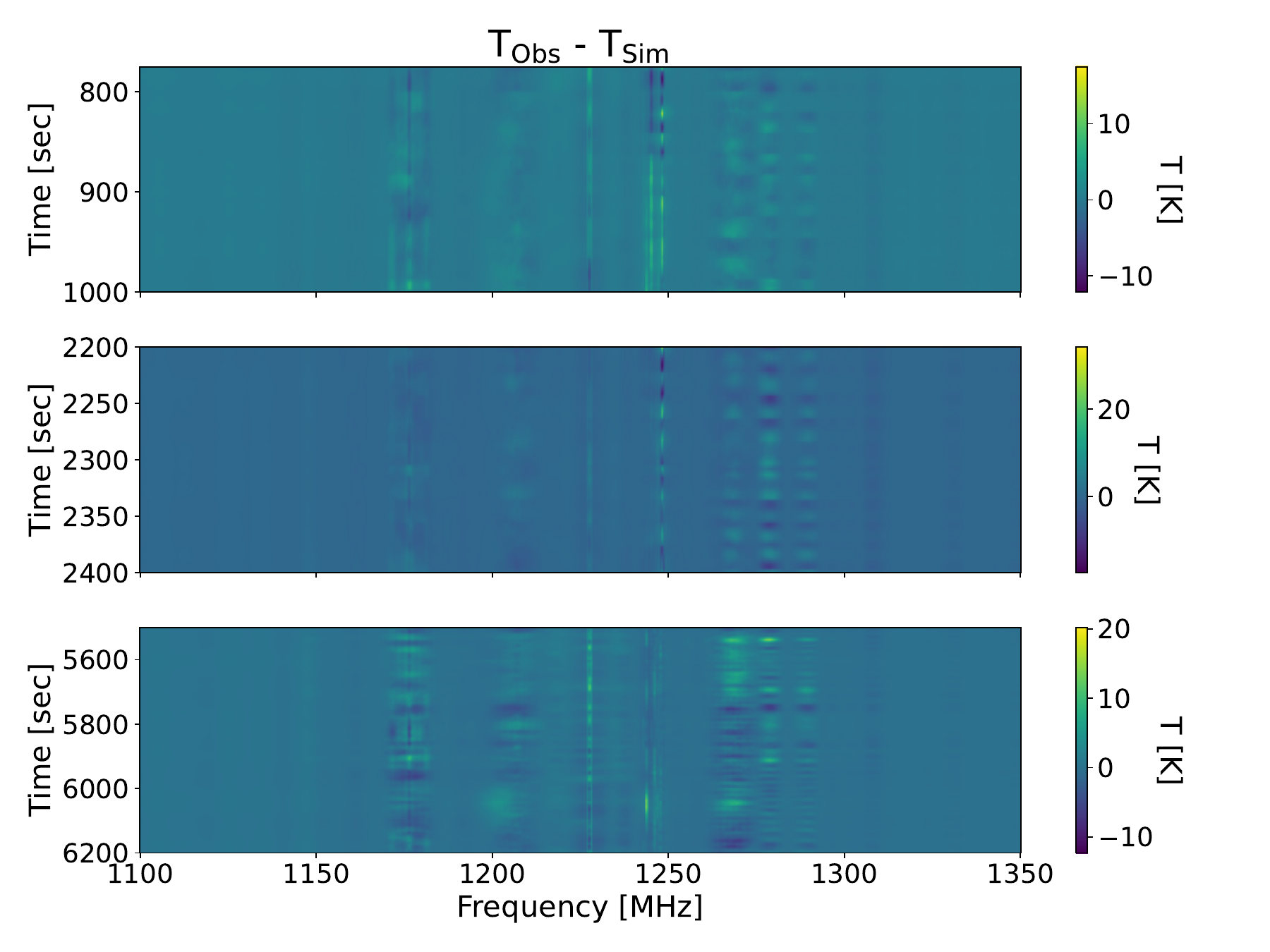}
                \label{fig:chi_fitting_temporal_residual_waterfall}
            \end{subfigure}
            \hfill
            \begin{subfigure}[b]{0.48\textwidth}
                \centering
                \includegraphics[width=\textwidth]{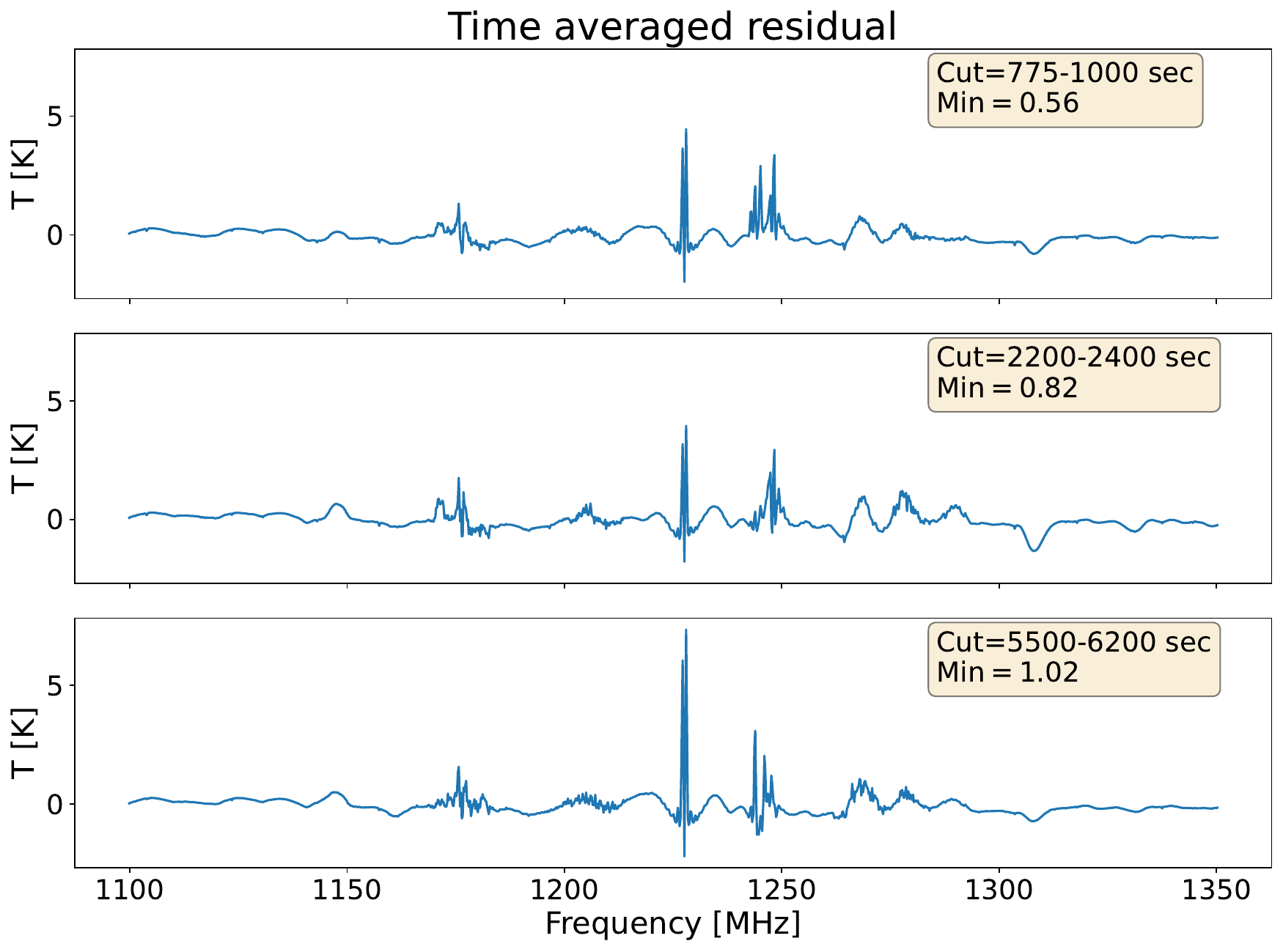}
                \label{fig:fitting_temporal_residual}
                \end{subfigure}
            \caption{The residual 2d and time-averaged residual 1d for the \emph{temporal masking scenario} (\autoref{fig:chi_fitting_temporal}). The various temporal periods are set at 775-1000 sec ($1^{\textnormal{st}}\ \&\ 4^{\textnormal{th}}$ row), 2200-2400 sec ($2^{\textnormal{nd}}\ \&\ 5^{\textnormal{th}}$ row), and 5500-6200 sec ($3^{\textnormal{rd}}\ \&\ 6^{\textnormal{th}}$ row).}
            \label{fig:chi_fitting_temporal_residual} 
        \end{figure}

         A better way to see the fitting accuracy is through the residuals as in \autoref{fig:chi_fitting_temporal_residual}.
          A reason for the observed difference could be that the frequency shapes used in the modelling (taken from literature, see \autoref{tab:sat_cat}) do not accurately follow the real satellite signal. When taking the time average of the residuals, the narrow spikes around $\approx$1225 MHz show the width of the peak is not captured perfectly in the model; this peak is associated with the GPS-L2 signal. Similarly, peaks at 1246 MHz \& 1248 MHz, belonging to Glonass-L2 signals, are not accurately accounted for by the simulation throughout the different time periods that are seen in the observational data. 
          Another example is the dip at 1305 MHz, indicating that the model has a peak that is not seen within the observational data; a possible explanation is that the out-of-band transmissions are being dampened; such dampening is not accounted for in our current modeling of the satellite signals.
         
        \subsubsection{Positional errors} \label{s:pos_errors}

            \begin{figure*}
                \centering
                \includegraphics[width=0.9\textwidth]{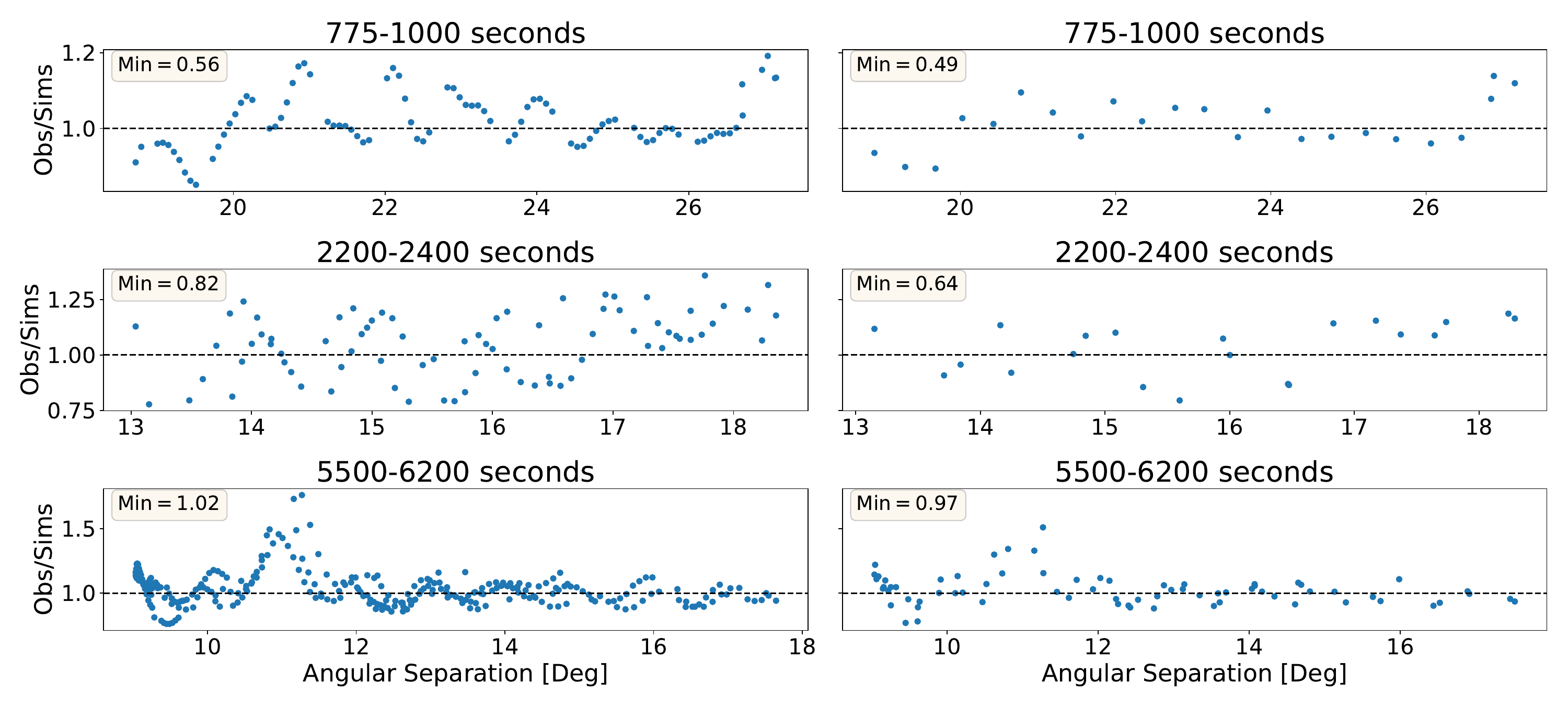}
                \caption{Ratio between the observational data and the model for the \emph{temporal masking scenario}). Isolating a single Galileo satellite operating at a frequency range (averaged) of  1278.6-1278.8 MHz as it moves with respect to the MeerKAT pointing. The $1^{\textnormal{st}}, 2^{\textnormal{nd}}\ \&\ 3^{\textnormal{rd}}$ rows represent the periods of 775-1000, 2200-2400 and 5500-6200 respectively. The left column is when no time average is applied, and the right is for a 10-second time average. The selected frequency is displayed with the CF for each panel.}
                \label{fig:termporal_resi_angular}
            \end{figure*}

        One last issue we will check with the fitting is the possibility of positional errors. As the dishes move back and forth, an error in the satellite position will translate to an amplitude difference that changes with time.
        To test this, we isolate the movement of one GAL satellite that gets very close to the telescope pointing and focus on 1278.75 MHz, corresponding to one of the peak frequency emissions from the GAL constellations. The immediate neighbouring channels to the peak frequency are averaged to increase the signal-to-noise.
        
        For this test, we use the data from above with temporal windows that do not have any satellites below $5^{\circ}$. Including data below the $5^{\circ}$ cutoff would make the result more obvious, but it would be affected by saturation and non-linearities.
         In \autoref{fig:termporal_resi_angular}, we show the signal ratio between the observation and the simulation. As the dishes and the satellite move, the same angular position of the satellite will be observed at different times. Therefore, we stack any time corresponding to the same angular position of the target satellite to increase the signal to noise. 
         The panels on the left reveal a ringing effect that oscillates around one, essentially following the shape of the telescope primary beam. The obvious implication is that there is an error in the angular position which translates to an error in the beam value we should use. There could also be smaller contributions from other satellites since we cannot fully isolate the signal. Those would further mix up the ratio as they will be at different positions for the same times considered here. 

         Even in the far sidelobes, the beam can change significantly in relative terms, with ratios of up to 100 between the minimum and maximum values as the beam oscillates, with a period of about 1 degree. The dishes are moving in azimuth at speeds of 5 arcmin/s, e.g. they would move 10 arcmin within the time resolution of 2 sec, which can lead to a mismatch even if the satellite is not moving.
         We therefore consider smoothing both the data and simulation over 10 seconds. The idea is that any errors in the position as the dishes move back and forth could be averaged out. The results are shown in \autoref{fig:termporal_resi_angular}. The oscillating structure seems to disappear, but part of that is simple because we now have fewer points. Overall, the ratio does get closer to 1, which improves the fitting, though not substantially.
     
    In summary, the satellite central frequencies are well represented in the observational data. The frequency structure of the different Global signals in particular GPS, GAL and Beidou were well described even with the overlapping contribution from GAL and Beidou at 1268 MHz. One concern comes from the GLONASS whose emitted power and gain are not well sourced in literature. Another issue of GLONASS is that the 1248 MHz had several signals with similar frequency structures which can cause difficulty in the fit given the uncertainties in the tabulated specs. Apart from the GLONASS issue, the models appear to have a good representation of the data, as long as saturation is kept to a minimum. The other challenge is whether satellite signals weaken faster the further we are from the central frequency.

    \section{Impact on the cosmological band} \label{s:application}
\begin{figure*}
                \centering
                \includegraphics[width=0.9\textwidth]{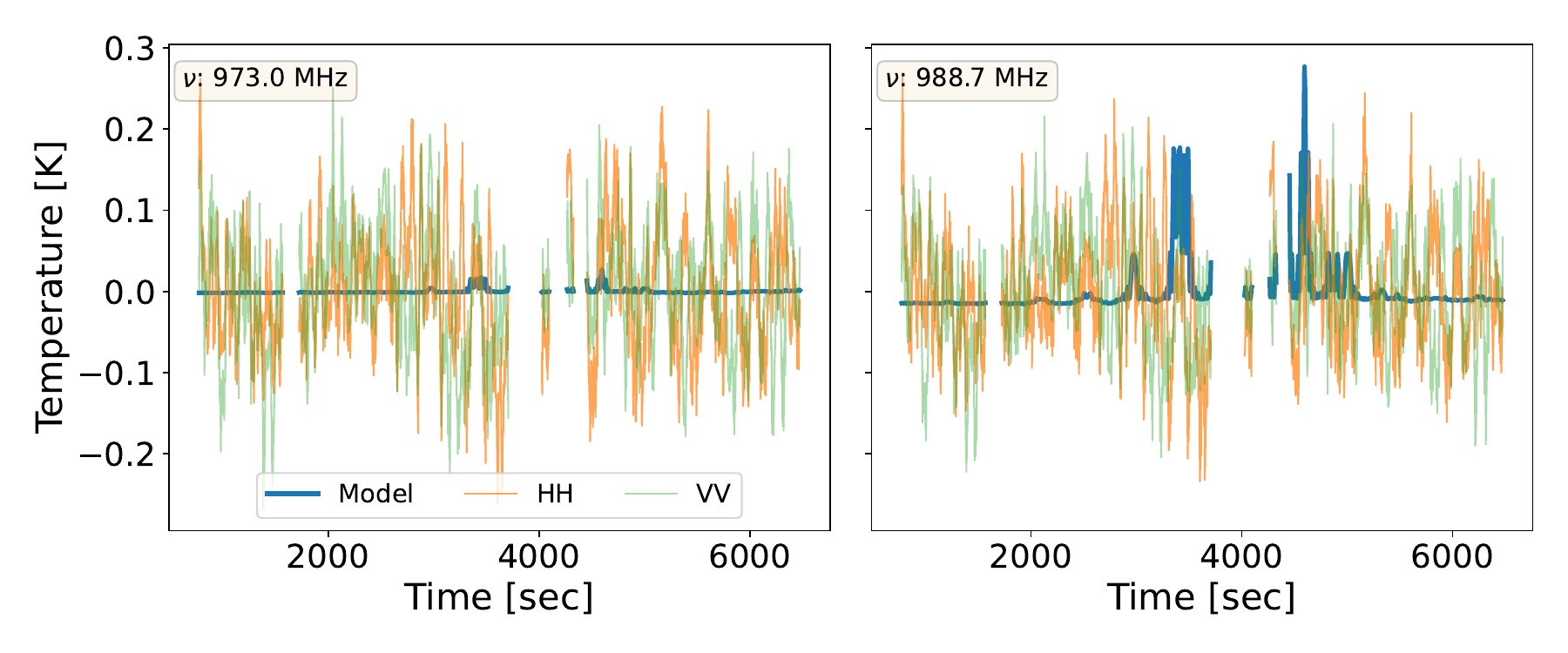}
                \caption{A temperature comparison across the observational scanning period between the satellite model (blue) and the residual calibrated observational data (W21) in the HH (orange) and VV (green) polarisation for the m000 antenna. Masks from W21 where included. The time average per channel is subtracted from each frequency channel. The frequency channels, 973.0 MHz (left panel) and 988.7 MHz (right panel), are selected as they reside in the cosmological band (973-1015 MHz) of the observation.}
                \label{fig:cosmo_band_together}           
            \end{figure*}

 So far, we have considered the impact of GNSS satellites around frequencies where they dominate. This allows for higher signal-to-noise when fitting the satellite simulations. These simulations can, in principle, be used to plan for observations that avoid strong contamination and also provide ways to flag the signal better. For cosmology, we have focused on a frequency band that we hope is completely free from satellite contamination, corresponding to  973-1015 MHz,  where \cite{cunnington_2023b} presented a measurement of the cross-correlation power spectrum. To further help clean the data, the flagging algorithms will usually completely remove time stamps where the signal saturates, even if the contamination is low at the target frequencies. \autoref{fig:cosmo_band_together} shows a comparison between the observation data and the simulation for two frequency channels within this "cosmology band". The same flags were applied to data and simulation, taken from \cite{Wang2020}. The $\alpha$s in the simulation were first fitted using data from the frequencies were we clearly see satellite peaks (1100 - 1350 MHz).
 
 Furthermore, to remove any background contamination, we have subtracted the time average for each channel, so the signal will fluctuate around zero. There is no obvious pattern in the data, which seems noise-dominated, and most of the satellite signal predicted by the simulations is within that noise. This makes it hard to try to fit for any out-of-band contamination from the satellites. However, it might happen that when we combine many observations together and calculate the {\HI} power spectrum, some out-of-band signals from the satellites will still contaminate the cosmological signal. Indeed, in \cite{cunnington_2023b}, we needed 30 PCA modes to clean the signal and achieve a detection. It's not clear what are the main reasons for such contamination. Probably a big part is related to issues in the bandpass calibration or other RFI. But it could be that there is some residual satellite contamination, although we do not see any obvious signature in the data. So in the next section, we use the fitted simulations to predict how much satellite contamination is expected in the cosmology band. We note that there is a fair amount of uncertainty in the out-of-band emission due to the lack of information in the literature.

        \subsection{Extracting the HI power spectrum in the presence of satellites}


            Our goal in this section is to test the impact of satellites on the ability of MeerKAT to measure the {\HI} power spectrum. To achieve this, we will combine the satellite simulations with simulations of the {\HI} signal and foregrounds. This will then be processed in the same way as we would do for the data, with foreground cleaning followed by a power spectrum estimator. The procedure is similar to what is described in \citet{2024MNRAS.527.4717I} where we go from time ordered data to data cubes and then power spectrum. To make the conclusions clearer we will neglect instrumental noise as this would tend to dilute any residual systematics.
                   
\begin{figure}
                \centering
                \includegraphics[width=0.48\textwidth]{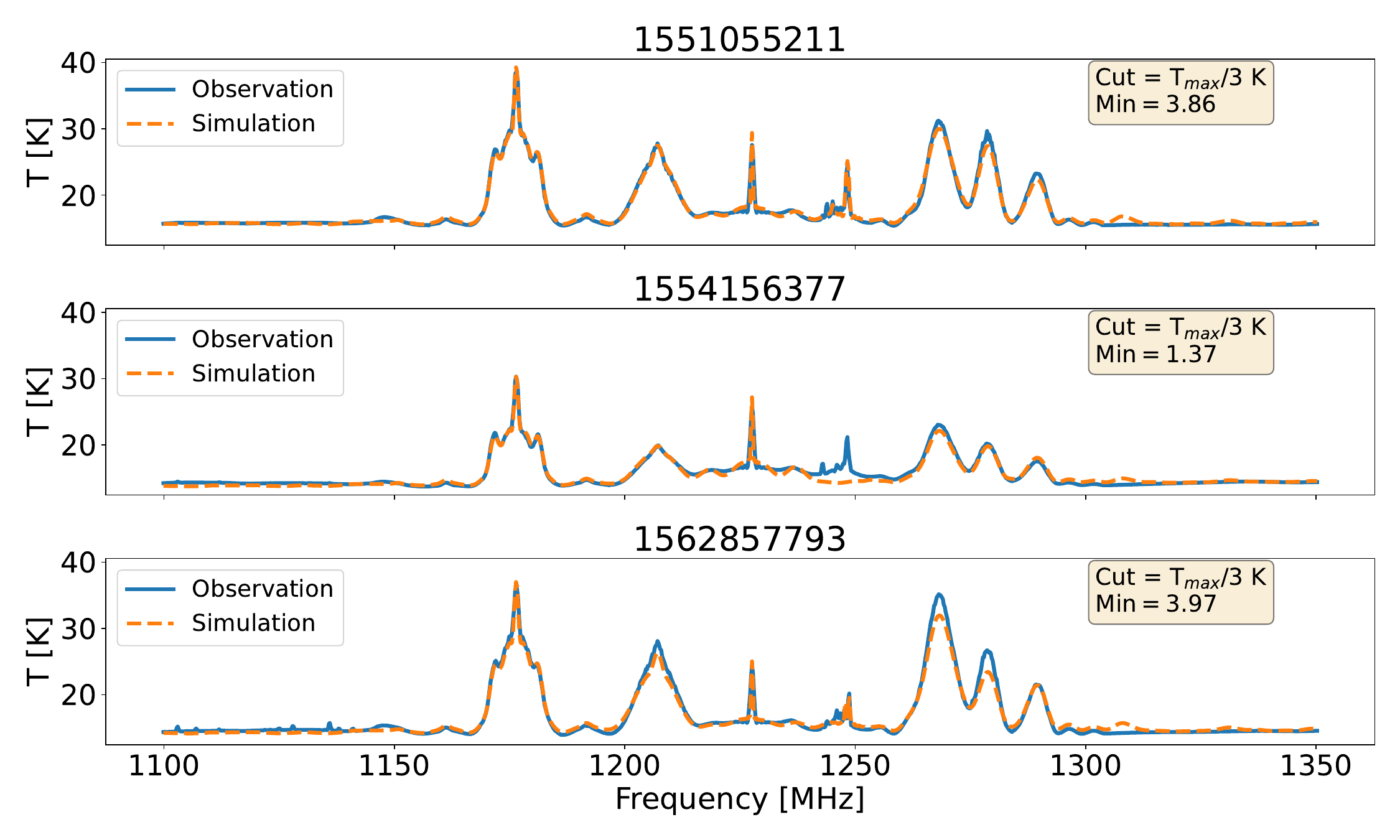}
                \caption{Best fit results for the \emph{full thermal masking scenario} with a 
                threshold ($T_{Th}$) = 33.33\% and three observation blocks.}
                \label{fig:3_block_pixthermal}
            \end{figure}

             We start by applying the same methodology as shown above to four more observational blocks. The inclusion of additional blocks (see \autoref{tab:observational_blocks}) is to improve the sky coverage and include a good mix of satellites that are expected to show up on different days. We fitted the simulation to each block and tried different masking scenarios.
             In \autoref{fig:3_block_pixthermal} we showcase three blocks (2 new additional blocks) using the best performing \emph{full thermal masking scenario} with a thermal threshold of 33\% of the maximum temperature to remove any saturated points. We decided to use these best fit models for further processing.
            The full simulated data blocks were then flagged. At this stage we opted to remove any data points with satellites closer than 5$^{\circ}$.  However, one of the blocks contains a GNSS satellite in geostationary orbit that is often below the 5$^{\circ}$ and would cause a large percentage of data to be masked. For this block, we applied a 1$^{\circ}$ flag instead.
            Since satellite movement differs across observations, the masks of the blocks do not overlap, and there are no gaps in the combined maps.

            For the {\HI} simulation, we used Gaussian density fields generated from a \texttt{CAMB} power spectrum \citep{lewis_2002} and generate HI maps at the sky positions we are observing. The Galactic emission component was generated assuming that the dominant contribution is from synchrotron emission and that synchrotron emission follows a power law parameterization:
            \begin{equation}
                T_\text{sync}(\nu, p) = T_\text{amp}(\nu, p)  \left(\frac{\nu}{408} \right)^{\beta_\text{sy}(p)},
            \end{equation} 
            where the amplitude template ($T_\text{amp}(\nu, p$)) at 408 MHz is the reprocessed, destriped Haslam 408 MHz map \citep{newhas_2015} with the combined CMB monopole and unresolved point source background constant offset removed \citep{wehus}. The spectral index template (${\beta_\text{sy}(p)}$) is the model for the spectral index map from \citet{mamd_2008}. 
            
            Both the {\HI} and Galactic foreground model required smoothing by the beam; whilst the EMSS beam is the optimal choice for modelling the satellite contributions, which are still of considerable magnitude up to several degrees away from the main lobe, a simple Gaussian beam model will suffice for smoothing the Galactic and {\HI} contributions as the aim of these simulations was to assess the impact of the satellite residuals on top of a smoothed sky signal \citep{Matsha2020}. The frequency-dependant Gaussian beam FWHM used was:
            \begin{equation}
                \theta_{{\rm{FWHM}}}(\nu) = 1.2^{\circ} \times \frac{1280\,{\rm{MHz}}} {\nu}.
            \end{equation}
            Finally, following again \citet{2024MNRAS.527.4717I}, we use the RA and DEC information from each data block to extract the corresponding foreground and {\HI} signal from the HEALPix \citep{healpix_2005} maps using tools from the \texttt{healpy} library. This way we generate equivalent TODs for the galactic synchrotron and {\HI}.

\begin{figure}
                \centering
                \includegraphics[width=0.5\textwidth]{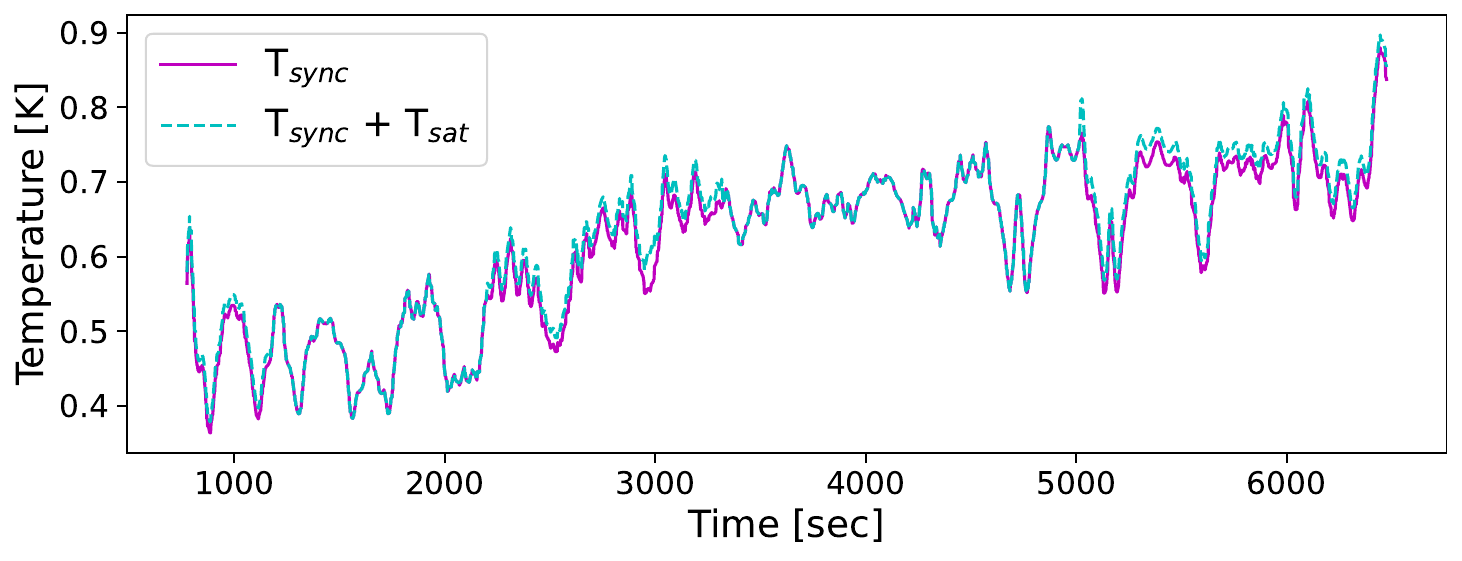}
                \caption{A comparison between the synchrotron emission only (magenta solid line) and the combined satellite and synchrotron emission (cyan dashed line) for the observational block of 155105211 after a $5^{\circ}$ angular cut is applied. )}
                \label{fig:temp_vs_time_211}
            \end{figure}

            For the five observation blocks, the residual satellite contributions are not visible by eye above the Galactic contributions within the TOD. \autoref{fig:temp_vs_time_211} gives an example of this by showing the frequency-averaged TOD for block 1551055211, both with and without the satellite contribution included in the simulation.  

            We now have 5 observation blocks (time ordered data), each including satellite signal, {\HI} and galactic synchrotron, convolved by the appropriate beam. As mentioned above, we do not include thermal noise so that we can test the results in the optimal case of high signal to noise, when small systematic residuals will become visible. A few ingredients are still missing: 
            the receiver temperature, the elevation dependant temperature (ground pickup and atmosphere) and the CMB monopole. However, these are constant or fairly constant over the 90 minutes blocks. Since we actually remove a smooth function of time in the processing of the real data, we can just neglect these contributions here. This means that we have also removed the time average of the combined signal at each frequency for each of the 5 blocks time ordered data.
            
\begin{figure}
                \centering
                \includegraphics[width=0.5\textwidth]{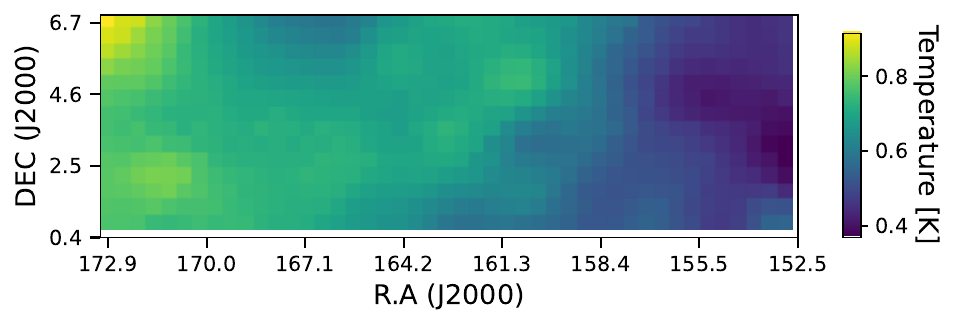}
                \caption{The combined system temperature map at $\nu = 983.5$ MHz for the five observation blocks. A pixel spacing of $0.4^{\circ}$ was applied.}
                \label{fig:map_example_211}
            \end{figure}

            The five observation blocks are then combined at each frequency to make a 2D map in RA and DEC. The TOD samples (Y) are averaged into map pixels (X) using a pointing matrix (A): 
            \begin{equation}
            X = ({\bf{A}}^{t} {\bf{A}})^{-1} {\bf{A}}^{t} Y, 
            \end{equation} 
            which is assembled using the \texttt{astropy} coordinate library and setting the pixel spacing to be $0.4^{\circ}$ (roughly two pixels per beam). Then, combining all the frequency maps, we have a 3D data cube. \autoref{fig:map_example_211} shows the total system temperature map at frequency 983.5 MHz.

            As the Galactic emission contribution dwarfs the {\HI} signal, the maps require foreground cleaning before power spectra can be made. We aim to determine how residual satellite contamination corrupts a measurement of the {\HI} signal in addition to standard problems of foreground cleaning, i.e. residual foreground contamination or conversely over-cleaning and removal of {\HI} signal \citep{cunnington_2023}. Therefore, we apply the same principal component analysis (PCA) foreground cleaning technique used in \cite{cunnington_2023b}. We first calculated the frequency-frequency covariance matrix of our data cube, then removed the largest eigenvalues under the assumption that the signal with the strongest frequency correlation is the Galactic signal. We increased the number of modes removed until it was clear we had strong signal loss.

            \begin{figure}
                \centering
                \includegraphics[width=0.48\textwidth]{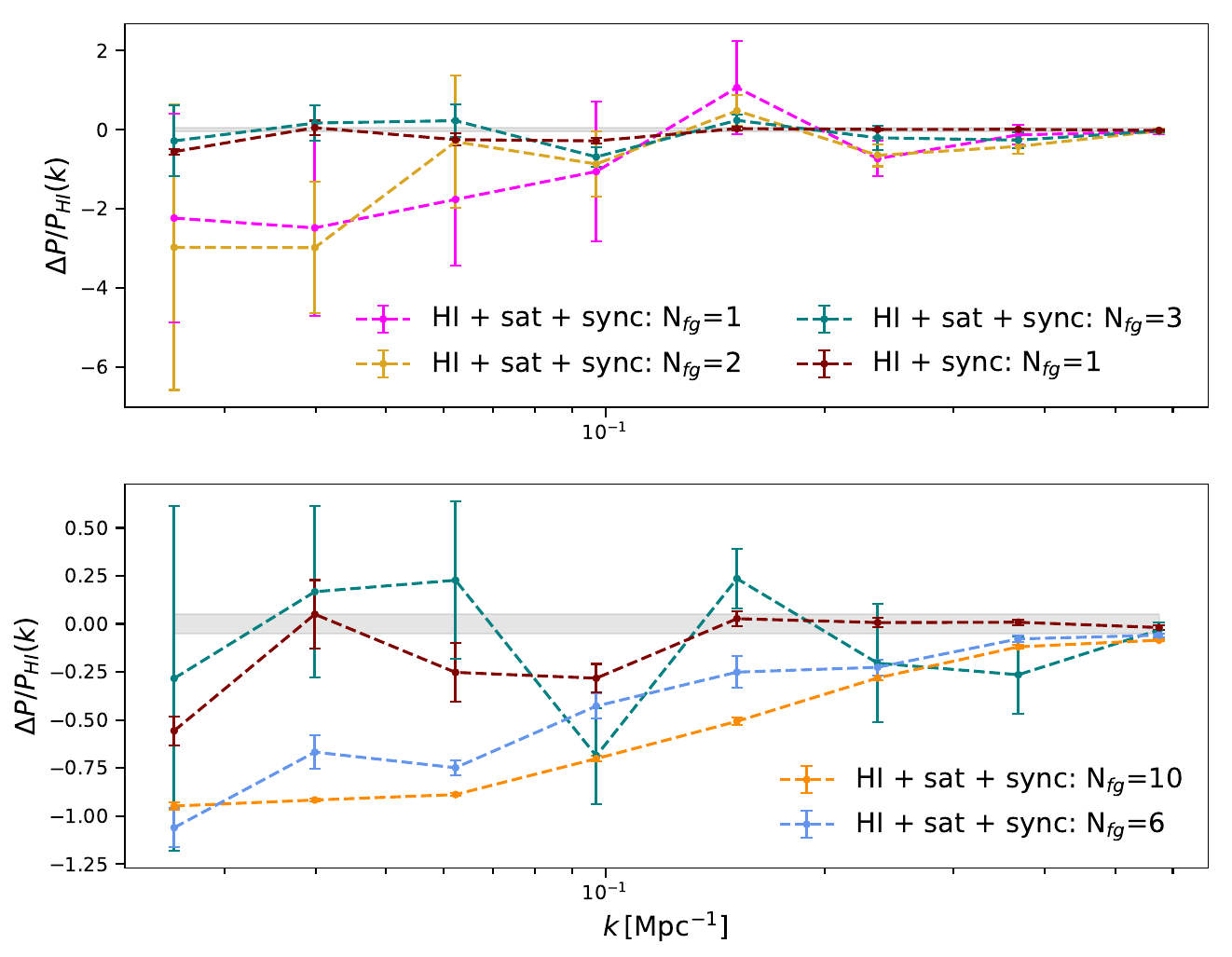}
                \caption{The percentage difference between our estimator and the {\HI} power spectrum. The estimator consists of the PCA cleaned cross-correlation power spectrum with the HI signal itself and is a proxy for cross-correlations with galaxy surveys.  The maroon curve represents {\HI} with synchrotron emission as the only contaminant and one PCA mode removed. The remaining curves contain residual satellite contamination with one (fuchsia), two (gold), 3 (teal), 6 (pastel blue) and 10 (orange) modes removed. The bottom panel highlights a zoomed-in version of the top panel with a larger number of modes removed.}
                \label{fig:5_avg_correlation}
            \end{figure}

            In order to better understand the residual contamination, we used an estimator based on the cross-correlation between the PCA-cleaned total emission maps and the input {\HI} maps (as convolved with our Gaussian beam model). The cross-correlation power spectra were calculated using functionality from the {\HI} simulations package \texttt{FastBox} \footnote{\url{https://github.com/philbull/FastBox}}, which creates a 3D Cartesian data cube. In reality, future data will require a more robust regridding of the maps \citep{2024MNRAS.528.5586C}. The power was averaged between 9 bins in $k-$space spanning from $k=0.02$ to  $k=0.7$. Twenty realisations of the {\HI} signal were made so that we can calculate the mean and standard deviation of the mean for the estimators.            

            \autoref{fig:5_avg_correlation} shows the percentage deviation of the cross-correlation power spectra with respect to the {\HI} signal, calculated as:
            \begin{equation}
                \dfrac{\Delta P(k)}{P_{\HI}(k)} = \dfrac{\rm{cross}(\rm{cleaned})}{\rm{auto}({\HI})} - 1
            \end{equation} 
            We considered two scenarios: the first where the only contaminant to the {\HI} signal is from diffuse Galactic synchrotron emission and the second where the contaminants are the combined Galactic and residual satellite temperatures. It can be seen that the removal of a single mode of frequency-correlated contaminants is enough for the scenario where only {\HI} and synchrotron emission are present in the maps. For the inclusion of residual satellite emission, we show the effect of different numbers of modes removed, choosing to zoom in on the most significant number of modes removed (bottom panel) due to the scales of the y-axis.
            
            We know that the satellite contamination is not smooth in frequency, so we do not necessarily expect PCA cleaning to work. Still, the power is reduced as we increase (slightly) the number of modes, showing that the PCA is able to distinguish the satellite signal from {\HI} up to a certain point.
            At the largest $k$-scales, the residual satellite emission always adds power regardless of the level of cleaning, although increasing the number of modes decreases power.  
            Conversely, at the smaller $k$-scales, the {\HI} emission is dominant, and so all the power spectra start to converge to zero. We highlight the line of $\Delta P(k)/P_{\HI}(k) = 0$ across all k-scales in grey as a visual aide.
            
            Between the largest and smallest scales, the residual satellite contamination can be seen to cause fluctuations in power, which seems to imply a level of mode mixing between the residual satellite contamination, residual foreground contamination and the {\HI} signal. Looking at the cross-power spectra for one mode removed compared to 10 modes removed, it can be seen that the level of mode mixing is inversely proportional to the number of modes removed, i.e: the more residual satellite contamination cleaned, the smoother the power spectrum ratio.
            
            Although removing ten modes from the total emission maps results in the smoothest cross-correlation, we have over-cleaned to such an extent that the {\HI} cannot be detected until $k > 0.3 $ Mpc$^{-1}$. Therefore we would argue that the optimum number of modes removed is three as this preserves {\HI}, albeit slightly contaminated, across the largest number of $k$ scales. We note that these results are based on the assumption that our simulations can be extended to these frequencies since no clear satellite signal is seen in this "cosmology" frequency band. Moreover, due to the intermittent nature of the satellite contamination, we expect this RFI contamination to average down as more blocks are combined.

\begin{figure}
                \centering
                \includegraphics[width=0.48\textwidth]{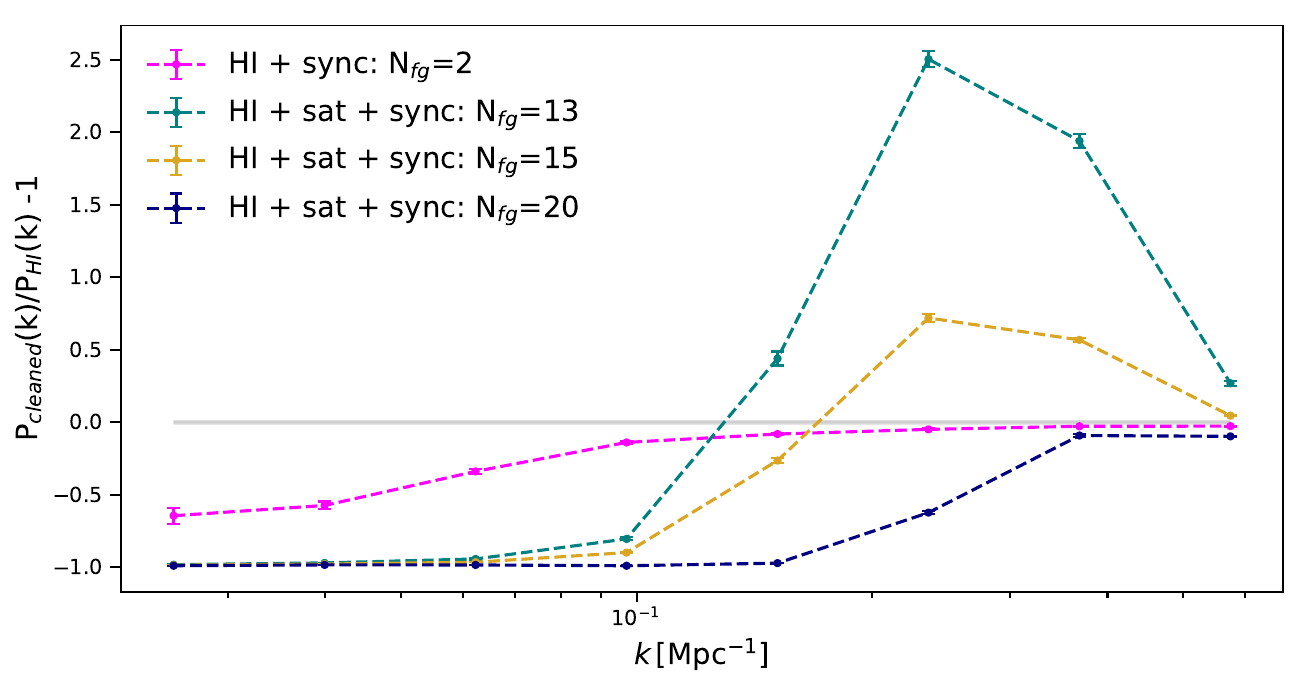}
                \caption{The auto-correlation for the {\HI} power spectrum with different modes cleaned with PCA.  The fuchsia curve represents {\HI} with synchrotron emission as the only contaminant and two modes removed. Meanwhile, the remaining curves contain residual satellite contamination with 13 (teal), 15 (gold), and 20 (navy blue) modes removed. The grey straight line is the ratio for the {\HI} only.}
                \label{fig:5_avg_aut_correlation}
            \end{figure}

            Finally, we also consider looking at the {\HI} in the autocorrelation (see \autoref{fig:5_avg_aut_correlation}), which required a larger number of modes to be removed. With 20 modes removed, we could detect {\HI} in the auto at the two largest k values but with strong signal loss across larger scales. Note as well that any robust conclusion on the number of modes removed requires an estimation of the signal loss, which is beyond the scope of this work.
            
\section{Conclusion} \label{s:con_fw}
    The ability to determine the effects of RFI from satellite emissions has been and will continue to be a growing concern as more satellites enter Earth's orbit and the sensitivity of radio telescopes increase. Until now, flagging has been the primary tool by radio astronomers to deal with unwanted RFI from both terrestrial and orbital sources. However, even with removing contaminated channels, we fear emissions from these sources would progress into channels that we assumed to be clean. To investigate these effects, we developed in this paper a set of detailed simulations capable of predicting the signal of GNSS satellites across the sky and fitted these to real data from MeerKAT single dish observations. Note that while we have not gone into detail concerning other telecommunication satellites, particularly those in a geostationary orbit, we have looked at the possible impact these satellites have on our data. In appendix~\ref{s:geo_sats}, we outlaid the movement of two satellites that operate in the L-band. However, we saw no changes in the data that would correspond to the nearby movement of the satellites. This reaffirms ITU regulations that this  L-sub-band is restricted to RNSS only.

    We constructed a method to calibrate the usually flagged satellite emission within the 1140-1310 MHz region and applied it to data from the MeerKAT single-dish {\HI} Intensity Mapping project (W21). We allowed the power of the different satellite signals to vary in the simulation and fitted their amplitude to the data. This is done for all frequencies and time-stamps. 
    We found that the fitting starts failing when the signal becomes too strong, which we attribute to the system going non-linear (possibly from the Analog-to-Digital Converter). This usually occurs when a satellite operating in the frequency range of interest breaches the 5$^{\circ}$ angular separation distance from the telescope's pointing. 
    To remedy this, we applied various masking scenarios ("angular", "thermal" and "full-thermal"). A $5^{\circ}$ cut seems to perform well but leads to a high percentage of flagging. We found that masking all the time stamps when the signal goes above a certain threshold performs as well or better while keeping more data. We also see some temporal variation of the fitted signal power. This is probably not due to a true time dependence of the power but because of other uncertainties that are absorbed by our fitted amplitudes. In particular the fact that satellites in a constellation may have different signals while we assume they all share the same signals. Nevertheless, our fitted simulations are able to capture most of the structures seen in the data.
    

    Finally, we looked at the effects of the satellite emission propagating to the lower frequency channels, mainly the 973-1015 MHz "Cosmology" frequency range. We showed that the simulated satellite signal is well within the observational data uncertainties for any given time dump and channel. 
    The next step was to examine how satellite emission would affect the auto \& cross-correlation of {\HI} power spectrum at this frequency range. To check this, we ran end-to-end simulations for a few observational blocks that included HI and foregrounds. We saw that the satellite signal strongly contaminates the HI power spectrum. Applying PCA cleaning, we were able to recover the power spectrum reasonably well in cross-correlation, when removing more than 3 modes.
    The auto-power spectrum however, was strongly contaminated, requiring up to 20 modes and leading to strong signal loss. This may indicate that satellite contamination might be the dominant contaminant in our current L-band analysis, although such conclusion requires some extrapolation in our knowledge of the satellite out-of-band emission.


\section*{Acknowledgements}

    We thank Chris Finlay for useful discussions throughout the work and Wenkai Hu for comments on the draft.
    BNE and MGS acknowledge support from the South African Radio Astronomy Observatory and National Research Foundation (Grant No. 84156).
    We acknowledge using the ILIFU\footnote{\url{www.life.ac.za}} cloud computing facility through the Inter-University Institute for Data Intensive Astronomy. The South African Radio Astronomy Observatory, a facility of the National Research Foundation, an agency of the Department of Science and Innovation, operates the MeerKAT telescope. JF thanks the support of Funda\c{c}\~{a}o para a Ci\^{e}ncia e a Tecnologia (FCT) through the research grants UIDB/04434/2020 and UIDP/04434/2020 and through the Investigador FCT Contract No. 2020.02633.CEECIND/CP1631/CT0002. 
    This result is part of a project that has received funding from the European Research Council (ERC) under the European Union's Horizon 2020 research and innovation programme (Grant agreement No. 948764).
    SC is supported by a UK Research and Innovation Future Leaders Fellowship grant [MR/V026437/1].
    IPC is supported by the European Union within the Next Generation EU programme [PNRR-4-2-1.2 project No. SOE\_0000136].
    AP is a UK Research and Innovation Future Leaders Fellow [grant MR/X005399/1].
    LW is a UK Research and Innovation Future Leaders Fellow [grant MR/V026437/1].




\bibliographystyle{mnras}
\bibliography{biblio}


\appendix

    \section{GNSS Power Spectrum Density} \label{app:psd}

    Below are various binary offset  carriers (BOC) used in this paper. We have incorporated them into our GNSS simulation and examples of their spectral shape we shown in \autoref{ss:psd}. 
    These are a combination of sine and cosine sub-carriers. Here the sub-carrier rate is denoted as $n_s$. A value $n$ is defined as $n=2 n_s/n_c$, which defines the number of half periods of the sub-carrier in one chip interval duration, where $n_c$ is denoted as the chip rate. We denote the sine and cosine sub-carrier for the BOC as BOC$_{\textnormal{sin}}$ and BOC$_{\textnormal{cos}}$, respectively. For simplicity we will set $w_c = \pi / n_c f_0$ and $w_s = \pi / n_s f_0$.  $f_0$ is the reference frequency of the signal and is set at 1.023 MHz to obtain the spectrum $S_{xx} = |P_{xx}|^2$ where $xx$ denotes the specific modulation in operation.

        \begin{itemize}
            \item BPSK:
            \begin{equation}
                P_{\textnormal{BPSK}_{(n_c)}}(\nu) = \dfrac{\textnormal{sinc} \big( \pi \nu /[n_c f_0] \big)}{\sqrt{n_c f_0}}.
            \end{equation}
            \item BOC$_{\textnormal{sin}}$ \& $n$ even:
            \begin{equation}
                P_{\textnormal{BOC}(n_s, n_c)}(\nu) = \sqrt{n_{c}f_{0}} \dfrac{\textnormal{sin} \bigg(w_c \nu \bigg)}{\pi \nu} \textnormal{tan} \bigg(\dfrac{w_s \nu}{2} \bigg)
                \label{eq:boc_s_even}
            \end{equation}
    
            \item BOC$_{\textnormal{sin}}$ \& $n$ odd:
            \begin{equation}
                P_{\textnormal{BOC}(n_s, n_c)}(\nu) = \sqrt{n_{c}f_{0}} \dfrac{\textnormal{cos} \bigg( w_c \nu \bigg)}{\pi \nu} \textnormal{tan} \bigg(\dfrac{w_s \nu}{2} \bigg)
                \label{eq:boc_s_odd}
            \end{equation}
            
            \item BOC$_{\textnormal{cos}}$ \& $n$ even:
            \begin{equation}
                P_{\textnormal{BOC}_{(n_s, n_c)}}(\nu) = \sqrt{n_c f_0} \dfrac{\textnormal{sin} \bigg(w_c \nu\bigg)}{\pi \nu} \dfrac{1 - \textnormal{cos} \bigg( \dfrac{w_s \nu}{2}\bigg)}{\textnormal{cos} \bigg(\dfrac{w_s \nu}{2} \bigg)}
                \label{eq:boc_c_even}
            \end{equation}
    
            \item BOC$_{\textnormal{cos}}$ \& $n$ odd:
            \begin{equation}
                P_{\textnormal{BOC}_{(n_s, n_c)}}(\nu) = \sqrt{n_c f_0} \dfrac{\textnormal{cos} \bigg(w_c \nu\bigg)}{\pi \nu} \dfrac{1 - \textnormal{cos} \bigg( \dfrac{w_s \nu}{2}\bigg)}{\textnormal{cos} \bigg(\dfrac{w_s \nu}{2} \bigg)}
                \label{eq:boc_c_odd}
            \end{equation}
            
            \item altBOC \& $n$ even \footnote{\url{https://gssc.esa.int/navipedia/index.php/AltBOC_Modulation}} :
            \begin{equation}
                S_{\textnormal{alt}_{(n_s, n_c)}}(\nu) = \dfrac{4 n_c f_0}{\pi^2} \dfrac{\textnormal{sin}^2 \bigg( w_c \nu \bigg)}{\textnormal{cos}^2 \bigg(\dfrac{w_s \nu}{2} \bigg)} \phi_{n_s, n_c}(\nu)
                \label{eq:altboc_even}
            \end{equation}
            
            \item altBOC \& $n$ odd:
            \begin{equation}
                S_{\textnormal{alt}_{(n_s, n_c)}}(\nu) = \dfrac{4 n_c f_0}{\pi^2} \dfrac{\textnormal{cos}^2 \bigg( w_c \nu \bigg)}{\textnormal{cos}^2 \bigg(\dfrac{w_s \nu}{2} \bigg)}
                \phi_{n_s, n_c}(\nu) 
                \label{eq:altboc_odd}
            \end{equation}
            with
            \begin{equation*}
                \phi_{n_s, n_c}(\nu) = \bigg[ \textnormal{cos}^2\bigg(\dfrac{w_s \nu}{2}\bigg) - \textnormal{cos}\bigg(\dfrac{w_s \nu}{2} \bigg) -2\textnormal{cos} \bigg(\dfrac{w_s \nu}{2}\bigg) \textnormal{cos}\bigg(\dfrac{w_s \nu}{4} \bigg) +2 \bigg]
            \end{equation*}
            
            \item TMBOC
            
            Note that Composite BOC (CBOC)\footnote{\url{https://www.unoosa.org/documents/pdf/icg/activities/2007/icg2/presentations/24.pdf}} has the same modulation structure as TMBOC.
            \begin{align}
                    S_{\textnormal{TMBOC(6,1,1/11)}}(\nu) & = \dfrac{10}{11}S_{\textnormal{BOC}_{\textnormal{sin}}(1,1)}(\nu) + \dfrac{1}{11} S_{\textnormal{BOC}_{\textnormal{sin}(6,1)}}(\nu) \\
                    & = \dfrac{f_0}{11 \pi^2 \nu^2} \textnormal{sin}^2 \bigg( \dfrac{\pi \nu}{f_0} \bigg) \bigg[ 10 \textnormal{tan}^2\bigg(\dfrac{\pi \nu}{2 f_0} + \textnormal{tan}^2 \bigg( \dfrac{\pi \nu}{1 f_0} \bigg) \bigg]
            \end{align}
        
        \end{itemize}

\section{GNSS signal information} \label{app:gnss_signal}

    \autoref{tab:number_of_satellites} shows the number of satellites above the horizon for the telescope during the observation. In \autoref{tab:sat_cat}, we summarise the details we used to model the observed emission of the satellite constellation's signals.

    \begin{table}
        \centering
        \begin{tabular}{c|c|c|c|c|c|c|c}
            Const. & GPS & GLO & GAL & BEI & IRNSS & SBAS & QZS \\
            \# & 15 & 13 & 13 & 18 & 5 & 9 & 0  
        \end{tabular}
        \caption{Number of satellites per constellation included, based on MeerKAT's field of view during the scanning period, for the observation on 25/02/2019 - 00:40:11 GMT.}
        \label{tab:number_of_satellites}
    \end{table}

        \begin{table*}
            \centering
            \begin{tabular}{cccccccccccc}
            \textbf{\#} & \textbf{Sys} &  \textbf{Band} &  \textbf{Signal} &  \textbf{Frequency [MHz]} & \textbf{Modulation} & \textbf{ Rate [MHz]} &  \textbf{P$_{\textnormal{t}}$ [dBW]} &  \textbf{G$_{\textnormal{t}}$ [dBi]}  \\  
            \hline
            \hline
            - & GPS & L1 & P(Y) & 1575.420 &  BPSK(10) & 10.2300 & 13.5 & 13.5   \\
            - & GPS & L1 & C/A & 1575.420 &   BPSK(1) &  1.0230 & 16.5 & 13.5   \\ 
            - & GPS & L1 & L1C-D & 1575.420 &  TMBOC(6,1,4/33) &  1.0230 & 10.0 & 10.0   \\ 
            - & GPS & L1 & M-D & 1575.420 & BOC$_{\textnormal{sin}}$(10,5) &  5.1150 & 18.2 & 13.5   \\ 
            \cellr
            1 & GPS & L2 & P(Y) & 1227.600 &  BPSK(10) & 10.2300 & 10.0 & 10.0   \\ 
            \cellr
            2 & GPS & L2 & L2CM & 1227.600 &   BPSK(1) &  0.5115 & 10.0 & 10.0   \\ 
            \cellr
            3 &  GPS & L2 & M-D & 1227.600 & BOC(10,5)$_{\textnormal{sin}}$ &  5.1150 & 16.0 & 13.5   \\ 
            \cellr
            4 & GPS & L5 & L5I & 1176.450 &  BPSK(10) & 10.2300 & 18.0 & 18.0   \\ 
             & GLO & L1 & L1SF(P) & 1602.000 &   BPSK(5) &  5.1100 & 10.0 & 10.0   \\ 
            \cellr
            5 & GLO & L2 & L2SF(P) & 1245.100 &   BPSK(5) &  5.1100 & 10.0 & 10.0   \\ 
            \cellr
            6 & GLO & L2 &  L2OF(C/A) & 1245.100 & BPSK(0.5) &  0.5110 & 10.0 & 10.0   \\ 
            \cellr
            7 & GLO & L3 &  L3OC-D & 1202.025 &  BPSK(10) & 10.2300 & 10.0 & 10.0   \\ 
            \cellr
            8 & GLO & L2 &  L2OC-D & 1248.300 &   BPSK(1) &  1.0230 & 13.0 & 12.0   \\ 
            \cellr
            9 & GLO & L2 &  L2OC-P & 1248.300 &  BOC(1,1) &  0.5115 & 5.0 & 5.0   \\ 
            - & GAL & E1 & OS-D(B) & 1575.420 &   CBOC(6,1,1/11) &  1.0230 & 10.0 & 10.0   \\ 
            \cellr
            10 & GAL & E6 & CS-P(C) & 1278.750 &   BPSK(5) &  5.1150 & 16.0 & 15.0   \\ 
            \cellr
            11 & GAL & E6 &  PRS(A) & 1278.750 &  BOCcos(10,5) &  5.1150 & 18.0 & 16.0   \\ 
            \cellr
            12 & GAL &  E5ab &  PRS(A) & 1191.795 & AltBOC(15,10) & 10.2300 & 10.0 & 10.0   \\ 
            \cellr
            13 & GAL &   E5a & E5a-D & 1176.450 & AltBOC(15,10) & 10.2300 & 6.0 & 6.0   \\ 
             - & BDS-2 &  B1-2 &  RS & 1561.098 &   BPSK(2) &  2.0460 & 10.0 & 10.0   \\ 
            \cellr
            14 & BDS-2 & B3 &  RS & 1268.520 &  BPSK(10) & 10.2300 & 16.0 & 16.0   \\ 
            \cellr
            15 & BDS-2 &   B2b &  OS & 1207.140 &   BPSK(2) &  2.0460 & 14.0 & 12.0   \\ 
            \cellr
            16 & BDS-2 &   B2b &  RS & 1207.140 &  BPSK(10) & 10.2300 & 18.0 & 18.0   \\ 
            - & BDS-3 &  B1-2 &  OS & 1561.098 &   BPSK(2) &  2.0460 & 10.0 & 10.0   \\ 
            - & BDS-3 & B1 &  B1C-Dl & 1575.420 &  TMBOC(6,1,4/33) &  1.0230 & 10.0 & 10.0   \\ 
            \cellr
            17 & BDS-3 & B3 &  B3C-Dm & 1268.520 &  BPSK(10) & 10.2300 & 15.0 & 13.5   \\ 
            \cellr
            18 & BDS-3 & B3 &  B3A-Dm & 1268.520 & BOC(15,2.5) &  2.5575 & 10.0 & 10.0   \\ 
            - & QZS-1 & L1 & C/A & 1575.420 &   BPSK(1) &  1.0230 & 10.0 & 10.0   \\ 
            - & QZS-1 & L1 & L1C-D & 1575.420 &  BOC(1,1) &  1.0230 & 10.0 & 10.0   \\ 
            - & QZS-1 & L1 & L1C-D & 1575.420 &  TMBOC(6,1,4/33) &  1.0230 & 10.0 & 10.0   \\ 
            - & QZS-1 & L1 & SAIF & 1575.420 &   BPSK(1) &  1.0230 & 10.0 & 10.0   \\ 
            \cellr
            - & QZS-1 & L2 & L2CL & 1227.600 &   BPSK(1) &  0.5115 & 10.0 & 10.0   \\ 
            \cellr
            - & QZS-1 & L6 &  L61(LEX)n & 1278.750 &   BPSK(5) &  5.1150 & 10.0 & 10.0   \\ 
            \cellr
            - & QZS-1 & L6 & L62o & 1278.750 &   BPSK(5) &  5.1150 & 10.0 & 10.0   \\ 
            \cellr
            - & QZS-1 & L5 & L5I & 1176.450 &  BPSK(10) & 10.2300 & 16.0 & 16.0   \\ 
            \cellr
            - & QZS-1 & L5 & L5Q & 1176.450 &  BPSK(10) & 10.2300 & 16.0 & 16.0   \\ 
            \cellr
            19 & IRNSS & L5 & SPS & 1176.450 &   BPSK(1) &  1.0230 & 16.0 & 14.0   \\ 
            \cellr
            20 & IRNSS & L5 & RS-D & 1176.450 &  BOC(5,2) &  2.0460 & 18.0 & 16.0   \\  
            - & SBAS & L1 & C/A & 1575.420 &   BPSK(1) &  1.0230 & 13.0 & 13.5   \\  
            \cellr
            21 & SBAS & L5 & L5I & 1176.450 &  BPSK(10) & 10.2300 & 18.0 & 16.0   \\
            \end{tabular}
            \caption{List of satellite signals used in the simulation. Only signals with an index were potentially present and therefore included (first columnn). 
            "Sys": constellation system; "Band": allocation of signal transmission; "Signal": name of radio transmission; "Frequency": peak frequency of transmission; "Modulation": frequency structure of signal; "Rate": n$_{\textnormal{c}} \times f_0$; "P$_{\textnormal{t}}$": transmitted power of transmission; "G$_{\textnormal{t}}$": antenna gain of transmission \citep{Harper2018,  g._montenbruck_2017_table}. The highlighted grey regions indicate the signals we considered for each of the constellations. All the satellites from a given constellation are assumed to have the same signals.} 
            \label{tab:sat_cat}
        \end{table*}

       \section{Geostationary satellites} \label{s:geo_sats}
        \begin{figure}
                    \centering
                    \includegraphics[width=0.5\textwidth]{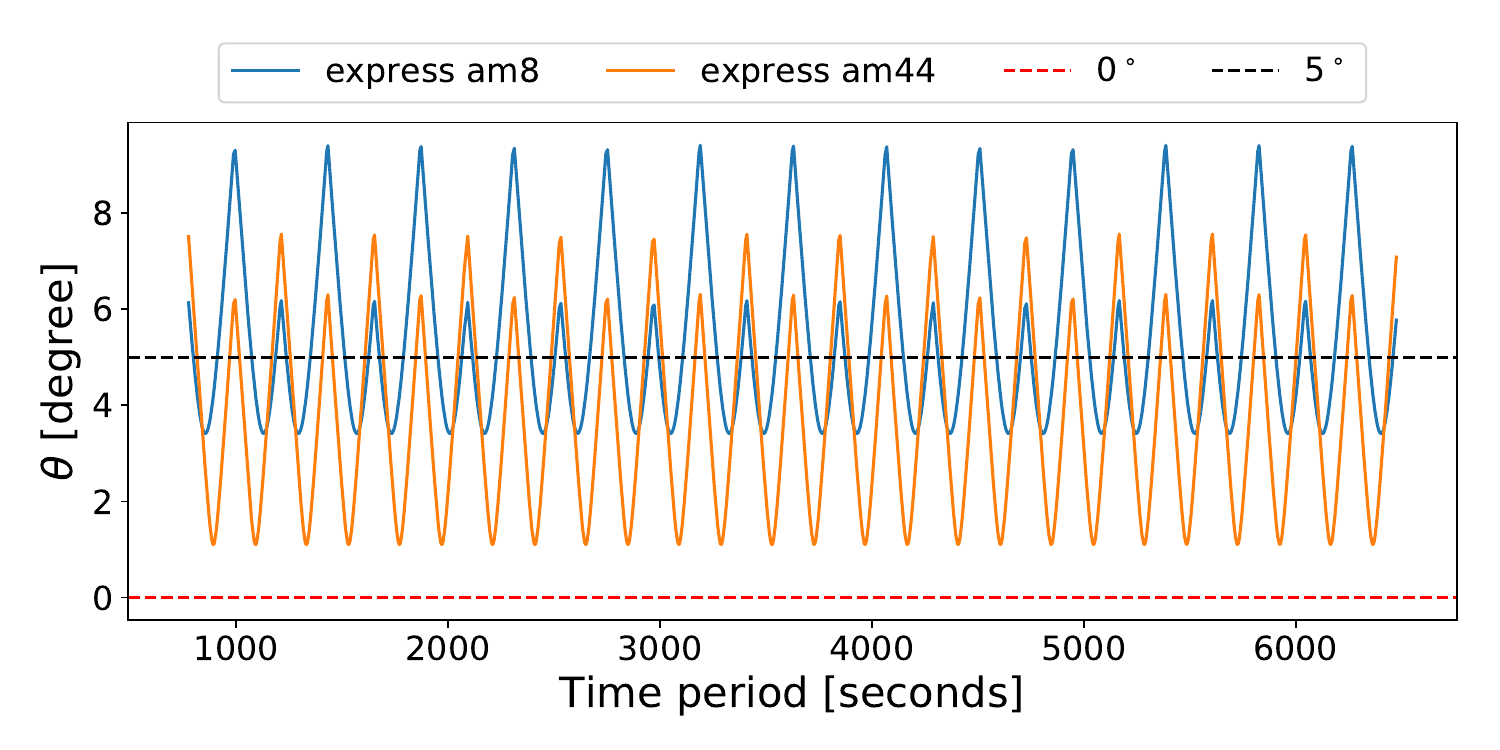}
                    \caption{The angular distance (\autoref{eq:angular_sep} of two Russian geostationary communication satellites that transmit in the L-band and cross below the $5^{\circ}$ boundary.}
                    \label{fig:ang_sep_geo}
                \end{figure}
                
            Besides the navigational satellites, many other satellites reside in a geostationary orbit around the Earth. These satellites generally operate at higher frequencies (K, KU, Ka-band) than their navigational counterparts and belong to not only government agencies but private companies such as Starlink\footnote{\url{https://www.starlink.com/}} OneWeb\footnote{\url{https://oneweb.net/}}, etc... for communication purposes.
        
            In our constructed satellite catalog, we excluded the telecommunication geostationary satellites due to the many unknown variables associated with them, in particular, their modulation and signal behavior. However, we can still track and locate satellites that enter the primary beam's $5^{\circ}$ boundary. In our analysis, we found 17 satellites passed into the $5^{\circ}$, of which two were actively transmitting in the L-band. We show their $\theta$ movement concerning MeerKAT's pointing in  \autoref{fig:ang_sep_geo}.
            
            With the additional positional information, we created a new mask targeting the timestamps $\leq 5^{\circ}$ and combined this mask with the already in use $5^{\circ}$ mask of the Galileo and GLONASS satellites. The additional masking removed $\sim$83\% of the data, which is less overall than the temporal masking but $\sim$56\% more than that of the $5^{\circ}$. In the case of the $\textnormal{FoM}$, we obtained a $C_1$ value of 30855.53 and a $C_2$ value of 0.54. This shows a factor of 3 improvements when compared to only the $5^{\circ}$ case (see  \autoref{tab:chi2_best_fit}).  

            However, we suspect that these two geostationary satellites do not emit within our frequency range of interest (1100-1350 MHz) for two reasons. (1) According to regulations the International Telecommunication Union sets, our frequency range is reserved for navigational satellite emission. (2) When comparing \autoref{fig:ang_sep_geo} with the bottom panel of \autoref{fig:angular_mask_case} and \autoref{fig:chi_fitting_temporal}, it seems that the model can account for all the peaks of emission and overlay and accurate fit. This implies that there is no need to include extra satellite emission in the model.

\bsp	
\label{lastpage}
\end{document}